\documentclass[apj]{emulateapj}
\usepackage{amsmath}
\pdfoutput=1
\newcommand{\etal}{et~al.\/}
\newcommand{\ie}{i.e.\/}
\newcommand{\eg}{e.g.\/}

\shorttitle{Formaldehyde Densitometry of Starburst Galaxies}
\shortauthors{Mangum \etal}

\begin{document}

\title{Formaldehyde Densitometry of Starburst Galaxies:
  Density-Independent Global Star Formation}

\author{Jeffrey G.~Mangum}
\affil{National Radio Astronomy Observatory, 520 Edgemont Road,
  Charlottesville, VA  22903, USA}
\email{jmangum@nrao.edu}

\author{Jeremy Darling}
\affil{Center for Astrophysics and Space Astronomy, Department of Astrophysical
and Planetary Sciences, Box 389, University of Colorado, Boulder, CO 
80309-0389, USA}
\email{jdarling@origins.colorado.edu}

\author{Christian Henkel\altaffilmark{1}}
\affil{Max Planck Instit\"ut f\"ur Radioastronomie, Auf dem H\"ugel
  69, 53121 Bonn, Germany}
\altaffiltext{1}{Also Astronomy Department, Faculty of Science, King Abdulaziz
  University, P.~O.~Box 80203, Jeddah, Saudi Arabia}
\email{chenkel@mpifr-bonn.mpg.de}

\and

\author{Karl M.~Menten}
\affil{Max Planck Instit\"ut f\"ur Radioastronomie, Auf dem H\"ugel
  69, 53121 Bonn, Germany}
\email{kmenten@mpifr-bonn.mpg.de}

\begin{abstract}
Accurate techniques that allow for the derivation of the spatial
density in star formation regions are rare.  A technique that has
found application for the derivation of spatial densities in Galactic
star formation regions utilizes the density-sensitive properties of
the K-doublet transitions of formaldehyde (H$_2$CO).  In this paper,
we present an extension of our survey of the formaldehyde $1_{10}-1_{11}$
($\lambda = 6.2$\,cm) and $2_{11}-2_{12}$ ($\lambda = 2.1$\,cm)
K-doublet transitions of H$_2$CO in a sample of 56 starburst systems
\citep{Mangum2008}.  We have extended the number of galaxies in which
both transitions have been detected from 5 to 13.  We have improved
our spatial density measurements by incorporating kinetic
temperatures based upon NH$_3$ measurements of 11 of the
galaxies with a total of 14 velocity components in our sample
\citep{Mangum2013}.  Our spatial density measurements 
lie in a relatively narrow range from 10$^{4.5}$ to 10$^{5.5}$
cm$^{-3}$.  This implies that the Schmidt-Kennicutt relation between
L$_{IR}$ and M$_{dense}$: (1) is an indication of the dense gas mass
reservoir available to form stars, and (2) is not directly dependent
upon a higher average density driving the star formation process in
the most luminous starburst galaxies.  
We have also used our H$_2$CO measurements to derive two separate
measures of the dense gas mass which are generally smaller, in many
cases by a factor of $10^2-10^3$, than those derived using HCN.  This
disparity suggests that H$_2$CO traces a denser, more compact,
component of the giant molecular clouds in our starburst galaxy
sample.

We also report measurements of the rotationally-excited $\lambda =
6.3$\,cm $^2\Pi_{1/2} J=1/2$ state of OH and the H111$\alpha$ radio
recombination line taken concurrently with our H$_2$CO $1_{10}-1_{11}$
measurements.   
\end{abstract}


\section{Introduction}
\label{intro}

Studies of the abundant CO molecule toward external galaxies have
revealed the large scale structure of the molecular mass in these
objects.  Unfortunately, high optical depth tends to limit the utility
of the abundant CO molecule as a probe of the spatial density and kinetic
temperature in dense gas environments, quantities necessary to assess
the possibility and conditions of star formation \citep{Liszt2010}.
Emission from less-abundant, higher-dipole moment molecules is
better-suited to the task of deriving the spatial density and kinetic
temperature within dense gas environments.  A variety of molecules
which trace dense molecular environments in mainly nearby galaxies have
been studied; see the reviews of \citet[][ for early work]{Henkel1991}
and \citet{Omont2007}.

Formaldehyde (H$_2$CO) has proven to be a reliable density
and kinetic temperature probe in Galactic molecular clouds
\citep{Mangum1993,Mangum1993a,Ginsburg2011}.  Existing measurements of
the H$_2$CO $1_{10}-1_{11}$ and $2_{11}-2_{12}$ K-doublet transitions
in a wide variety of galaxies by \citet{Baan1986,Baan1990,Baan1993},
and \citet{Araya2004} have mainly concentrated on measurements of the 
$1_{10}-1_{11}$ transition.  By collecting a consistent set of
measurements of \textit{both} K-doublet transitions we are able to
derive the spatial density within the extragalactic star formation
regions which comprise this study.  Using the unique density
selectivity of the K-doublet transitions of H$_2$CO we have measured
the spatial density in a sample of galaxies exhibiting starburst
phenomena and/or high infrared luminosity \citep{Mangum2008}.  In the
present 
work we have nearly tripled the number of galaxies within which the
spatial density can be derived from five to thirteen.  In eleven of
these galaxies with a total of fourteen velocity components we have
further improved the quality of our spatial density determination by
applying kinetic temperatures derived from NH$_3$ measurements
\citep{Mangum2013}.  In
\S\ref{H2coProbe} we discuss the specific properties of the H$_2$CO
molecule that make it a good probe of spatial density.
\S\ref{Observations} presents a summary of our observations; \S\ref{Results}
our H$_2$CO, OH, H111$\alpha$, and continuum emission measurement
results; \S\ref{Analysis} analyses of our H$_2$CO, OH, and
H111$\alpha$ measurements, including Large Velocity Gradient (LVG)
model fits and dense gas mass calculations based on our H$_2$CO
measurements.

\section{Formaldehyde as a High Density Probe}
\label{H2coProbe}

The ubiquitous and readily-observed formaldehyde (H$_2$CO) molecule
is a reliable probe of the high density environs of star
formation regions.  Pervasive in the interstellar medium, H$_2$CO
exhibits a moderate abundance which appears to vary by less than an
order-of-magnitude within star formation regions in our own Galaxy
\citep{Mangum1990}.  As a slightly asymmetric rotor molecule, most
H$_2$CO rotational transitions are split into two sub levels.  These
energy levels are designated by a total angular momentum quantum
number, J, the projection of J along the symmetry axis for a limiting
prolate symmetric top, K$_{-1}$, and the projection of J along the
symmetry axis for a limiting oblate symmetric top, K$_{+1}$.  This
splitting leads to two types of transitions: the high-frequency
$\Delta$J = 1, $\Delta$K$_{-1}$ = 0, $\Delta$K$_{+1}$ = $-$1
``P-branch'' transitions and the lower-frequency $\Delta$J = 0,
$\Delta$K$_{-1}$ = 0, $\Delta$K$_{+1}$ = $\pm$1 ``Q-branch''
transitions, popularly known as the ``K-doublet'' transitions 
\citep[see discussion in][]{Mangum1993,McCauley2011}.  The P-branch
transitions are only seen in emission in regions where n($H_2$)
$\gtrsim$ 10$^{4}$ cm$^{-3}$.  In the K-doublet transitions, for
n($H_2$) $\lesssim$ 10$^{5.5}$ cm$^{-3}$, the lower energy states of
the $1_{10}-1_{11}$ through $5_{14}-5_{15}$ transitions become
overpopulated due to a collisional selection effect
\citep{Evans1975,Garrison1975}.  This overpopulation cools the J$\leq$
5 K-doublets to excitation temperatures lower than that of the cosmic
microwave background, causing them to appear in absorption.  For n($H_2$) 
$\gtrsim$ $10^{5.5}$ cm$^{-3}$ and an assumed kinetic temperature of
40 K, this collisional pump is quenched and the
J $\leq$ 5 K-doublets are driven into emission over a wide range of
kinetic temperatures and abundances \citep[see Figure\,1 in][]{Mangum2008}.

Measurements of the spatial density in galactic and extragalactic star
formation regions often rely upon only order-of-magnitude or critical
density estimates.  The collisional excitation of H$_2$CO provides a
direct translation to density that is fairly kinetic temperature
independent.  The use of H$_2$CO as a density probe offers
unprecedented improvement to these more approximate density
determinations, allowing for an actual \textit{measurement} of this
important physical quantity.

\section{Observations}
\label{Observations}

The measurements reported here were made using the National Radio
Astronomy Observatory (NRAO\footnote{The National Radio Astronomy
  Observatory is a facility of the National Science Foundation
  operated under cooperative agreement by Associated Universities,
  Inc.}) Green Bank Telescope (GBT) during the 
periods 2006/06/07-14 \citep[reported in ][]{Mangum2008}, 2007/01/16-04/22,
2007/11/05-11/26, 2008/04/14-09/22, and 2011/05/04-07/16.  Using exactly
the same observing setup as was used for the observations presented in
\citet{Mangum2008}, single-pointing measurements were obtained of the
$1_{10}-1_{11}$ 
\citep[4.829660 GHz;][]{Tucker1971} and $2_{11}-2_{12}$
\citep[14.488479 GHz;][]{Tucker1972} K-doublet 
transitions of H$_2$CO, the H111$\alpha$ radio recombination line
(RRL) at 4.744183\,GHz, and two ($F=1-0$ and $1-1$) of the three 6\,cm 
lines of the rotationally-excited $^2\Pi_{1/2}
J=1/2$ state of OH at 4.750656(3) and 4.765562(3)\,GHz
\citep{Radford1968,Lovas1992} toward a sample of 56
infrared-luminous and/or starburst galaxies (Table~\ref{tab:galaxies}).
Our galaxy sample was chosen to represent both galaxies with measured
H$_2$CO emission or absorption \citep{Baan1993,Araya2004} and
other galaxies with substantial molecular emission deduced from HCN
measurements \citep{Gao2004a}.  The single-beam H$_2$CO 
$1_{10}-1_{11}$, H111$\alpha$, and OH $^2\Pi_{1/2} J=1/2$
transitions were measured simultaneously using 4 spectral windows each
with 50\,MHz of 
bandwidth sampled by 16384 channels.  All of the 4.8\,GHz measurements
($\theta_{B} = 153^{\prime\prime}$) utilized the position switching 
technique with reference position located 30 arcmin west in azimuth
from each galaxy position.  Galaxy and reference integrations were 3
minutes each at 4.8\,GHz.  The H$_2$CO $2_{11}-2_{12}$ transition was
measured using a dual-beam ($\theta_{B} = 51^{\prime\prime}$; beam
separation 330$^{\prime\prime}$ in 
cross-elevation) receiver over 50\,MHz of bandwidth sampled by 16384
channels.  The dual-beam system at 14.5\,GHz allowed for both position
switching or beam nodding.  Galaxy and reference integrations were 2
minutes each at 14.5\,GHz.  The correlator configurations produced
a spectral channel width of 3.052\,kHz, which is approximately 0.2 and
0.08\,km\,s$^{-1}$ at 4.8 and 14.5\,GHz, respectively.

For eighteen of the galaxies listed in Table~\ref{tab:galaxies}
measurements of one or both K-doublet H$_2$CO transitions were
presented in \citet{Mangum2008}.  Additional observations and/or
updated analysis 
(\ie\ modified spectral smoothing) of the K-doublet H$_2$CO
measurements of these eighteen galaxies are included in the results
presented here.

\begin{deluxetable*}{lcccccl} 
\tablewidth{475pt}
\tablecolumns{7}
\tablecaption{Extragalactic Formaldehyde Survey Source List\label{tab:galaxies}}
\tablehead{
\colhead{Galaxy\tablenotemark{a}} & 
\colhead{$\alpha$} & 
\colhead{$\delta$} & 
\colhead{v$_{hel}$\tablenotemark{b}} & 
\colhead{D\tablenotemark{c}} &
\colhead{T$_{dust}$\tablenotemark{d}} &
\colhead{Classification\tablenotemark{e}} \\
& \colhead{(J2000)} &\colhead{(J2000)}& 
\colhead{(km s$^{-1}$)} & \colhead{(Mpc)} & \colhead{(K)} & 
}
\startdata
NGC\,55        & 00:14:54.5  & $-$39:11:19 & 129 & $1.5\pm0.1$ & \nodata &
SB(s)m: edge-on \\
NGC\,253       & 00:47:33.1  & $-$25:17:18 & 251    & $3.4\pm0.2$  & 34 & SAB(s)c \\
IC\,1623       & 01:07:47.2  & $-$17:30:25 & 6028   & $80.7\pm5.7$  & 39 & LIRG \\
NGC\,520       & 01:24:35.3  & $+$03:47:37 & 2281   & $30.5\pm2.1$  & 38 & 
Pec, Pair, Sbrst \\
NGC\,598       & 01:33:54.0  & $+$30:40:07 & $-$179 & $0.9\pm0.3$ & \nodata & SA(s)cd \\
NGC\,660       & 01:43:01.7  & $+$13:38:36 & 856    & $12.2\pm0.9$  & 37 & SB(s)a:pec, HII LINER\\
IR\,01418+1651 & 01:44:30.5  & $+$17:06:09 & 8101   & $109.7\pm7.7$ & \nodata & LIRG\\
NGC\,695       & 01:51:14.9  & $+$22:34:57 & 9769   & $130.2\pm9.1$ & 34 & SO pec:LIRG\\
Mrk\,1027      & 02:14:05.6  & $+$05:10:24 & 9061   & $120.8\pm8.5$ & 37 & I:LIRG\\
NGC\,891       & 02:22:33.4  & $+$42:20:57 & 529    & $9.4\pm0.7$  & 28 & SA(s)b?:sp \\
NGC\,925       & 02:27:16.9  & $+$33:34:35 & 553    & $9.3\pm0.7$  & \nodata & SAB(s)d\\
NGC\,1022      & 02:38:32.7  & $-$06:40:39 & 1503   & $19.2\pm1.3$  & 39 & SB(s)a\\
NGC\,1055      & 02:41:45.2  & $+$00:26:35 & 996    & $13.4\pm0.9$  & 29 & SBb:sp:Sy2 LINER\\
Maffei\,2      & 02:41:55.1  & $+$59:36:15 & $-$17  & $3.1\pm0.2$   & \nodata & SAB(rs)bc\\
NGC\,1068      & 02:42:40.7  & $-$00:00:48 & 1136   & $15.2\pm1.1$  & 40 & SA(rs)b:Sy1/2\\
UGC\,02369     & 02:54:01.8  & $+$14:58:15 & 9262   & $124.8\pm8.8$ & \nodata & DBL\\
NGC\,1144      & 02:55:12.2  & $-$00:11:01 & 8750   & $115.3\pm8.1$ & 32 & S pec\\
NGC\,1365      & 03:33:36.4  & $-$36:08:25 & 1652   & $21.5\pm1.5$  & 32 & SBb(s)b\\
IR\,03359+1523 & 03:38:47.1  & $+$15:32:53 & 10507  & $142.1\pm10.0$ & \nodata & LIRG\\
IC\,342        & 03:46:49.7  & $+$68:05:45 & 31     & $3.8\pm0.3$  & 30 & SAB(rs)cd \\
NGC\,1614      & 04:33:59.8  & $-$08:34:44 & 4847   & $64.2\pm4.5$  & 46 & SB(s)c:pec\\
VIIZw31       & 05:16:46.4  & $+$79:40:13 & 16290  & $220.8\pm15.5$ & 34 & \nodata \\
NGC\,2146      & 06:18:37.7  & $+$78:21:25 & 918    & $16.7\pm1.2$  & 38 & SB(s)ab:pec\\
NGC\,2623      & 08:38:24.1  & $+$25:45:17 & 5535   & $79.4\pm5.6$ & \nodata & LIRG\\
Arp\,55        & 09:15:55.1  & $+$44:19:55 & 11957  & $164.3\pm11.6$ & 36 & Pair \\
NGC\,2903      & 09:32:10.1  & $+$21:30:02 & 556    & $7.4\pm0.5$  & 29 & SB(s)d \\
UGC\,5101      & 09:35:51.6  & $+$61:21:11 & 11810  & $164.3\pm11.5$ & 36 & Sy1.5, LINER \\
M\,82          & 09:55:52.2  & $+$69:40:47 & 203    & $5.9\pm0.4$  & 45 & I0,Sbrst \\
M\,82SW        & 09:55:50.0  & $+$69:40:43 & 203    & \nodata  & \nodata & \nodata  \\
NGC\,3079      & 10:01:57.8  & $+$55:40:47 & 1150   & $20.7\pm1.5$  & 32 & SB(s)c, LINER \\
IR\,10173+0828 & 10:19:59.9  & $+$08:13:34 & 14716  & $206.7\pm14.5$ & \nodata &
Sbrst \\
NGC\,3227      & 10:23:30.7  & $+$19:52:46 & 1111   & $20.3\pm1.4$  & \nodata & SAB(s):pec\\
NGC\,3627      & 11:20:15.0  & $+$12:59:30 & 727    & $6.5\pm0.5$  & 30 & SAB(s)b:LINER \\
NGC\,3628      & 11:20:17.2  & $+$13:35:20 & 847    & $8.5\pm0.6$  & 30 & Sb:pec:sp \\
NGC\,3690      & 11:28:32.2  & $+$58:33:44 & 3121   & $48.5\pm3.4$  & \nodata & Merger \\
NGC\,4631      & 12:42:08.0  & $+$32:32:29 & 606    & $7.6\pm0.5$  & 30 & SB(s)d \\
NGC\,4736      & 12:50:53.0  & $+$41:07:14 & 308    & $4.8\pm0.3$  & \nodata & SA(r)ab;Sy2;LINER \\
Mrk\,231       & 12:56:14.2  & $+$56:52:25 & 12642  & $178.1\pm12.5$ & \nodata & SA(rs)c:pec \\
NGC\,5005      & 13:10:56.2  & $+$37:03:33 & 946    & $19.3\pm1.4$  & 28 & SAB(rs)bc\\
IC\,860        & 13:15:04.1  & $+$24:37:01 & 3866   & $53.8\pm3.8$  & \nodata & Sa, Sbrst \\
NGC\,5194      & 13:29:52.7  & $+$47:11:43 & 463    & $9.1\pm0.6$  & \nodata & SA(s)bc:pec \\
M\,83          & 13:37:00.9  & $-$29:51:57 & 518    & $4.0\pm0.3$  & 31 & SAB(s)c \\
Mrk\,273       & 13:44:42.1  & $+$55:53:13 & 11324  & $160.5\pm11.2$ & 48 & LINER\\
NGC\,5457      & 14:03:12.6  & $+$54:20:57 & 241    & $6.2\pm0.4$  & \nodata & SAB(rs)cd \\
IR\,15107+0724 & 15:13:13.1  & $+$07:13:27 & 3897   & $61.9\pm4.4$  & \nodata & Sbrst \\
Arp\,220       & 15:34:57.1  & $+$23:30:11 & 5434   & $82.9\pm5.8$  & 44 & Pair, Sbrst \\
NGC\,6240      & 16:52:59.0  & $+$02:24:02 & 7339   & $108.8\pm7.6$  & 41 & I0:pec, LINER, Sy2 \\
IR\,17208-0014 &	17:23:21.9  & $-$00:17:00 & 12834  & $183.0\pm12.8$ & 46 & ULIRG\\
IR\,17468+1320 & 17:49:06.7  & $+$13:19:54 & 4881   & $74.1\pm5.2$  & \nodata & LIRG \\
NGC\,6701      & 18:43:12.4  & $+$60:39:12 & 3950   & $59.1\pm4.1$ & 32 & SB(s)a\\
NGC\,6921      & 20:28:28.8  & $+$25:43:24 & 4399   & $63.1\pm4.4$ & 34 & SA(r)0/a\\
NGC\,6946      & 20:34:52.3  & $+$60:09:14 & 48     & $5.5\pm0.4$  & 30 & SAB(rs)cd \\
IC\,5179       & 22:16:09.1  & $-$36:50:37 & 3447   & $48.8\pm3.4$  & 33 & SA(rs)bc\\
NGC\,7331      & 22:37:04.1  & $+$34:24:56 & 821    & $14.4\pm1.0$ & 28 & SA(s)b;LINER\\
NGC\,7479      & 23:04:56.6  & $+$12:19:22 & 2385   & $33.7\pm2.4$  & 36 & SB(s)c;LINER\\
IR\,23365+3604 & 23:39:01.3  & $+$36:21:09 & 19330  & $262.5\pm18.4$ & 45 & ULIRG\\
Mrk\,331       & 23:51:26.7  & $+$20:35:10 & 5422   & $74.9\pm5.2$ & 41 & LIRG\vspace{2pt}
\enddata
\tablenotetext{a}{IR $\equiv$ IRAS (Infrared Astronomical Satellite)
  throughout this paper.}
\tablenotetext{b}{Heliocentric velocity drawn from the
  literature.}
\tablenotetext{c}{NED$^f$ Hubble flow distance corrected for Virgo
  cluster, Great Attractor, and Shapley supercluster.  For NGC\,598 no
  Hubble flow distance available, so NED ``redshift-independent'' distance
assumed.} 
\tablenotetext{d}{From \citet{Gao2004b}, who used IRAS 60 and
  100$\mu$m dust continuum emission ratios with an assumed dust
  emissivity $\propto\nu^{-\beta}$ with $\beta = 1.5$.}
\tablenotetext{e}{From NED, Sbrst = starburst galaxy.}
\tablenotetext{f}{The NASA/IPAC Extragalactic Database (NED) is
  operated by the Jet Propulsion Laboratory, California Institute of
  Technology, under contract with the National Aeronautics and Space
  Administration.}
\end{deluxetable*} 

To calibrate the intensity scale of our measurements, several
corrections need to be considered:
\begin{description}
\item[Opacity:] Historical opacity estimates based on atmospheric
  model calculations using ambient pressure, temperature, and relative
  humidity measurements indicated that $\tau$ at 4.8 and 14.5 GHz was
  $\sim$ 0.01 and 0.05 during our observations (assuming elevation
  $\gtrsim$ 30 degrees).  The opacity corrections
  $\exp(\tau_0\csc(EL))$ are $\lesssim 1.02$ and $\lesssim 1.05$,
  respectively.
\item[Flux:] Assuming point-source emission, one can use the
  current relation \citep[derived from point-source radiometric
    continuum measurements;][]{Maddalena2008} for the aperture 
  efficiency $\eta_A = 0.71 \exp\left(-\left(0.0163
      \nu\mathrm{(GHz)}\right)^2\right)$ to convert antenna
      temperature to flux density.  At 4.8 and 14.5 GHz this yields
      $\eta_A$ = 0.71 and 0.67, respectively. For elevation 90 degrees
      and zero atmospheric opacity, 
  T$_A$/S = $2.846\eta_A$ = 2.0 and 1.9 for 4.8 and 14.5 GHz,
  respectively.  For elevation $\gtrsim 30^\circ$, T$_A$/S = 1.97 and
  1.80 at 4.8 and 14.5 GHz.  These are the K/Jy calibration factors
  used to convert our spectra to flux density assuming point-source
  emission.  Measurements of the standard flux calibration sources
  3C\,48 and 3C\,286 yielded T$_A$/S = $1.95\pm0.04$ and
  $1.83\pm0.04$ at 4.8 and 14.5\,GHz, consistent with the standard
  empirical values.  Note also that since the opacity correction is small,
  $T^*_A = T_A\exp(A\tau_0) 
  \simeq \frac{T_A}{\eta_l} \simeq T_A$, where $\eta_l = 0.99$ for the
  GBT.  Using $\eta_{mb} \simeq 1.32 \eta_A$, we can write the main
  beam brightness temperature as $T_{mb} \simeq
  \frac{T^*_A}{\eta_{mb}}$.
\item[Galaxy Structure:] Several of our galaxies are suspected to have
  high-density structure (measured with high-dipole moment molecules
  like HCN) on scales approaching the size of our beam at 14.5 GHz
  (51$^{\prime\prime}$).  The beam coupling correction necessary to
  account for structure in our 14.5 GHz measurements, relative to our
  point-source assumption, is given by:
\begin{equation}
  f_{coupling} = \frac{\theta^2_B + \theta^2_S}{\theta^2_B}
\label{eq:fcoupling}
\end{equation}
  This correction factor is less than 20\% for $\theta_S \leq
  23^{\prime\prime}$.  With the exception of M\,82, none of the galaxies
  in our sample have measured dense molecular gas structure larger
  than $\sim 20^{\prime\prime}$.  We therefore assume that, with the
  exception of M\,82\footnote{As was discussed in \citet{Mangum2008},
    the source 
    coupling correction for the extended nature of M\,82 results in a
    20\% \textit{decrease} in the measured H$_2$CO
    $1_{10}-1_{11}$/$2_{11}-2_{12}$ K-doublet transition ratio,
    which leads to a corresponding \textit{increase} in the derived
    spatial density.}, all of the H$_2$CO emission reported in these
  measurements is from structures smaller than the primary beams of
  our measurements.  
\item[Absolute Amplitude Calibration:] The GBT absolute amplitude
  calibration is reported to be accurate to 10--15\% at all
  frequencies, limited mainly by temporal drifts in the noise diodes
  used as absolute amplitude calibration standards.  As noted above,
  measurements of the standard flux calibration sources 3C\,48 and
  3C\,286 yielded T$_A$/S = $1.95\pm0.04$ and $1.83\pm0.04$ at 4.8 and
  14.5\,GHz, suggesting that our absolute flux calibration is good to
  $\sim 5$\%.  Relative calibration between our 4.8 and 14.5 GHz
  measurements is assumed to be $\sim 5\%$, which produces a
  negligible impact on density measurements obtained from H$_2$CO line
  ratios. 
\end{description}

\section{Results}
\label{Results}

\subsection{H$_2$CO}
\label{FormResults}

Table\,\ref{tab:h2comeasurements} lists
our H$_2$CO $1_{10}-1_{11}$ and $2_{11}-2_{12}$ measurement results,
including updated results for the eighteen galaxies in our present
sample that were presented in \citet{Mangum2008}.  Since
additional observations and/or updated analysis of these eighteen
galaxies are included in the analysis presented in this work, we list
these updated H$_2$CO measurement results in italics in
Table~\ref{tab:h2comeasurements}.  H$_2$CO measurements for
thirty-eight galaxies and one galaxy offset position (M\,82SW) are
completely new measurements.  The new H$_2$CO measurements presented
in this paper have nearly tripled the number of galaxies within 
which both H$_2$CO K-doublet transitions have been detected from five
to thirteen.  For each detection we list the peak intensity,
heliocentric central line velocity, velocity width (FW(ZI/HM)), and
integrated intensity derived from direct channel-by-channel
integration of and gaussian fits to each of the line profiles.
Uncertainties expressed as one-sigma are listed in parentheses for
each quantity in Tables~\ref{tab:h2comeasurements}.

\begin{figure*}
\centering
\includegraphics[scale=0.75,trim=50pt 40pt 20pt 60pt,clip=true]{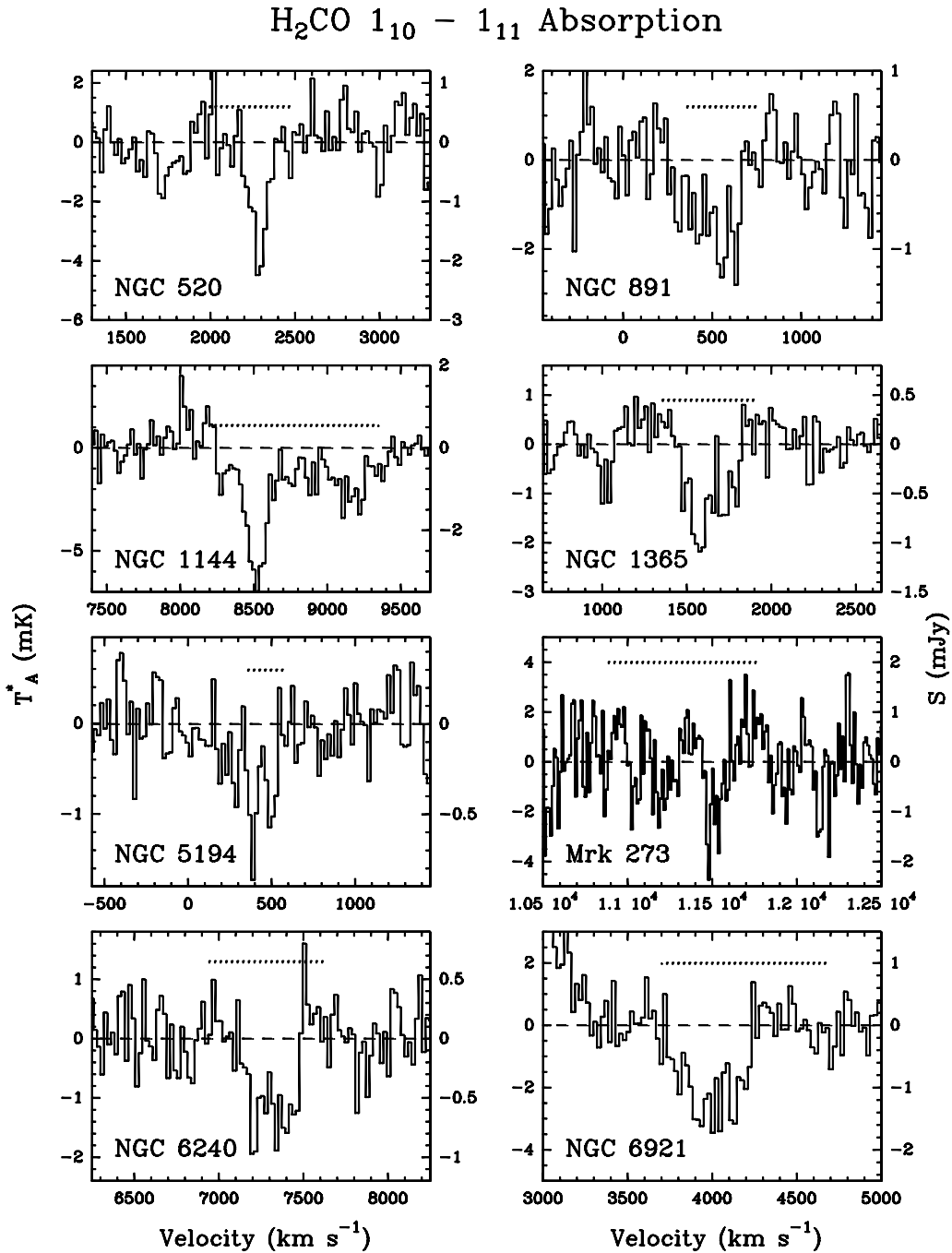}
\caption{H$_2$CO $1_{10}-1_{11}$ spectra for galaxies
  toward which the $2_{11}-2_{12}$ was not detected.  The dotted line
  above each spectrum indicates the full-width zero intensity (FWZI)
  CO linewidth.}
\label{fig:ExgalFormSingles-OnePage}
\end{figure*}

\begin{figure*}
\resizebox{\hsize}{!}{
\includegraphics[trim=15mm 15mm 15mm 30mm, clip, scale=0.40]{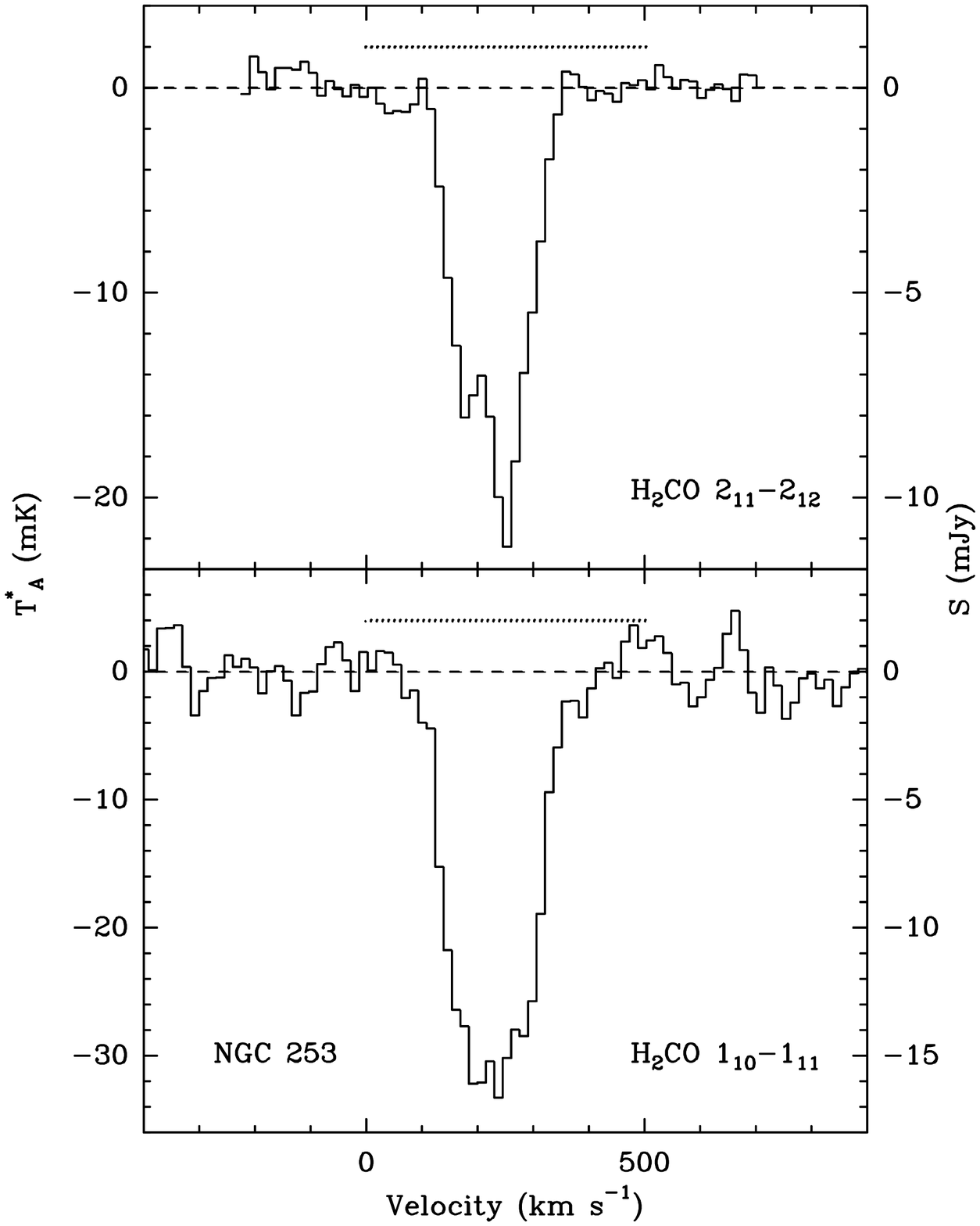}
\includegraphics[trim=15mm 15mm 15mm 30mm, clip, scale=0.40]{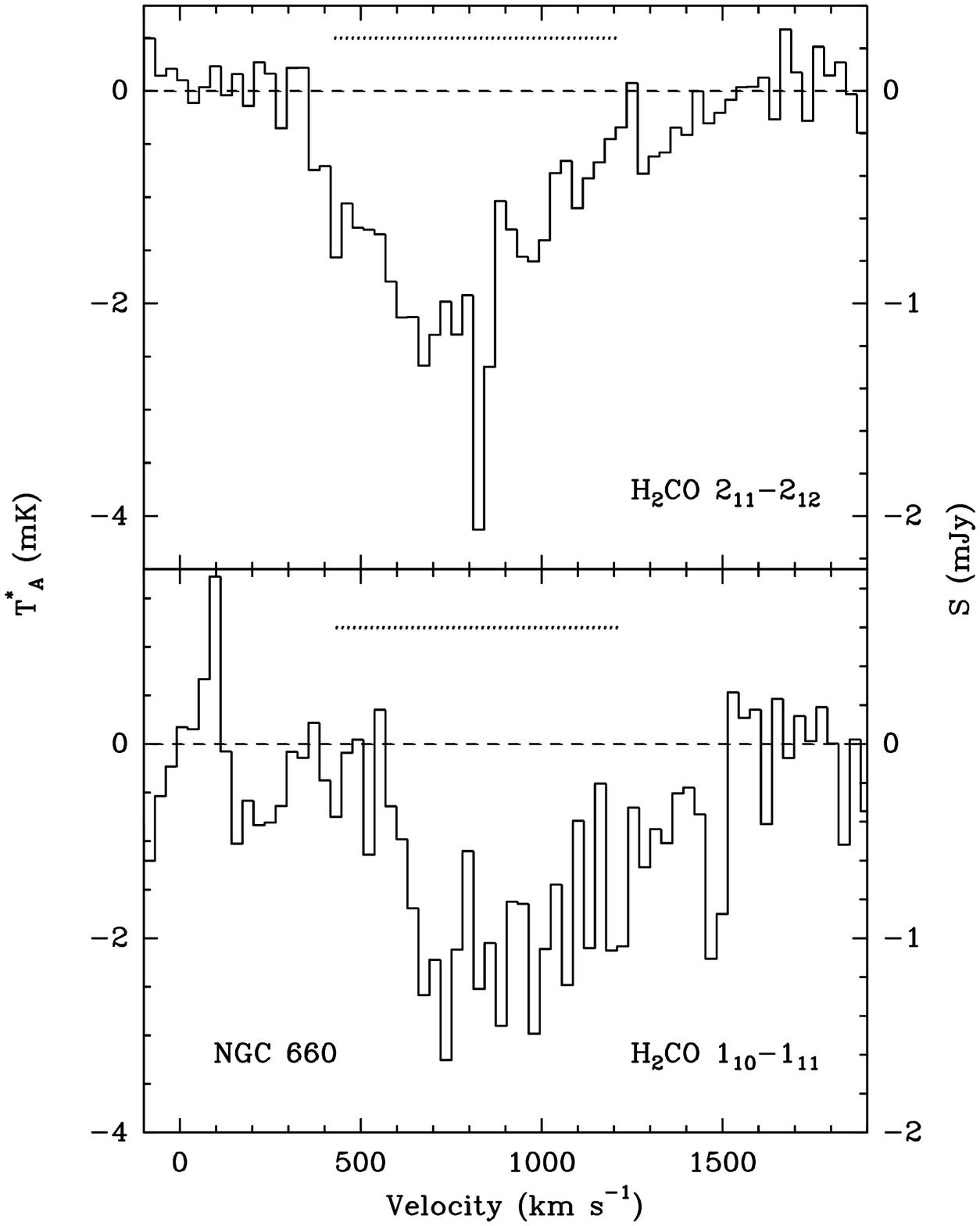}} \\
\resizebox{\hsize}{!}{
\includegraphics[trim=15mm 15mm 15mm 30mm, clip, scale=0.40]{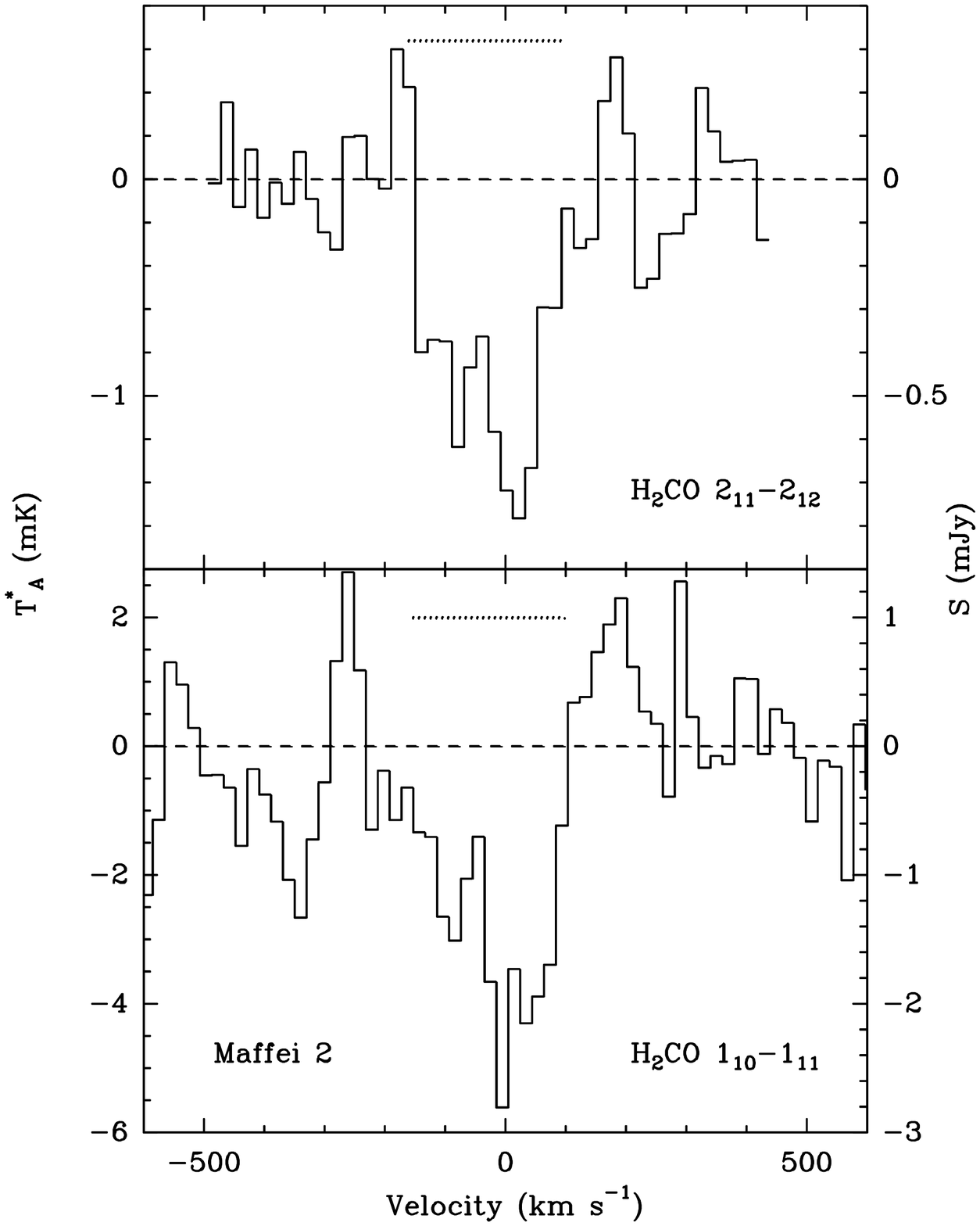}
\includegraphics[trim=15mm 15mm 15mm 30mm, clip, scale=0.40]{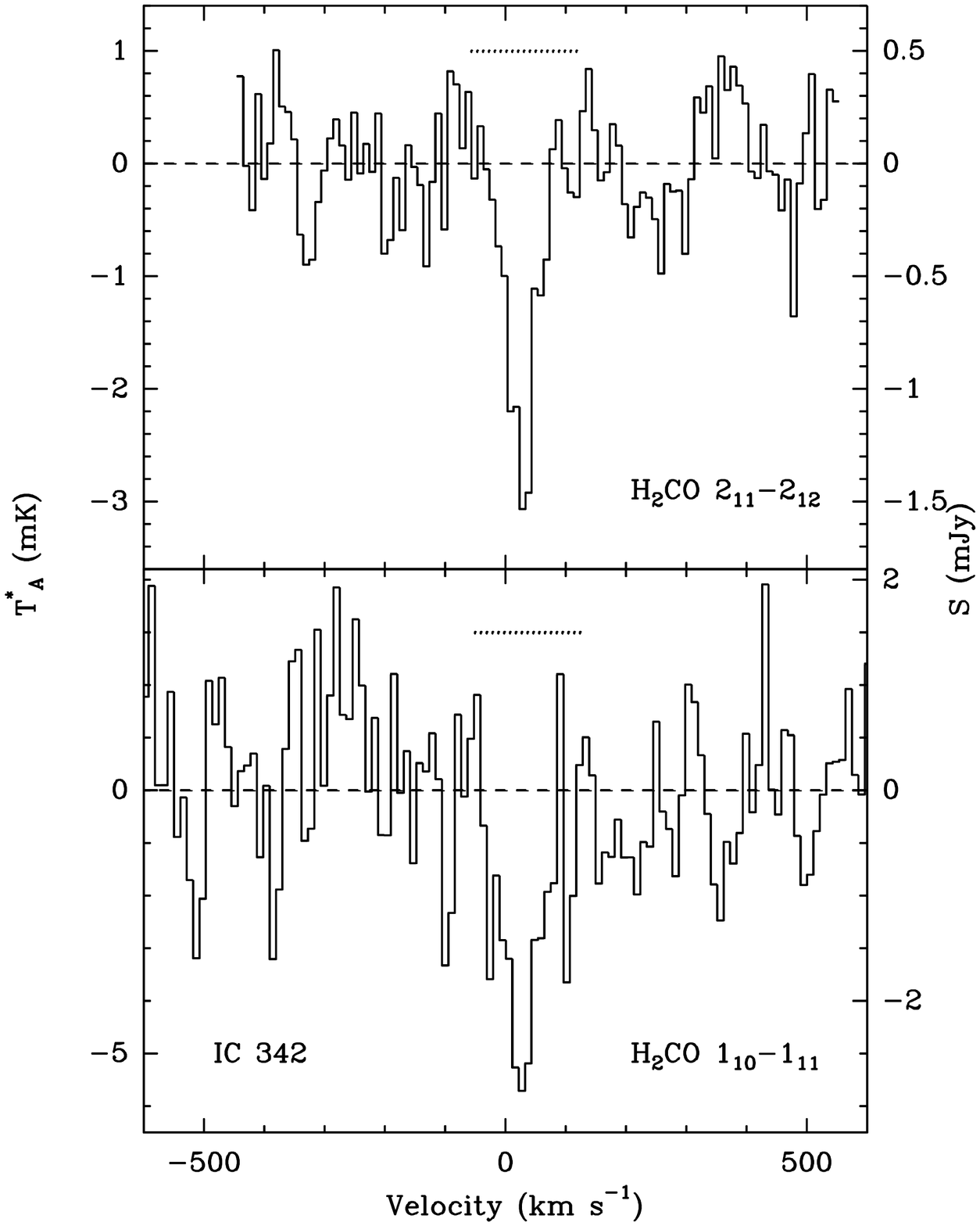}}
\caption{H$_2$CO $2_{11}-2_{12}$ (panel top) and
  $1_{10}-1_{11}$ (panel bottom) spectra of NGC\,253 (top left),
  NGC\,660 (top right), Maffei\,2 (bottom left), and IC\,342 (bottom
  right).  The dotted line within each spectrum indicates the FWZI CO
  linewidth.} 
\label{fig:NGC253NGC660Maffei2IC342FormSpec}
\end{figure*}

\begin{figure*}
\resizebox{\hsize}{!}{
\includegraphics[trim=15mm 15mm 15mm 30mm, clip, scale=0.40]{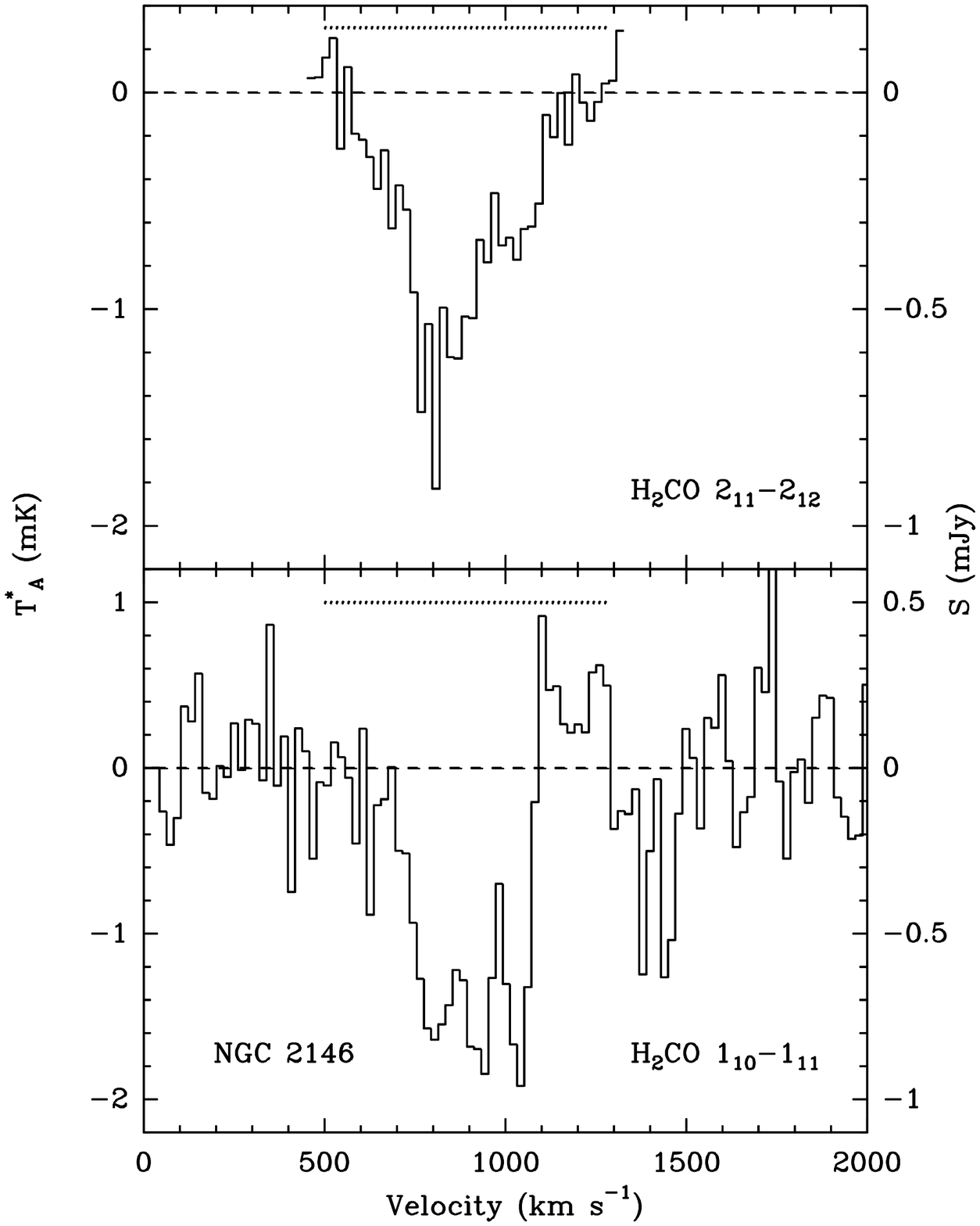}
\includegraphics[trim=15mm 15mm 15mm 30mm, clip, scale=0.40]{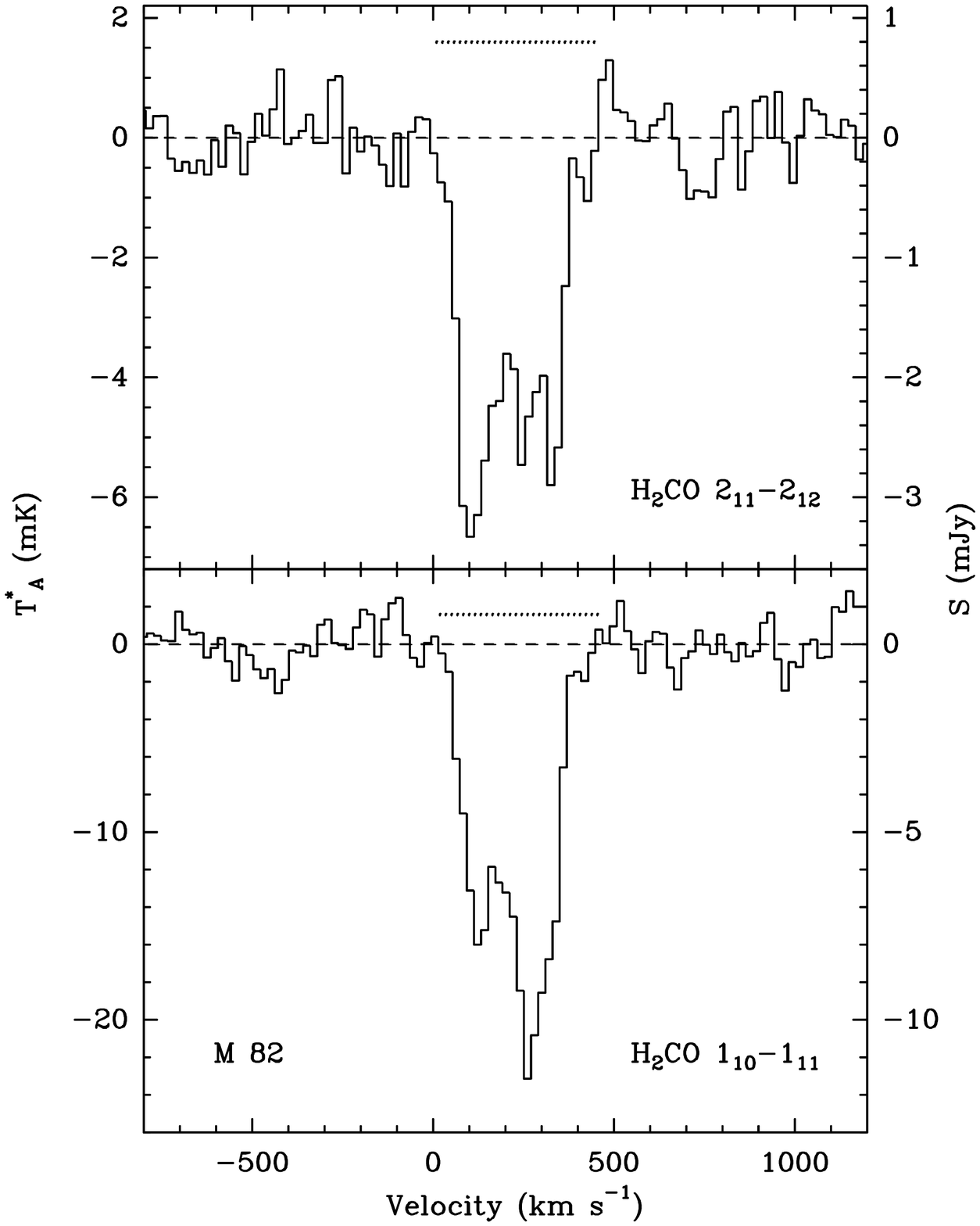}} \\
\resizebox{\hsize}{!}{
\includegraphics[trim=15mm 15mm 15mm 30mm, clip, scale=0.40]{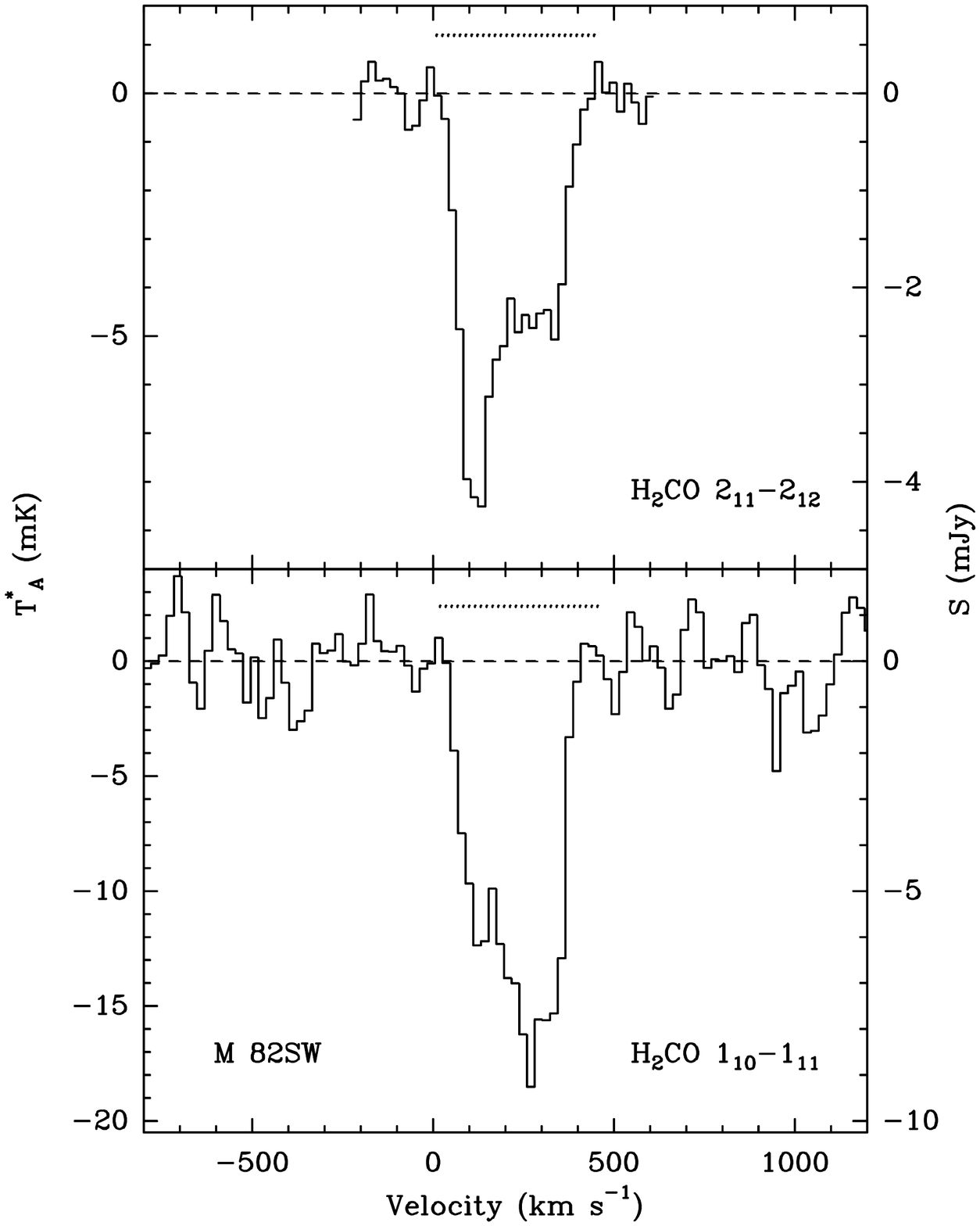}
\includegraphics[trim=15mm 15mm 15mm 30mm, clip, scale=0.40]{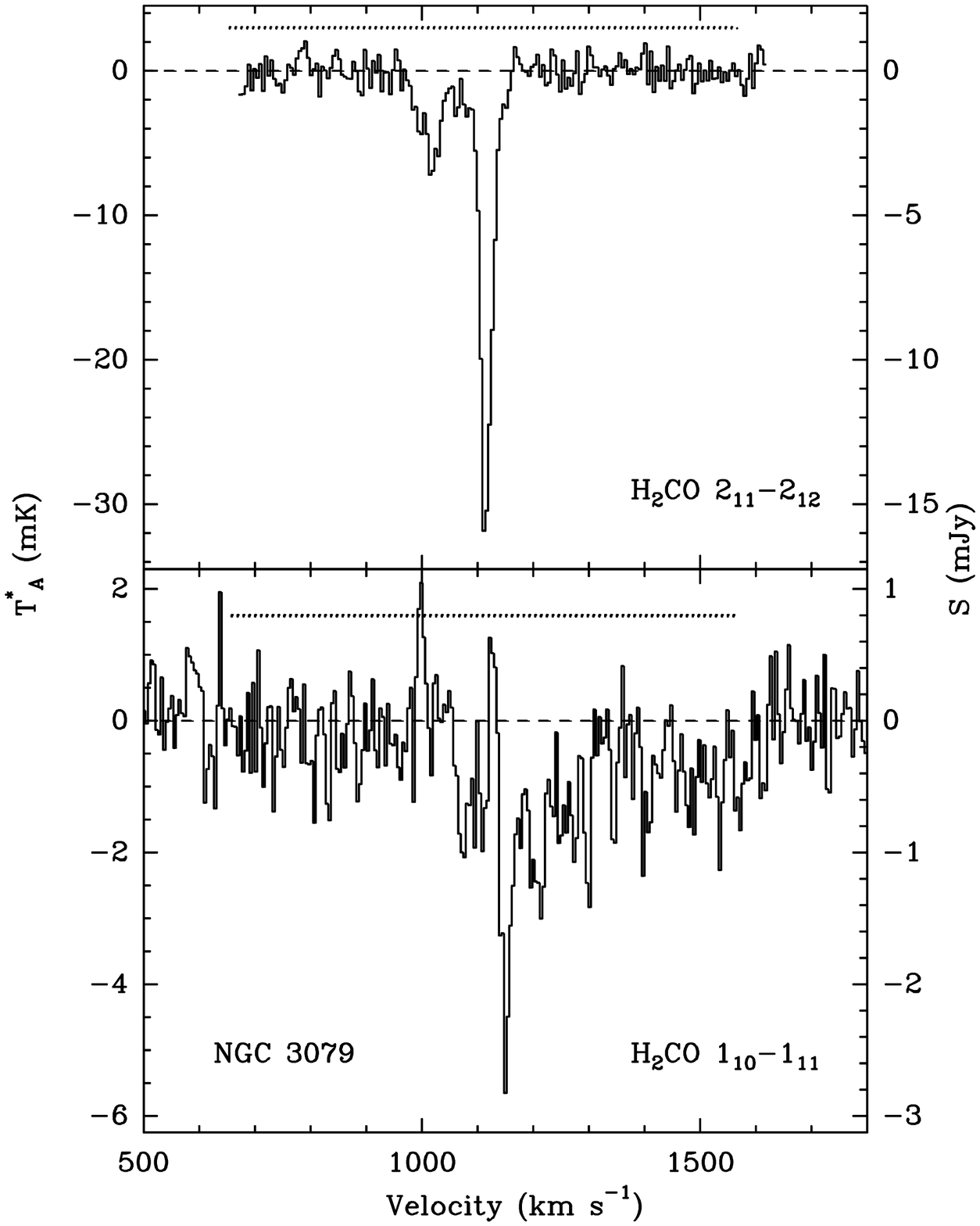}}
\caption{H$_2$CO $2_{11}-2_{12}$ (panel top) and
  $1_{10}-1_{11}$ (panel bottom) spectra of NGC\,2146 (top left),
  the kinematical center of M\,82 (top right), M\,82SW, which is the
  ($-12^{\prime\prime}$, $-4^{\prime\prime}$) offset position from the
  kinematical center of M\,82 (bottom left), and NGC\,3079 (bottom
  right).  The dotted line within each spectrum indicates the FWZI CO
  linewidth.} 
\label{fig:NGC2146M82M82SWNGC3079FormSpec}
\end{figure*}

\begin{figure*}
\resizebox{\hsize}{!}{
\includegraphics[trim=15mm 15mm 15mm 30mm, clip, scale=0.40]{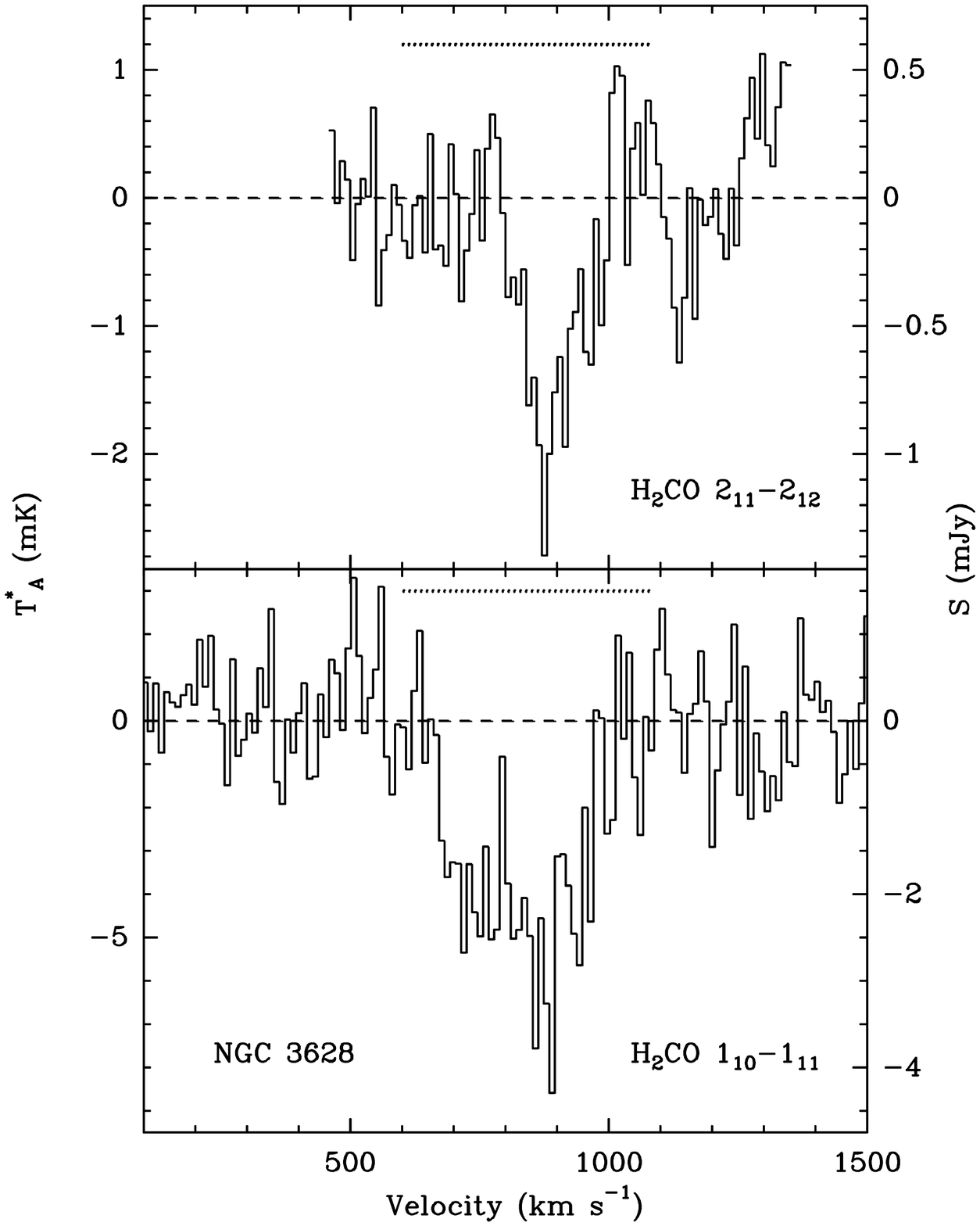}
\includegraphics[trim=15mm 15mm 15mm 30mm, clip, scale=0.40]{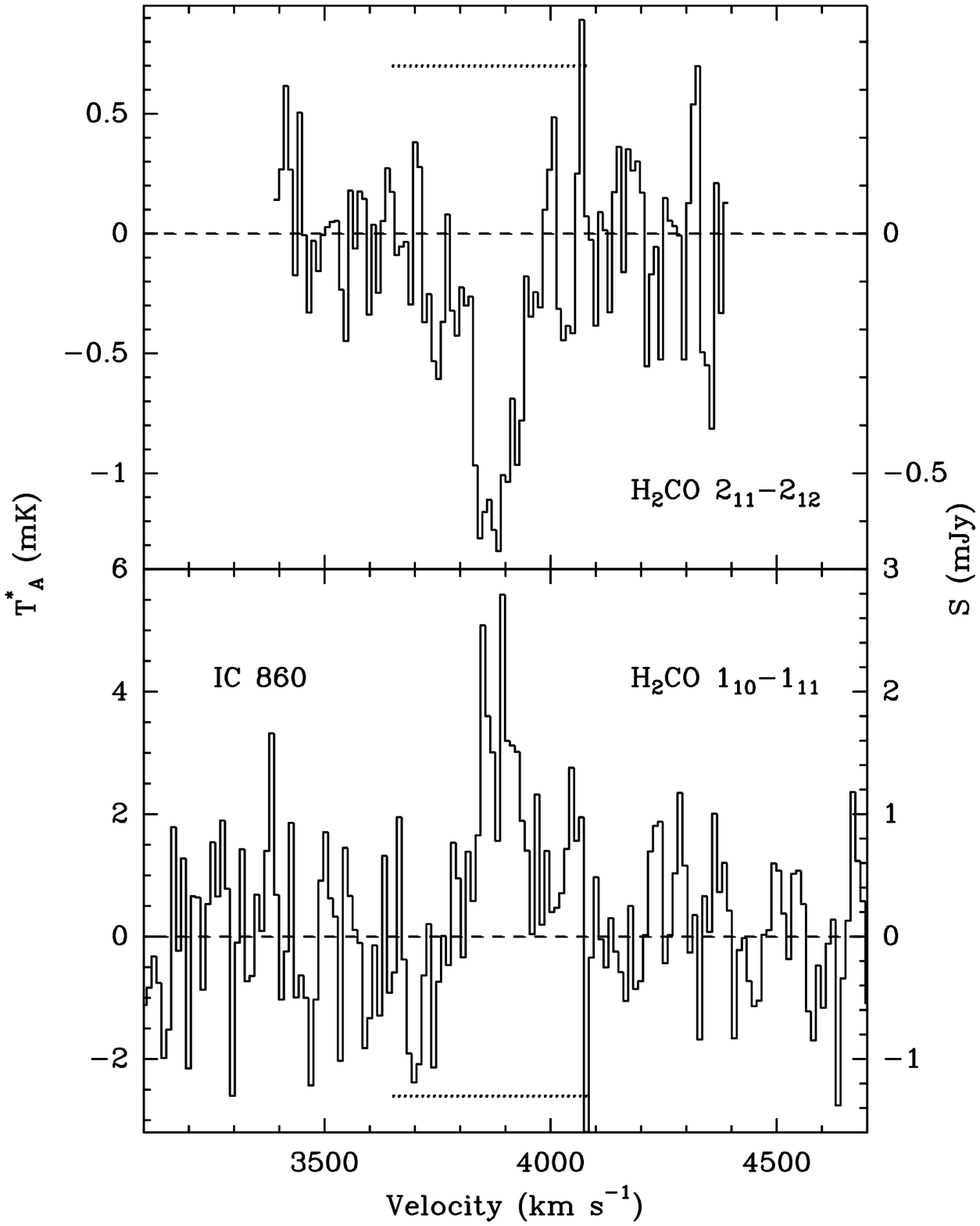}} \\
\resizebox{\hsize}{!}{
\includegraphics[trim=15mm 15mm 15mm 30mm, clip, scale=0.40]{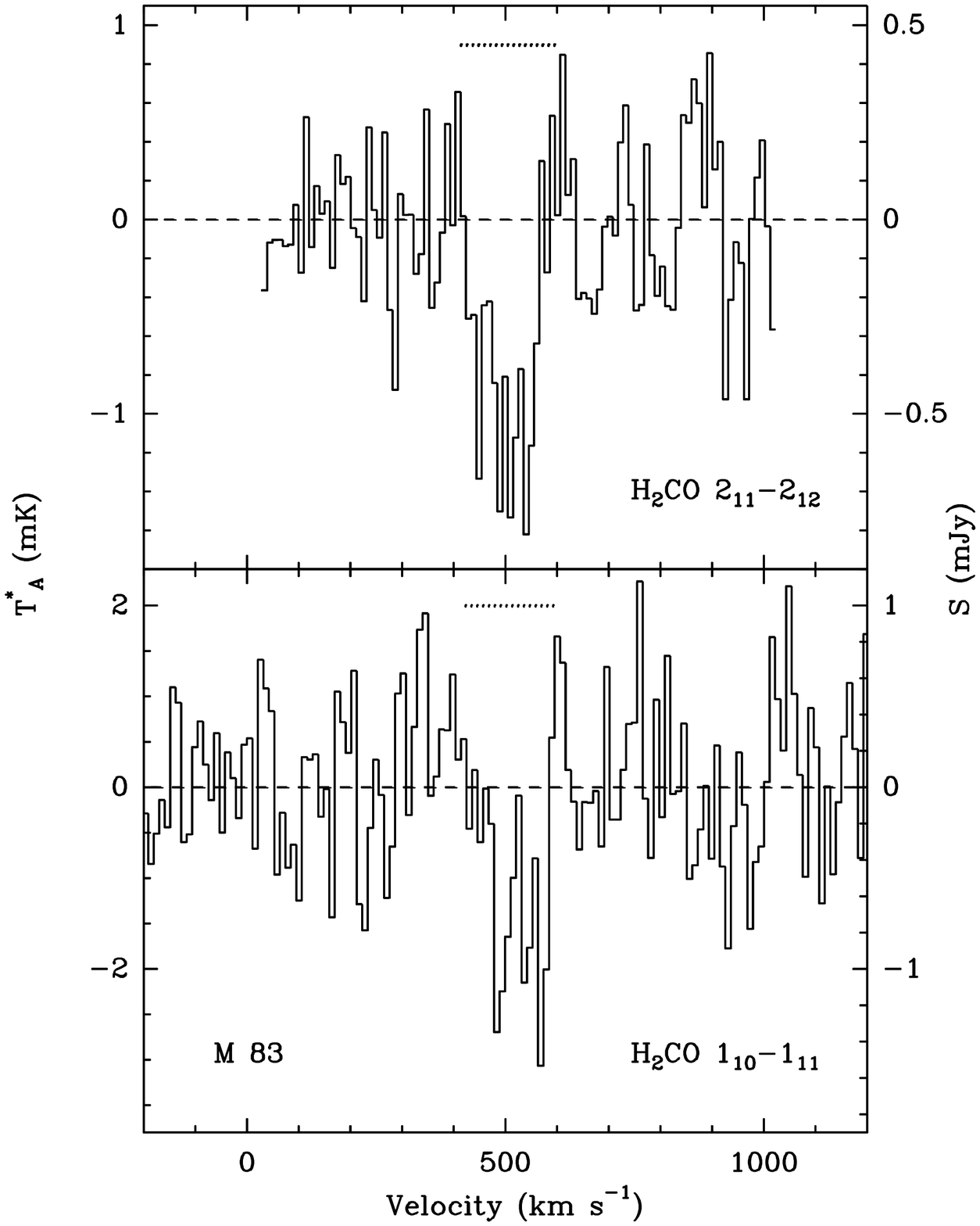}
\includegraphics[trim=15mm 15mm 15mm 30mm, clip, scale=0.40]{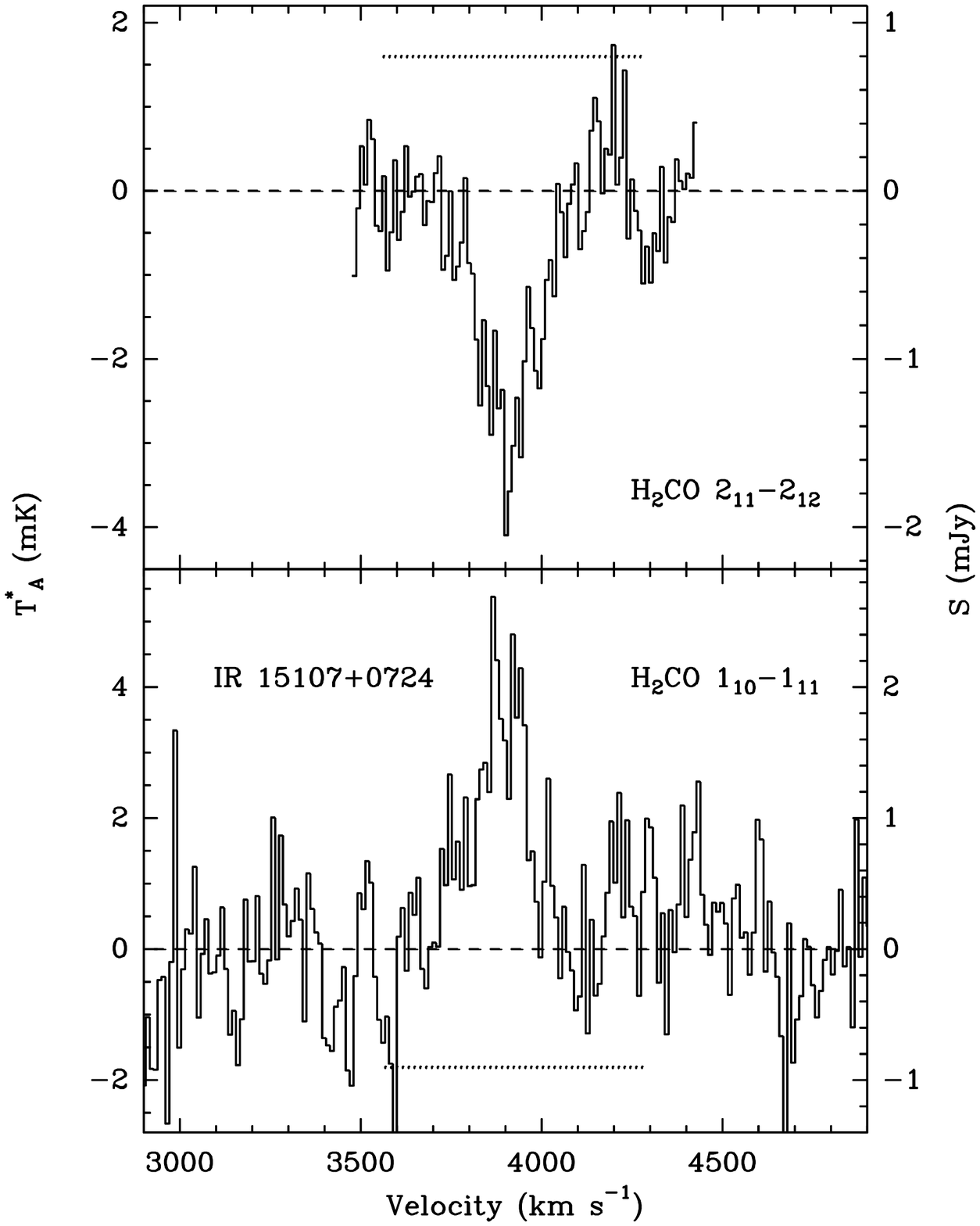}}
\caption{H$_2$CO $2_{11}-2_{12}$ (panel top) and
  $1_{10}-1_{11}$ (panel bottom) spectra of NGC\,3628 (top left),
  IC\,860 (top right), M\,83 (bottom left), and IR\,15107+0724 (bottom
  right).  The dotted line within each spectrum indicates the FWZI CO
  linewidth.} 
\label{fig:NGC3628IC860M83IR15107FormSpec}
\end{figure*}

\begin{figure*}
\centering
\includegraphics[trim=15mm 15mm 15mm 30mm, clip, scale=0.40]{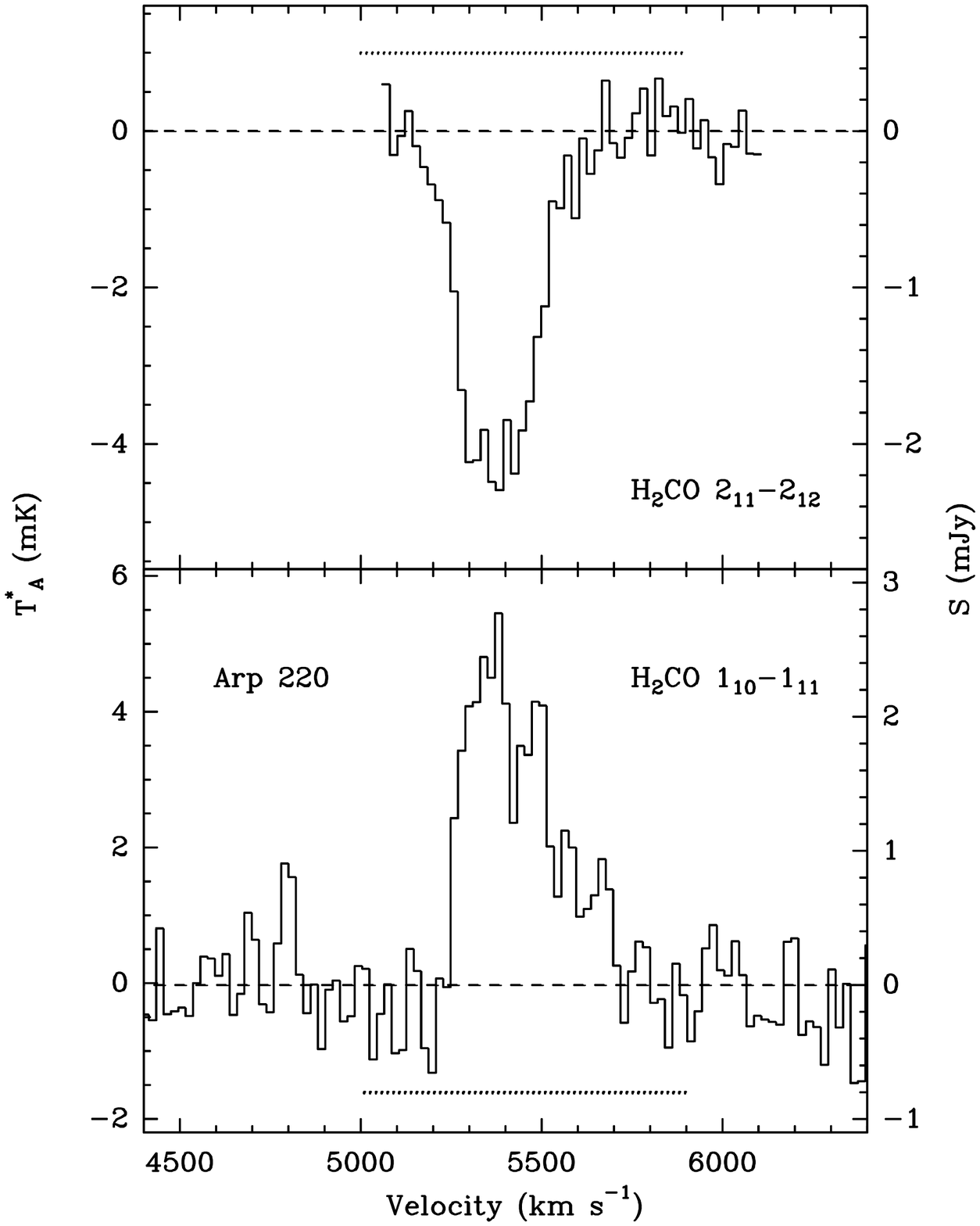}
\includegraphics[trim=15mm 15mm 15mm 30mm, clip, scale=0.40]{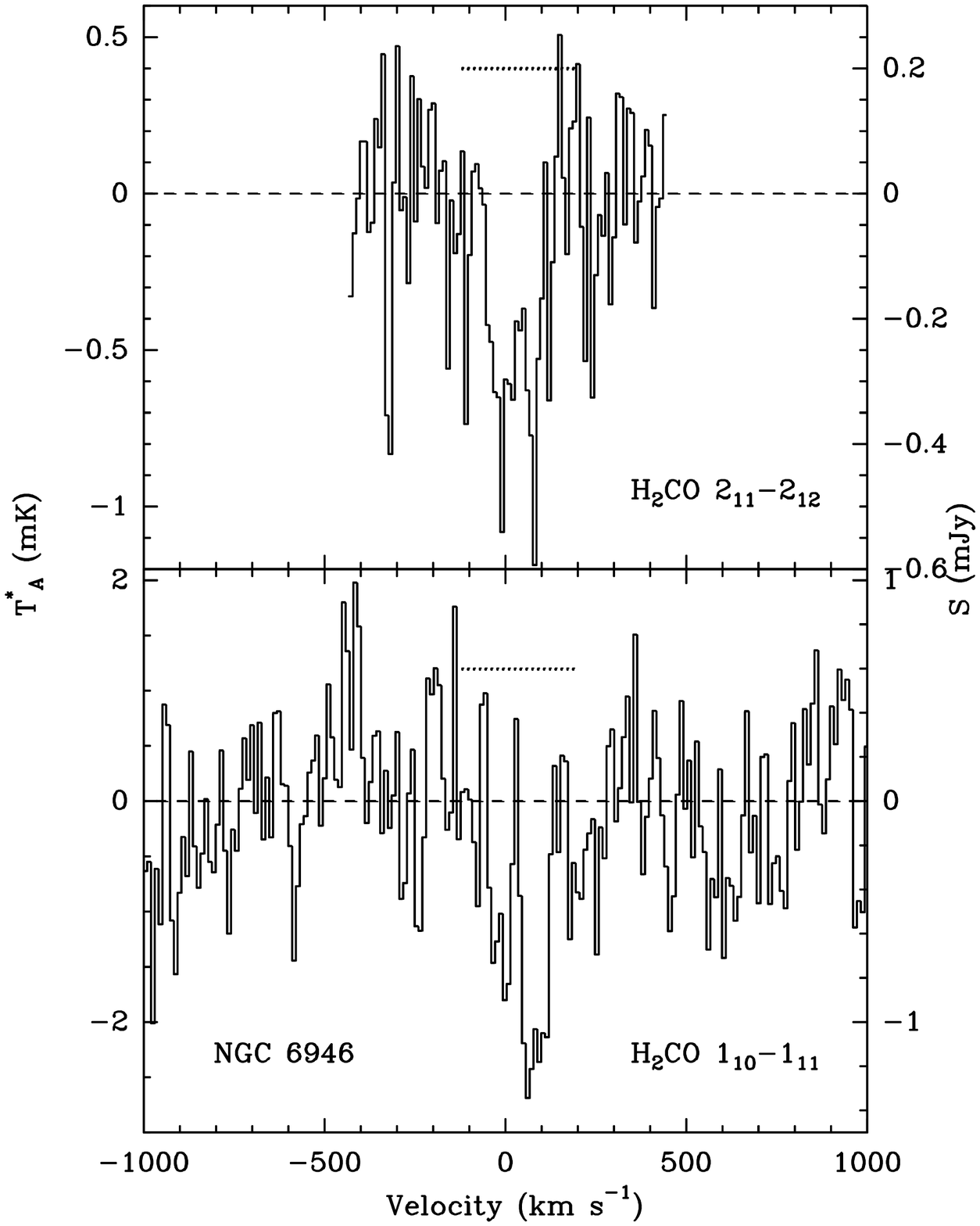}
\caption{H$_2$CO $2_{11}-2_{12}$ (panel top) and
  $1_{10}-1_{11}$ (panel bottom) spectra of Arp\,220 (left) and
  NGC\,6946 (right).  The dotted line within each spectrum indicates
  the FWZI CO linewidth.} 
\label{fig:Arp220NGC6946FormSpec}
\end{figure*}

Spectra for the galaxies detected in only the
$1_{10}-1_{11}$ or $2_{11}-2_{12}$ transition are displayed in
Figure~\ref{fig:ExgalFormSingles-OnePage}.
NGC\,253, NGC\,660, Maffei\,2, IC\,342 (all in
Figure~\ref{fig:NGC253NGC660Maffei2IC342FormSpec}), NGC\,2146, M\,82,
M\,82SW, NGC\,3079 (all in
Figure~\ref{fig:NGC2146M82M82SWNGC3079FormSpec}), NGC\,3628, IC\,860,
M\,83, IR\,15107+0724 (all in
Figure~\ref{fig:NGC3628IC860M83IR15107FormSpec}), Arp\,220, and
NGC\,6946 (both in Figure~\ref{fig:Arp220NGC6946FormSpec}) were all
detected in both H$_2$CO transitions.  These spectra have been
gaussian smoothed to the spectral channel widths quoted in
Table~\ref{tab:h2comeasurements} to both increase the channel-specific
signal-to-noise ratio of our measurements and more closely match other
molecular spectral line measurements of these galaxies
\citep[\ie\ ][]{Gao2004a}.

%
%

\tabletypesize{\scriptsize}

\begin{deluxetable*}{lllrrrr} 
\tablewidth{415pt}
\tablecolumns{7}
\tablecaption{H$_2$CO Measurements\tablenotemark{a}\label{tab:h2comeasurements}}
\tablehead{
\colhead{Galaxy} & 
\colhead{Transition\tablenotemark{b}} & 
\colhead{Fit\tablenotemark{c}} &
\colhead{$T^*_A$} & 
\colhead{v$_{hel}$\tablenotemark{d,e}} & 
\colhead{FW(HM/ZI)\tablenotemark{e,f}} & 
\colhead{$\int$T$^*_A$dv\tablenotemark{g}} \\
&&& \colhead{(mK)} & 
\colhead{(km s$^{-1}$)} & 
\colhead{(km s$^{-1}$)} & 
\colhead{(mK km s$^{-1}$)}
}
\startdata
NGC\,55        & $2_{11}-2_{12}$ & D10 & (0.8) & \nodata & \nodata & 
              \nodata \\

NGC\,253       & $\mathit{1_{10}-1_{11}}$ & D15 & $-$32.8(1.8) &
              230.1 & 349.1 & $-$5773.7(299.5) \\
              & & G15 & $-$34.3(1.3) & 227.0(3.0) & 163.0(7.0) &
              $-$5947.9(339.4) \\ 

              & $2_{11}-2_{12}$   & D15 & $-$20.2(0.4)  & 237.8
              & 242.9    & $-$2542.3(51.6) \\
              & & G15 & $-$11.4(1.0)  &
              169.1(3.1) & 58.3(6.3) & $-$708.2(99.5) \\
              &                & G15 & $-$18.7(0.6) &
              252.6(2.9) & 93.1(6.3) & $-$1847.8(136.4) \\

IC\,1623       & $1_{10}-1_{11}$ & D10 & (1.3) & \nodata & \nodata & 
              \nodata \\ 

NGC\,520       & $\mathit{1_{10}-1_{11}}$ & D20 & $-$4.4(0.9) & 2301.2
              & 215.4 & $-$449.1(93.9)  \\
              & & G20 & $-$4.0(0.4) & 2286.6(4.8) & 106.8(11.6) &
              $-$451.9(64.3) \\ 

              & $\mathit{2_{11}-2_{12}}$ & D10 & (0.5) &
              \nodata & \nodata & \nodata \\

NGC\,598       & $1_{10}-1_{11}$ & D10 & (0.7) & \nodata & \nodata & 
              \nodata \\ 

              & $2_{11}-2_{12}$ & D10 & (0.3) & \nodata & \nodata 
              & \nodata \\

NGC\,660       & $\mathit{1_{10}-1_{11}}$ & D30 & $-$3.2(0.6) & 1041.6
              & 976.6 & $-$1511.9(261.7) \\ 
              & & G30 & $-$2.2(0.2) & 923.3(47.3) & 765.0(135.5) &
              $-$1758.2(366.4) \\ 

              & $\mathit{2_{11}-2_{12}}$ & D30 & $-$3.7(0.4) & 871.6 & 1060.8 &
              $-$1280.8(223.5) \\
              & & G30 & $-$2.2(0.2) & 754.3(19.1) & 547.2(47.8) &
              $-$1257.6(140.3) \\

IR\,01418$+$1651& $1_{10}-1_{11}$ & D20 & (0.6) & \nodata & \nodata &
              \nodata \\ 

              & $2_{11}-2_{12}$ & D10 & (0.3) & \nodata & \nodata &
              \nodata \\ 

NGC\,695       & $1_{10}-1_{11}$ & D10 & (0.8) & \nodata & \nodata &
              \nodata \\  

              & $2_{11}-2_{12}$ & D10 & (0.5) & \nodata & \nodata &
              \nodata \\ 

Mrk\,1027      & $1_{10}-1_{11}$ & D10 & (1.3) & \nodata & \nodata &
              \nodata \\ 

NGC\,891       & $\mathit{1_{10}-1_{11}}$ & D20 & $-$2.8(0.8) & 476.3 &
              434.8 & $-$538.9(160.1) \\ 
              & & G20 & $-$1.7(0.3) & 535.0(48.8) & 413.6(145.2) &
              $-$726.6(280.4) \\ 

              & $2_{11}-2_{12}$ & D10 & (0.3) & \nodata & \nodata 
              & \nodata \\ 

NGC\,925       & $1_{10}-1_{11}$ & D10 & (1.4) & \nodata & \nodata &
              \nodata \\ 

NGC\,1022      & $1_{10}-1_{11}$ & D10 & (1.2) & \nodata & \nodata &
              \nodata \\

              & $2_{11}-2_{12}$ & D10 & (0.7) & \nodata & \nodata &
              \nodata \\

NGC\,1055      & $1_{10}-1_{11}$ & D10 & (1.2) & \nodata & \nodata &
              \nodata \\

              & $2_{11}-2_{12}$ & D10 & (0.3) & \nodata & \nodata &
              \nodata \\ 

Maffei\,2      & $1_{10}-1_{11}$ & D20 & $-$5.5(0.9) & $-$64.0 & 354.5
              & $-$756.5(144.7) \\ 
              & & G20 & $-$2.1(0.6) & $-$98.1(45.5) & 127.5(85.2) &
              $-$284.9(208.7) \\
              & & G20 & $-$4.5(1.0) & 24.7(17.1) & 102.1(30.3) &
              $-$493.2(180.2) \\

              & $2_{11}-2_{12}$ & D20 & $-$1.4(0.2) & $-$15.5 & 303.1
              & $-$220.5(24.7) \\ 
              & & G20 & $-$0.9(0.2) & $-$90.7(12.8) & 78.9(29.0) &
              $-$78.1(33.2) \\
              & & G20 & $-$1.4(0.2) & 18.3(9.1) & 91.3(21.8) &
              $-$136.0(36.9) \\

NGC\,1068      & $1_{10}-1_{11}$ & D10 & (1.1) & \nodata & \nodata &
              \nodata \\ 

              & $2_{11}-2_{12}$ & D10 & (0.4) & \nodata & \nodata &
              \nodata \\ 

UGC\,02369     & $1_{10}-1_{11}$ & D10 & (1.1) & \nodata & \nodata &
              \nodata \\ 

NGC\,1144      & $1_{10}-1_{11}$ & D20 & $-$6.9(0.7) & 8838.7 & 1214.4
              & $-$2251.6(417.6) \\
              & & G20 & $-$2.1(0.7) & 8281.8(8.6) & 51.6(20.1) &
              $-$116.8(60.4) \\
              & & G20 & $-$6.2(0.4) & 8517.4(5.4) & 158.7(13.2) &
              $-$1044.8(113.0) \\
              & & G20 & $-$1.8(0.2) & 9088.9(39.0) & 609.0(119.6) &
              $-$1186.0(275.9) \\

              & $2_{11}-2_{12}$ & D10 & (0.4) & \nodata & \nodata &
              \nodata \\

NGC\,1365      & $1_{10}-1_{11}$ & D20 & $-$2.1(0.4) & 1637.6 & 398.4 &
              $-$410.0(77.1) \\
              & & G20 & $-$1.6(0.3) & 1608.1(21.8) & 266.9(57.9) &
              $-$449.3(121.8) \\

              & $2_{11}-2_{12}$ & D10 & (2.0) & \nodata & \nodata &
              \nodata \\

IR\,03359$+$1523 & $1_{10}-1_{11}$ & D10 & (1.3) & \nodata & \nodata &
                \nodata \\ 

IC\,342        & $\mathit{1_{10}-1_{11}}$ & D10 & $-$5.6(1.7) & 22.1 &
              116.7 & $-$391.1(94.8) \\ 
              & & G10 & $-$5.2(0.8) & 25.0(4.9) & 67.9(11.5) &
              $-$372.2(83.4) \\

              & $\mathit{2_{11}-2_{12}}$ & D10 & $-$2.8(0.4) & 23.5 &
              109.8 & $-$138.8(22.7) \\ 
              & & G10 & $-$2.6(0.2) & 27.4(2.1) & 51.1(4.9) &
              $-$139.1(17.6) \\ 

NGC\,1614      & $1_{10}-1_{11}$ & D10 & (1.5) & \nodata & \nodata &
              \nodata \\

VIIZw31       & $1_{10}-1_{11}$ & D10 & (1.1) & \nodata & \nodata &
              \nodata \\

              & $2_{11}-2_{12}$ & D10 & (0.3) & \nodata & \nodata &
              \nodata \\ 
NGC\,2146      & $1_{10}-1_{11}$ & D20 & $-$1.9(0.3) & 883.9 & 436.0 &
              $-$483.2(68.7) \\  
              & & G20 & $-$1.7(0.2) & 867.3(13.4) & 239.8(36.4) &
              $-$430.3(77.3) \\
              & & G20 & $-$1.6(0.4) & 1038.5(5.3) & 49.6(14.5) &
              $-$83.8(31.4) \\

              & $2_{11}-2_{12}$ & D20 & $-$1.7(0.1) & 869.4 & 569.1 &
              $-$366.0(19.6) \\
              & & G20 & $-$1.2(0.1) & 824.3(8.9) & 231.8(23.3) &
              $-$296.8(35.2) \\
              & & G20 & $-$0.6(0.1) & 1046.9(12.3) & 109.5(29.4) &
              $-$65.3(22.1) \\

NGC\,2623      & $1_{10}-1_{11}$ & D10 & (1.0) & \nodata & \nodata 
              & \nodata \\

              & $2_{11}-2_{12}$ & D10 & (0.4) & \nodata & \nodata
              & \nodata \\ 

Arp\,55        & $\mathit{1_{10}-1_{11}}$ & D10 & (2.6) &
              \nodata & \nodata & \nodata \\  

NGC\,2903      & $\mathit{1_{10}-1_{11}}$ & D10 & (0.3) & \nodata & \nodata &
              \nodata \\

              & $2_{11}-2_{12}$ & D10 & (0.2) & \nodata & \nodata &
              \nodata \\

UGC\,05101    & $\mathit{1_{10}-1_{11}}$ & D20 & (0.6) & \nodata &
              \nodata & \nodata \\

              & $2_{11}-2_{12}$ & D10 & (0.3) & \nodata & \nodata &
              \nodata \\

M\,82          & $\mathit{1_{10}-1_{11}}$ & D20 & $-$22.8(1.1) & 231.7 & 453.8 &
              $-$4600.1(237.0) \\
              & & G20 & $-$14.7(0.9) & 124.6(4.1) & 98.1(9.0) &
              $-$1531.6(167.6) \\
              & & G20 & $-$21.7(0.7) & 275.0(3.3) & 135.3(8.0) &
              $-$3128.0(212.8) \\

              & $\mathit{2_{11}-2_{12}}$ & D20 & $-$6.0(0.4) & 234.6 &
              465.5 & $-$1440.9(92.7) \\ 
              & & G20 & $-$5.7(0.6) & 110.2(5.8) & 97.4(13.1) &
              $-$595.2(99.4) \\ 
              & & G20 & $-$4.7(0.4) & 278.6(10.3) & 177.1(26.6) &
              $-$883.5(151.2) \vspace{2pt}
\enddata
\end{deluxetable*} 
 
\addtocounter{table}{-1}
\begin{deluxetable*}{lllrrrr} 
\tablewidth{415pt}
\tablecolumns{7}
\tablecaption{\vspace{-8.3pt} \hspace{80pt} --- {\it Continued}}
\tablehead{
\colhead{Galaxy} & 
\colhead{Transition\tablenotemark{b}} & 
\colhead{Fit\tablenotemark{c}} &
\colhead{$T^*_A$} & 
\colhead{v$_{hel}$\tablenotemark{d,e}} & 
\colhead{FW(HM/ZI)\tablenotemark{e,f}} & 
\colhead{$\int$T$^*_A$dv\tablenotemark{g}} \\
&&& \colhead{(mK)} & 
\colhead{(km s$^{-1}$)} & 
\colhead{(km s$^{-1}$)} & 
\colhead{(mK km s$^{-1}$)}
}
\startdata
M\,82SW        & $1_{10}-1_{11}$ & D20 & $-$18.2(1.8)
              & 228.0 & 424.4 & $-$4040.0(320.0) \\
              & & G20 & $-$11.2(1.4) & 132.0(12.6) & 116.4(24.6) &
              $-$1384.6(342.9) \\
              & & G20 & $-$18.0(1.1) & 282.7(9.1) & 144.2(19.2) &
              $-$2767.9(406.9) \\

              & $2_{11}-2_{12}$ & D20 & $-$7.7(0.4) 
              & 225.2 & 464.8 & $-$1632.9(98.1) \\
              & & G20 & $-$7.3(0.5) & 116.4(3.5) & 96.4(7.9) &
              $-$743.3(80.2) \\
              & & G20 & $-$4.7(0.3) & 274.9(9.3) & 185.4(22.7) &
              $-$925.2(125.0) \\

NGC\,3079      & $\mathit{1_{10}-1_{11}}$ & G05 & $+$1.0(0.4) & 1006.8(12.0) &
              55.2(33.6) & 56.0(41.4) \\
              & & G05 & $-$2.0(0.2) & 1186.7(11.4) & 200.9(35.0) &
              $-$436.1(87.9) \\
              & & G05 & $-$0.9(0.2) & 1501.9(25.6) & 213.5(78.8) &
              $-$210.6(90.3) \\

              & $2_{11}-2_{12}$ & G05 & $-$5.0(0.6) & 1017.8(3.4) &
              56.6(8.2) & $-$300.9(57.0) \\
              & & G05 & $-$27.6(0.9) & 1115.8(0.4) & 25.5(1.0) &
              $-$748.3(37.4) \\

IR\,10173+0828 & $\mathit{1_{10}-1_{11}}$ & D10 & (2.2) & \nodata &
              \nodata & \nodata \\  

NGC\,3227      & $1_{10}-1_{11}$ & D10 & (2.1) & \nodata & \nodata &
              \nodata \\ 

NGC\,3627      & $1_{10}-1_{11}$ & D10 & (0.8) & \nodata & \nodata &
              \nodata \\ 

              & $2_{11}-2_{12}$ & D10 & (0.3) & \nodata & \nodata
              \nodata \\

NGC\,3628      & $\mathit{1_{10}-1_{11}}$ & D10 & $-$8.5(1.1) & 826.4 &
              384.0 & $-$1295.6(191.7) \\  
              & & G10 & $-$4.1(0.9) & 725.9(11.5) & 87.7(25.5) &
              $-$381.7(137.8) \\ 
              & & G10 & $-$5.8(0.6) & 875.4(11.4) & 153.8(29.0) &
              $-$954.9(206.4) \\ 

              & $2_{11}-2_{12}$ & D10 & $-$2.5(0.4) & 900.7 & 210.8 &
              $-$209.4(43.2) \\
              & & G10 & $-$1.7(0.2) & 886.5(6.4) & 119.1(16.0) &
              $-$217.1(37.6) \\ 

NGC\,3690      & $1_{10}-1_{11}$ & D10 & (0.8) & \nodata & \nodata
              \nodata \\

              & $2_{11}-2_{12}$ & D10 & (0.4) & \nodata & \nodata
              \nodata \\

NGC\,4631      & $2_{11}-2_{12}$ & D10 & (0.3) & \nodata & \nodata
              \nodata \\

NGC\,4736      & $2_{11}-2_{12}$ & D10 & (0.8) & \nodata & \nodata
              \nodata \\

Mrk\,231       & $1_{10}-1_{11}$ & D10 & (0.7) & \nodata & \nodata &
              \nodata \\

              & $2_{11}-2_{12}$ & D10 & (0.4) & \nodata & \nodata &
              \nodata \\

NGC\,5005      & $2_{11}-2_{12}$ & D10 & (0.7) & \nodata & \nodata  &
              \nodata \\ 

IC\,860        & $\mathit{1_{10}-1_{11}}$ & D10 & $+$5.5(1.2) & 3921.1
              & 315.8 & 501.5(185.3) \\
              & & G10 & $+$3.6(0.6) & 3889.1(9.3) & 117.5(22.0) &
              449.5(111.3) \\ 

              & $2_{11}-2_{12}$ & D10 & $-$1.2(0.2) & 3879.8 & 215.0 &
              $-$128.9(17.3) \\
              & & G10 & $-$1.1(0.1) & 3877.9(3.6) & 108.8(8.6) &
              $-$130.9(13.6) \\

NGC\,5194      & $1_{10}-1_{11}$ & D20 & $-$1.7(0.3) & 369.7 & 434.5 &
              $-$222.8(67.3) \\ 
              & & G20 & $-$0.8(0.2) & 403.9(30.9) & 267.7(81.1) &
              $-$237.2(89.8) \\

              & $2_{11}-2_{12}$ & D10 & (0.3) & \nodata & \nodata &
              \nodata \\

M\,83          & $\mathit{1_{10}-1_{11}}$ & D10 & $-$3.0(0.8) & 510.0 &
              181.0 & $-$167.6(67.1) \\
              & & G10 & $-$1.7(0.5) & 522.6(15.6) & 104.3(39.2) &
              $-$192.3(93.1) \\ 

              & $\mathit{2_{11}-2_{12}}$ & D10 & $-$1.5(0.3) & 494.6 & 
              152.1 & $-$117.6(21.1) \\
              & & G10 & $-$1.1(0.2) & 507.2(8.7) & 109.2(24.0) &
              $-$129.9(34.9) \\ 

Mrk\,273       & $1_{10}-1_{11}$ & D10 & $-$4.7(1.5) & 11521.7 & 171.6
              & $-$216.6(120.3) \\
              & & G10 & $-$2.5(0.8) & 11505.0(15.5) & 95.3(38.2) &
              $-$253.0(131.2) \\

              & $2_{11}-2_{12}$ & D10 & (0.4) & \nodata & \nodata &
              \nodata \\
NGC\,5457      & $1_{10}-1_{11}$ & D10 & (0.9) & \nodata & \nodata
              & \nodata \\

              & $2_{11}-2_{12}$ & D10 & (0.4) & \nodata & \nodata 
              & \nodata \\

IR\,15107+0724 & $\mathit{1_{10}-1_{11}}$ & D10 & $+$5.3(1.0) & 3888.1
              & 391.9 & 714.0(190.1) \\ 
              & & G10 & $+$3.7(0.3) & 3888.4(8.0) & 182.0(19.1) &
              713.5(98.5) \\ 

              & $\mathit{2_{11}-2_{12}}$ & D10 & $-$3.7(0.4) & 3912.3
              & 368.6 & $-$506.7(75.5) \\
              & & G10 & $-$2.7(0.2) & 3908.6(5.7) & 179.3(13.7) &
              $-$508.9(51.2) \\ 

Arp\,220       & $\mathit{1_{10}-1_{11}}$ & D20 & $+$5.5(0.7) & 5483.5
              & 531.2 & 1323.3(171.3) \\
              & & G20 & $+$5.1(0.4) & 5352.3(7.0) & 154.1(18.2) &
              842.1(116.7) \\ 
              & & G20 & $+$3.2(1.0) & 5487.2(6.7) & 59.5(19.9) &
              200.4(92.0) \\ 
              & & G20 & $+$1.7(0.4) & 5596.0(38.8) & 175.6(86.4) &
              316.9(168.8) \\ 

              & $\mathit{2_{11}-2_{12}}$ & D20 & $-$4.1(0.4) & 5405.0
              & 544.5 & $-$1012.2(89.7) \\
              & & G20 & $-$3.5(0.9) & 5318.0(24.7) & 136.9(29.6) &
              $-$505.4(167.7) \\ 
              & & G20 & $-$3.2(0.7) & 5443.1(25.5) & 137.6(42.7) &
              $-$474.5(181.5) \\ 
              & & G20 & $-$0.5(0.2) & 5607.4(29.1) & 85.4(76.7) &
              $-$41.1(42.9) \\

NGC\,6240      & $\mathit{1_{10}-1_{11}}$ & D20 & $-$1.9(0.6) & 7309.8
              & 392.8 & $-$354.3(101.8) \\
              & & G20 & $-$1.5(0.3) & 7309.5(26.7) & 260.4(70.2) &
              $-$401.1(136.7) \\

              & $2_{11}-2_{12}$ & D10 & (0.4) & \nodata & \nodata 
              & \nodata \\ 

IR\,17208-0014 & $1_{10}-1_{11}$ & D10 & (1.2) & \nodata & \nodata &
              \nodata \\

              & $2_{11}-2_{12}$ & D10 & (0.9) & \nodata & \nodata &
              \nodata \\

IR\,17468+1320 & $\mathit{1_{10}-1_{11}}$ & D10 & (1.5)
              & \nodata & \nodata & \nodata \\ 

NGC\,6701      & $1_{10}-1_{11}$ & D10 & (1.4) & \nodata & \nodata &
              \nodata \\

              & $2_{11}-2_{12}$ & D10 & (0.5) & \nodata  & \nodata  &
              \nodata \\ 

NGC\,6921      & $1_{10}-1_{11}$ & D20 & $-$3.4(0.6) & 3960.7 & 609.9 &
              $-$1012.7(182.8) \\
              & & G20 & $-$2.8(0.3) & 3995.4(20.1) & 364.4(52.0) &
              $-$1087.9(196.5) \\

              & $2_{11}-2_{12}$ & D10 & (0.4) & \nodata & \nodata &
              \nodata \\

NGC\,6946      & $\mathit{1_{10}-1_{11}}$ & D10 & $-$2.7(0.7) & 51.2 &
              275.9 & $-$269.2(95.5) \\
              & & G10 & $-$1.7(0.5) & $-$11.6(6.8) & 44.4(16.4) &
              $-$78.5(37.4) \\
              & & G10 & $-$2.7(0.4) & 81.2(5.0) & 64.5(12.4) &
              $-$185.0(45.9) \\

              & $\mathit{2_{11}-2_{12}}$ & D10 & $-$1.1(0.3) & 36.0 & 209.6 &
              $-$94.0(28.0) \\
              & & G10 & $-$0.7(0.1) & $-$4.7(5.7) & 74.5(15.2) &
              $-$58.0(14.6) \\
              & & G10 & $-$0.9(0.2) & 77.9(3.2) & 34.4(7.7) &
              $-$32.4(9.2) \\

IC\,5179       & $1_{10}-1_{11}$ & D10 & (1.9) & \nodata & \nodata  &
              \nodata \vspace{2pt}

\enddata
\end{deluxetable*} 
 
\addtocounter{table}{-1}
\begin{deluxetable*}{lllrrrr} 
\tablewidth{415pt}
\tablecolumns{7}
\tablecaption{\vspace{-8.3pt} \hspace{80pt} --- {\it Continued}}
\tablehead{
\colhead{Galaxy} & 
\colhead{Transition\tablenotemark{b}} & 
\colhead{Fit\tablenotemark{c}} &
\colhead{$T^*_A$} & 
\colhead{v$_{hel}$\tablenotemark{d,e}} & 
\colhead{FW(HM/ZI)\tablenotemark{e,f}} & 
\colhead{$\int$T$^*_A$dv\tablenotemark{g}} \\
&&& \colhead{(mK)} & 
\colhead{(km s$^{-1}$)} & 
\colhead{(km s$^{-1}$)} & 
\colhead{(mK km s$^{-1}$)}
}
\startdata
NGC\,7331      & $1_{10}-1_{11}$ & D10 & (0.7) & \nodata & \nodata  &
              \nodata \\ 

              & $2_{11}-2_{12}$ & D10 & (0.5) & \nodata & \nodata  &
              \nodata \\

NGC\,7479      & $1_{10}-1_{11}$ & D10 & (1.1) & \nodata & \nodata  &
              \nodata \\ 

IR\,23365$+$3604 & $1_{10}-1_{11}$ & D10 & (1.0) & \nodata & \nodata  &
                \nodata \\ 

                & $2_{11}-2_{12}$ & D10 & (0.3) & \nodata & \nodata  &
                \nodata \\  

Mrk\,331         & $1_{10}-1_{11}$ & D10 & (1.2) & \nodata & \nodata  &
                \nodata \vspace{2pt}

\enddata
\tablenotetext{a}{~Table entries in parentheses are standard
  deviations, while entries that only list the RMS noise are
  non-detections.}
\tablenotetext{b}{~Transitions in italics reanalyzed and/or amended with
  new measurements from those presented in \citet{Mangum2008}.}
\tablenotetext{c}{~Gnn / Dnn $\equiv$ Gaussian / Direct measurement
  results with nn km/s gaussian spectral smoothing.}
\tablenotetext{d}{~Heliocentric optical velocity frame.}
\tablenotetext{e}{~Uncertainty in direct measurement v$_{hel}$ and
  FWZI assumed to be 2\,km/s, or 1/5 to 1/10 of a smoothed channel
  width, for all measurements.}
\tablenotetext{f}{~Full-width half maximum (FWHM) given for gaussian
  fits; full-width zero intensity (FWZI) given for direct 
  measurements.}
\tablenotetext{g}{~Derived from direct integration of line profile.}
\end{deluxetable*} 

\subsection{OH and H111$\alpha$}
\label{OHH111Results}

\begin{figure*}
\centering
\includegraphics[trim=50pt 30pt 20pt 60pt,clip=true,scale=0.60]{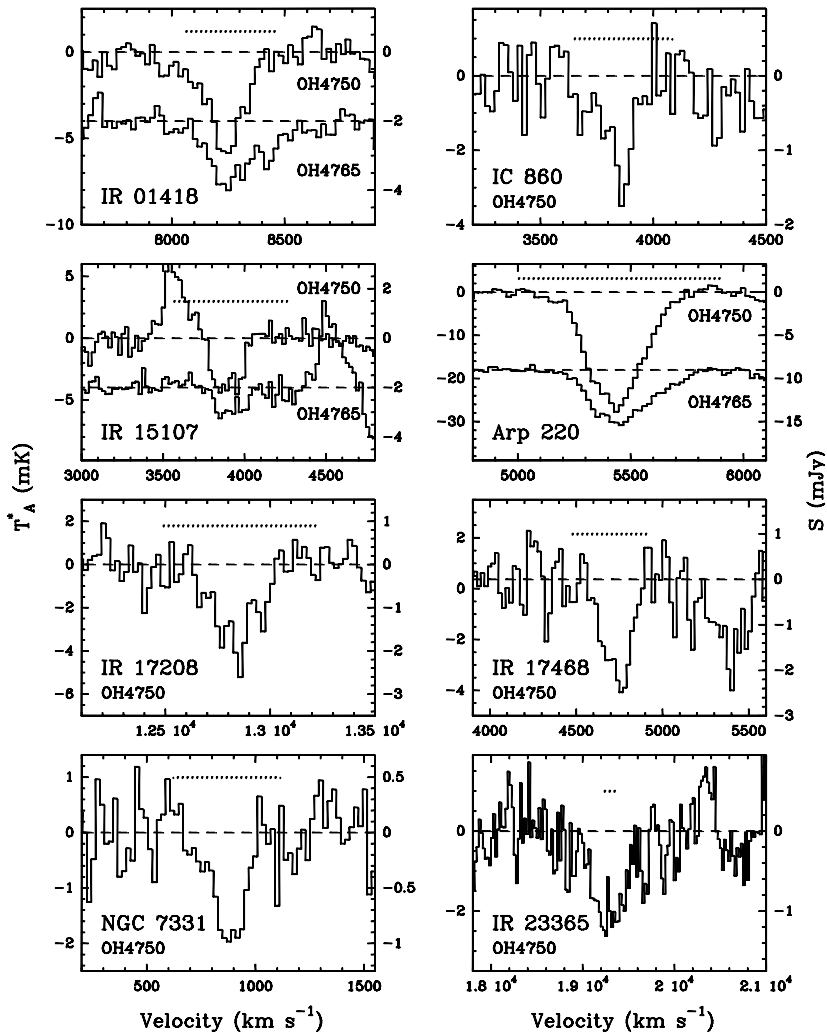}
\caption{OH4750 and OH4765 MHz spectra.  With the exception of
  IR\,17468 the dotted line within each spectrum indicates the FWZI CO
  linewidth.  The dotted line for IR\,17468 indicates the FWZI OH
  1667\,MHz linewidth \citep{Baan1993}.  The anomalous ``peaks'' in the
  IR\,15107 OH spectra at 3550 and 4500\,km\,s$^{-1}$ are due to a
  fixed-frequency receiver resonance.  Note that the OH4750  
  spectrum of Arp\,220 shown here differs from that shown in
  \citet{Mangum2008} due to a scaling error of a factor 1.97 in the
  \citet{Mangum2008} spectrum.}
\label{fig:ExgalOH}
\end{figure*}

\begin{figure}
\resizebox{\hsize}{!}{
\includegraphics[scale=0.50,trim=60pt 90pt 40pt 270pt,clip=true]{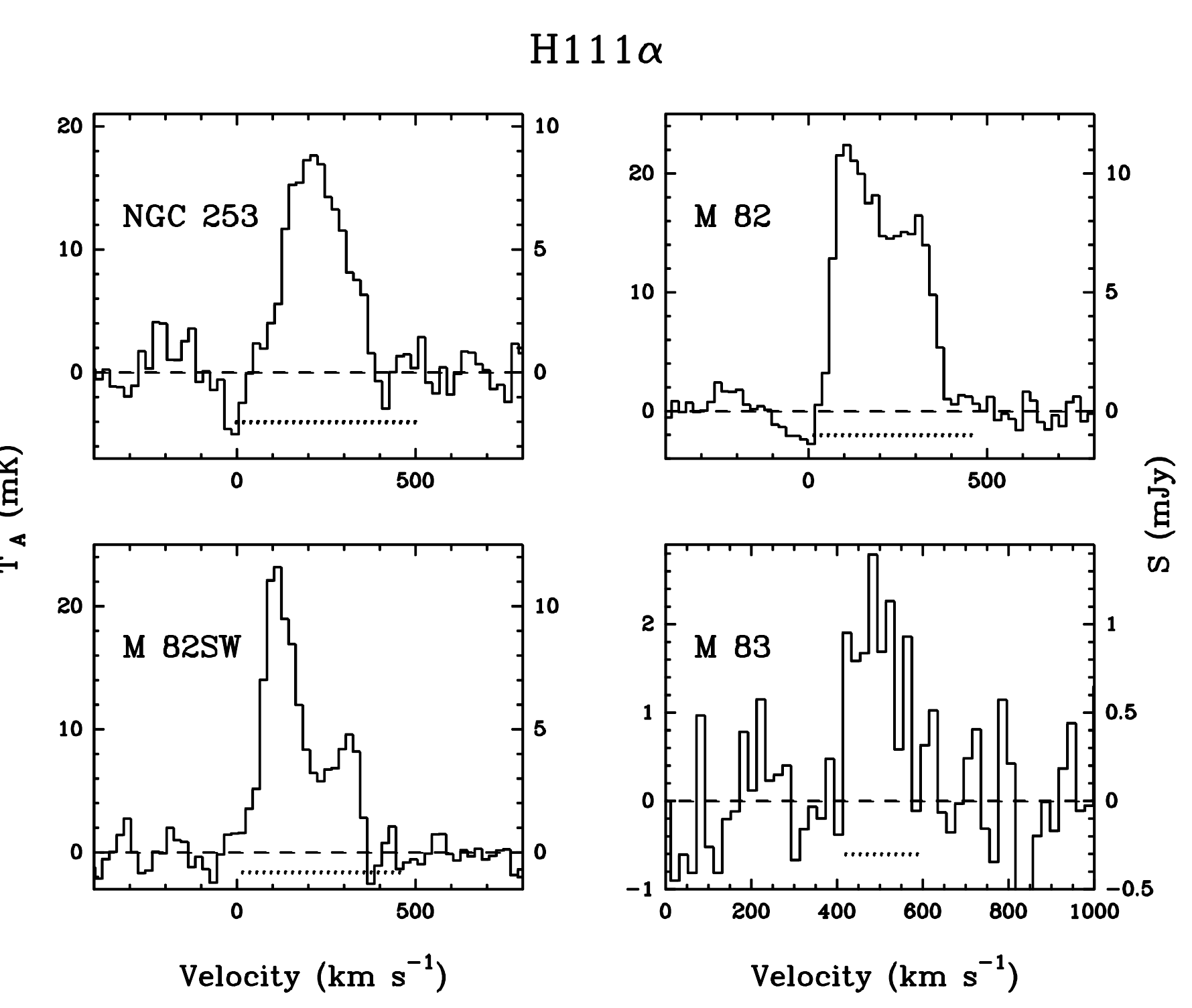}}
\caption{H111$\alpha$ spectra.  The dotted line within each spectrum
  indicates the FWZI CO linewidth.}
\label{fig:ExgalH111}
\end{figure}

Table~\ref{tab:ohh111results} lists and Figures~\ref{fig:ExgalOH} and
\ref{fig:ExgalH111} show our measured OH $^2\Pi_{1/2} J=1/2$ $F=1-0$
and $1-1$ ($E_u \sim 182$\,K above ground) and H111$\alpha$ radio
recombination line results. We report new detections of these OH and
H111$\alpha$ transitions toward nine galaxies, while updated analysis
is presented for OH and H111$\alpha$ detections quoted for four
galaxies presented in \citet{Mangum2008}.  For each OH and
H111$\alpha$ detection we list the same quantities in
Table~\ref{tab:ohh111results} as derived for our H$_2$CO spectra in
Table~\ref{tab:h2comeasurements}.

Toward most of our galaxies we did not detect the OH or H111$\alpha$ 
transitions.  The RMS noise values for our OH 4765 and 4750 MHz and
H111$\alpha$ measurements of the undetected galaxies are listed in
Table~\ref{tab:nondetections}.

Rotationally excited OH has been used as a tracer of the molecular
environment within AGN \citep{Henkel1986, Henkel1987, Henkel1990,
  Impellizzeri2006}.  AGN come in two main 
types: those with (type 1) and those without (type 2) broad optical
atomic line emission. In the unified scheme of active galactic nuclei, all
AGN are intrinsically similar, with the observed differences in type
due to orientation/observer perspective.  In the framework of a common
paradigm a significant column density of 
molecular material, in the form of a parsec-scale torus, obscures our
view of the AGN in type 2 objects.  Attempts to detect the obscuring
molecular material using measurements of molecular absorption or
emission have yielded few confirmations \citep[\eg\ ][]{Schmelz1986,
  Baan1992, StaveleySmith1992}.  While the 
existence of molecular AGN tori is confirmed by H$_2$O megamaser emission in
some objects \citep{Lo2005,Reid2009}, other tracers like OH
in the ground rotational transitions at 1.7\,GHz trace gas at larger
galactocentric distance
\citep[\eg\ ][]{Pihlstrom2001,Klockner2003,Klockner2004}. Before
concluding that parsec-scale tori are rarely molecular, however, we
should note that radiative excitation effects, in which coupling to
the nonthermal 
continuum can suppress the opacity in the lowest transitions, may
strongly affect the shape of a molecular spectral energy distribution
(SED). To explore this effect, detailed measurements of
rotationally-excited OH lines \citep[see also][]{Impellizzeri2006} may
be worthwhile. Before obtaining interferometric high resolution
measurements, however, it first has to be demonstrated that the lines are
detectable. This is shown in Figure~\ref{fig:ExgalOH}. The observed OH
line widths for the nine galaxies detected in our sample are similar,
though slightly larger, than those obtained in H$_2$CO, suggesting a
similar dynamical origin for the OH and H$_2$CO emitting 
regions.  Note, though, that similarity in galaxy-scale (\ie large)
line width does not uniquely associate the physical regions from which
these molecular emission lines originate.  Viable alternate mechanisms
which can produce similarly large line widths include shocked gas,
turbulence, and mixture of outflowing material with gas motions.
Apparent optical depths for each of the OH and H$_2$CO line
region measurements range from 0.04 (Arp\,220) to 0.22 (IC\,860),
suggesting optically thin absorption.

Among the eight sources with detected rotationally-excited OH
absorption, a total of five (IR\,01418$+$1651 (alias 
IIIZw\,35), IC\,860, IR\,15107$+$0724, Arp\,220, and IR\,17208$-$0014)
are well known (Ultra)Luminous Infrared Galaxies ((U)LIRGs) with
prominent OH maser or megamaser emission 
\citep[\eg\ ][]{Baan1989, Henkel1990, Baan1992a}. The combination of
ground state maser emission and OH absorption in the rotationally
excited $^2\Pi_{1/2}$ level is likely a consequence of the strong
infrared radiation field \citep[\eg\ ][]{Henkel1986, Henkel1987,
  Lockett2008, Willett2011}. We
are not aware of any 18\,cm OH detections towards IR\,17468$+$1320,
NGC\,7331, and IR\,23365$+$3604 \citep[\eg\ ][]{Baan1992a} so that
these cases are less constrained.

\begin{deluxetable*}{llrrr}
\tablewidth{0pt}
\tablecolumns{5}
\tablecaption{OH and H111$\alpha$ Detected Galaxies\label{tab:ohh111results}}
\tablehead{
\colhead{Galaxy} & 
\colhead{Transition\tablenotemark{a}} & 
\colhead{$T^*_A$\tablenotemark{b}} & 
\colhead{v$_{hel}$} & 
\colhead{FWZI} \\
&& \colhead{(mK)} & 
\colhead{(km s$^{-1}$)} & 
\colhead{(km s$^{-1}$)}
}
\startdata
NGC\,253       & H111$\alpha$ & 17.6(1.7) & 219.3(3.8) & 183.5(9.0) \\
IR\,01418+1651 & OH4750 & $-5.4(0.7)$ & 8236.8(5.3) & 198.6(12.5) \\
              & OH4765 & $-3.2(0.6)$ & 8284.1(10.0) & 312.5(23.9) \\
M\,82          & $\mathit{H111\alpha}$ & 21.6(2.0) & 119.0(6.5) & 107.5(13.2) \\
              && 16.4(1.2) & 271.3(11.8) & 161.8(27.1) \\
M\,82SW        & H111$\alpha$ & 22.3(1.2) & 120.5(3.2) & 109.0(8.1) \\
              && 9.0(1.3) & 291.4(7.9) & 108.2(20.8) \\
IC\,860        & \textit{OH4750} & $-2.7(0.8)$ & 3856.0(7.1) & 132.0(17.1) \\
              & \textit{OH4765} & (0.8) & \nodata & \nodata \\
M\,83          & H111$\alpha$ & 2.2(0.6) & 491.8(12.3) & 135.9(30.1) \\
IR 15107+0724 & \textit{OH4750} & $-4.3(0.6)$ & 3891.4(9.6) & 190.3(24.4) \\
              & \textit{OH4765} & $-2.1(0.6)$ & 3910.5(13.2) & 203.6(35.9) \\
Arp 220       & \textit{OH4750}\tablenotemark{c} & $-26.3(0.7)$ & 5432.5(2.2) & 277.6(5.4) \\
              & \textit{OH4765} & $-11.9(0.4)$ & 5458.5(4.5) & 314.9(11.2) \\
IR\,17208-0014 & OH4750 & $-3.6(0.7)$ & 12835.6(12.4) & 250.1(32.1) \\
              & OH4765 & (0.7) & \nodata & \nodata \\
IR\,17468+1320 & OH4750 & $-4.7(1.3)$ & 4737.9(7.0) & 341.5(16.7)\\
               & OH4765 & (1.34) & \nodata & \nodata \\
NGC\,7331      & OH4750 & $-2.0(0.6)$ & 881.1(6.1) & 153.8(14.6) \\
               & OH4765 & (0.64) & \nodata & \nodata \\
IR\,23365+3604 & OH4750 & $-2.3(0.7)$ & 19284.6(12.7) & 398.2(35.1) \\
               & OH4765 & (0.66) & \nodata & \nodata \vspace{2pt}
\enddata
\tablenotetext{a}{~Transitions in italics reanalyzed from measurements
  presented in \citet{Mangum2008}.}
\tablenotetext{b}{~RMS noise levels are for 20 km~s$^{-1}$ channels.}
\tablenotetext{c}{~Arp\,220 OH4750 spectrum shown in \citet{Mangum2008}
  was mis-scaled by a factor of 1.97.}
\end{deluxetable*}

\begin{deluxetable}{cccc}
\tabletypesize{\scriptsize}
\tablewidth{0pt}
\tablecolumns{4}
\tablecaption{OH and/or H111$\alpha$ Results\label{tab:nondetections}}
\tablehead{
\colhead{Galaxy} & \multicolumn{3}{c}{RMS\tablenotemark{a,b}} \\
\cline{2-4}
& \colhead{OH 4765} & \colhead{OH 4750}
& \colhead{H111$\alpha$} \\
& \colhead{(mK)} & \colhead{(mK)} & \colhead{(mK)}}
\startdata
NGC\,253	         & 3.15 & 1.95 & Emis \\
IC\,1623          & 0.91 & 0.80 & 0.94 \\
NGC\,520	         & 0.81 & 0.78 & 0.74 \\
NGC\,598	         & 0.68 & 0.67 & 0.70 \\
NGC\,604	         & 1.22 & 1.13 & 1.23 \\
NGC\,660	         & 0.79 & 0.80 & 0.76 \\
IR\,01418+1651    & Abs & Abs & 0.85 \\
NGC\,695	         & 0.54 & 0.80 & 0.55 \\
Mrk\,1027         & 0.97 & 0.95 & 1.02 \\
NGC\,891	         & 0.94 & 0.90 & 0.82 \\
NGC\,925          & 0.99 & 1.02 & 1.07 \\
NGC\,1022         & 0.96 & 0.90 & 0.99 \\
NGC\,1055         & 0.80 & 0.80 & 0.91 \\
Maffei\,2         & 1.24 & 1.12 & 1.19 \\
NGC\,1068         & 2.73 & 3.72 & 4.22 \\
UGC\,02369        & 0.96 & 0.96 & 0.89 \\
NGC\,1144         & 0.78 & 0.90 & 0.73 \\
NGC\,1365         & 1.53 & 0.55 & 1.44 \\
IR\,03359+1523    & 1.17 & 1.22 & 1.10 \\
IC\,342	         & 1.66 & 1.24 & 1.90 \\
NGC\,1614         & 1.14 & 1.26 & 1.09 \\
VIIZw31          & 1.03 & 0.71 & 1.07 \\
NGC\,2146         & 2.33 & 0.97 & 3.18 \\
NGC\,2623         & 0.83 & 0.82 & 0.91 \\
NGC\,2903         & 0.42 & 0.46 & 0.52 \\
Arp\,55           & 1.67 & 1.63 & 1.68 \\
UGC\,05101        & 1.09 & 1.09 & 1.15 \\
M\,82             & 3.24 & 3.81 & Emis \\
M\,82SW           & 3.63 & 3.63 & Emis \\
NGC\,3079         & 2.55 & 1.00 & 7.61 \\
IR\,10173+0828    & 1.61 & 1.68 & 1.81 \\
NGC\,3227         & 2.42 & 1.86 & 3.28 \\
NGC\,3627         & 0.55 & 0.70 & 0.71 \\
NGC\,3628         & 0.94 & 0.96 & 0.83 \\
NGC\,3690         & 2.11 & 1.87 & 0.49 \\
Mrk\,231          & 1.39 & 1.42 & 1.26 \\
IC\,860           & Abs & 0.83 & 0.90 \\
NGC\,5194         & 0.36 & 0.36 & 0.39 \\
M\,83	         & 0.70 & 0.66 & Emis \\
Mrk\,273          & 3.63 & 2.15 & 4.05 \\
NGC\,5457         & 0.54 & 0.70 & 0.55 \\
IR\,15107+0724    & Abs & Abs & 0.63 \\
Arp\,220          & Abs & Abs & 0.91 \\
NGC\,6240	 & 0.95 & 0.95 & 0.89 \\
IR\,17208-0014    & Abs & 0.66 & 0.81 \\
IR\,17468+1320    & 1.34 & Abs & 1.28 \\
NGC\,6701         & 0.87 & 0.84 & 0.60 \\
NGC\,6921         & 0.80 & 0.93 & 0.88 \\
NGC\,6946	 & 0.55 & 0.56 & 0.52 \\
IC\,5179          & 1.50 & 1.87 & 1.86 \\
NGC\,7331         & Abs & 0.64 & 0.70 \\
NGC\,7479         & 1.17 & 0.87 & 1.10 \\
IR\,23365+3604    & Abs & 0.66 & 0.67 \\
Mrk\,331          & 0.68 & 0.72 & 0.64 \vspace{2pt}
\enddata
\tablenotetext{a}{RMS noise levels are for 20 km~s$^{-1}$ channels.} 
\tablenotetext{b}{An ``Abs'' entry indicates that the line was detected in 
absorption, ``Emis'' in emission (see Table~\ref{tab:ohh111results}).} 
\end{deluxetable}

\subsection{Continuum Emission}
\label{ContResults}

Table~\ref{tab:contmeas} lists our measured continuum
levels derived from the mean zero-level offset of our spectroscopic
measurements.  Note that for galaxies whose flux measurements were
presented in \citet{Mangum2008} we have reanalyzed those flux measurements and
have taken a mean of the baseline rather than just the central channel
offset\footnote{In \citet{Mangum2008} the ``central channel offset''
  was derived by fitting a polynomial to the spectral baseline and
  capturing the central channel value from that baseline (continuum) fit.} for the flux measurement.  Differences between our derived
continuum fluxes and those quoted in \citet{Mangum2008} are with one exception
less than 2\% for our 4.8 GHz measurements, the one exception being
NGC\,3079 where the difference is 15\%.  For our 14.5 GHz continuum
measurements the differences are less than 15\%.

For most of the galaxies in our sample the measured 
continuum fluxes agree with those quoted in \citet{Araya2004},
\citet{Baan1990}, and the NED archive.  Exceptions to this agreement
are shown in \textit{italic} in Table~\ref{tab:contmeas}.  Continuum
emission from starburst galaxies at these frequencies emits as a power
law with spectral index $\alpha$ (S$_\nu \propto \nu^{-\alpha}$).
Likely emission mechanisms at the frequencies of our GBT observations
include synchrotron emission ($\alpha \simeq 0.7-0.8$) and free-free
emission ($\alpha \simeq -2.0$ to $+0.1$).  The GBT fluxes
listed in Table~\ref{tab:contmeas} suggest that the
centimeter-wavelength emission from the majority of the starburst
galaxies in our sample is produced via synchrotron emission, though
several galaxies, including NGC\,1055 and UGC\,05101, appear to have
flat spectra suggestive of free-free emission.

\begin{deluxetable*}{llll|lll}
\tabletypesize{\scriptsize}
\tablewidth{0pt}
\tablecolumns{7}
\tablecaption{Measured Continuum Levels\label{tab:contmeas}}
\tablehead{
\colhead{Galaxy} & \multicolumn{3}{c}{4.8~GHz}&
\multicolumn{3}{|c}{14.5 GHz}\\ 
\cline{2-4}  \cline{5-7}
& \colhead{GBT\tablenotemark{a,b}} & \colhead{Arecibo\tablenotemark{c},140ft\tablenotemark{d}} & 
\colhead{NED} & \colhead{GBT\tablenotemark{a,b}} & \colhead{140ft\tablenotemark{d}} 
& \colhead{NED} \\
& \colhead{(Jy)} & \colhead{(Jy)} & \colhead{(Jy)} & \colhead{(Jy)} & 
\colhead{(Jy)} & \colhead{(Jy)}
}
\startdata
NGC\,55        & PM &\nodata& 0.150--0.197 & 0.003 &\nodata&\nodata\\
NGC\,253       & 1.771 & 1.20 & 2.0--2.5 & \itshape 0.522 & 0.37 &\nodata\\
IC\,1623       & 0.077 &\nodata& 0.096 & \nodata&\nodata&\nodata\\
NGC\,520       & 0.081 & 0.079 & 0.087--0.126 & 0.033 &\nodata&\nodata\\
NGC\,598       & PM &\nodata& 1.300 & 0.002 &\nodata&\nodata\\
NGC\,660       & 0.151 & 0.140 & 0.156-0.232  & 0.288 &\nodata&\nodata\\
IR\,01418+1651 & 0.031 & 0.028 &\nodata& 0.018 &\nodata&\nodata\\
NGC\,695       & \itshape 0.024 &\nodata& 0.041 & 0.011 &\nodata&\nodata\\
Mrk\,1027      & 0.021 &\nodata&\nodata&\nodata&\nodata&\nodata\\
NGC\,891       & 0.319 &\nodata& 0.194--0.342 & 0.101 &\nodata&\nodata \\
NGC\,925       & 0.030 &\nodata&\nodata&\nodata&\nodata&\nodata\\
NGC\,1022      & 0.031 &\nodata&\nodata& 0.171 &\nodata&\nodata\\
NGC\,1055      & 0.068 &\nodata& 0.063--0.086 & 0.058 &\nodata&\nodata\\
Maffei\,2      & 0.276 &\nodata& 0.243--0.375 &\nodata&\nodata&\nodata\\
NGC\,1068      & 1.680 &\nodata& 1.770--2.187 & 0.555 &\nodata& 0.680 \\
UGC\,02369     & 0.112 &\nodata&\nodata&\nodata&\nodata&\nodata\\
NGC\,1144      & 0.062 &\nodata& 0.052 & 0.011 &\nodata&\nodata\\
NGC\,1365      & 0.235 &\nodata& 0.180--0.230 & 0.052 &\nodata&\nodata\\
IR\,03359+1523 & 0.016 &\nodata&\nodata&\nodata&\nodata&\nodata\\
IC\,342        & 0.086 &\nodata& 0.080--0.277 & 0.050 &\nodata&\nodata\\
NGC\,1614      & 0.072 &\nodata& 0.063 &\nodata&\nodata&\nodata\\
VIIZw31       & 0.032 &\nodata&\nodata& 0.015 &\nodata&\nodata\\
NGC\,2146      & 0.389 &\nodata&\nodata& 0.223 &\nodata& 0.30 \\
NGC\,2623      & 0.061 & 0.064 & 0.057--0.073 & 0.009 &\nodata&\nodata\\
Arp\,55        & 0.039 &\nodata&\nodata&\nodata&\nodata&\nodata\\
NGC\,2903      & 0.115 & 0.041 & 0.118--0.180 & 0.019 &\nodata&\nodata\\
UGC\,05101     & 0.054 &\nodata& 0.077 & 0.048 &\nodata&\nodata\\
M\,82          & 3.233 & 3.55 & 3.67--3.96 & \itshape 1.304 & 1.79 & 1.73 \\
M\,82SW        &\nodata&\nodata&\nodata& 1.128 &\nodata&\nodata\\
NGC\,3079      & 0.342 & 0.33 & 0.32  & 0.159 & 0.14  &\nodata\\
IR\,10173+0828 & PM & 0.020&\nodata&\nodata&\nodata&\nodata \\
NGC\,3227      & \itshape 0.559 & 0.008 & 0.045 &\nodata&\nodata&\nodata\\
NGC\,3627      & 0.121 &\nodata&\nodata&\nodata&\nodata&\nodata\\
NGC\,3628      & 0.186 & 0.131,0.33 & 0.222--0.283 & \itshape 0.043 & 0.11 &\nodata\\
NGC\,3690      & 0.321 &\nodata& 0.362--0.407 & 0.115 &\nodata&\nodata\\
NGC\,4631      &\nodata&\nodata& 0.299-0.438 & 0.033 &\nodata&\nodata\\
NGC\,4736      &\nodata&\nodata& 0.090--0.119 & 0.022 &\nodata&\nodata\\
Mrk\,231       & \itshape 0.321 &\nodata& 0.417 & 0.129 &\nodata&\nodata\\
NGC\,5005      &\nodata&\nodata& 0.060 & 0.001 &\nodata&\nodata\\
IC\,860        & \itshape 0.010 &\nodata& 0.042 & 0.031 &\nodata&\nodata\\
NGC\,5194      & \itshape 0.150 &\nodata& 0.215--0.436 & 0.023 &\nodata&\nodata\\
M\,83          & \itshape 0.322 &\nodata& 0.648--0.780 & 0.071 &\nodata&\nodata\\
Mrk\,273       & \itshape 0.012 &\nodata& 0.070--0.103 & 0.028 &\nodata&\nodata\\
NGC\,5457      & 0.023 &\nodata&\nodata& 0.013 &\nodata&\nodata\\
IR\,15107+0724 & \itshape 0.017 & 0.031 &\nodata& 0.021 &\nodata&\nodata\\
Arp\,220       & 0.217 & 0.172,0.22 & 0.22 & \itshape 0.082 & 0.11  &\nodata\\
NGC\,6240      & 0.141 &\nodata& 0.178 & 0.035 &\nodata&\nodata\\
IR\,17208-0014 & 0.071 & 0.062 & 0.068 & 0.028 &\nodata&\nodata\\
IR\,17468+1320 & \itshape 0.190 &\nodata& 0.047 &\nodata&\nodata&\nodata\\
NGC\,6701      & 0.025 &\nodata& 0.022 & 0.008 &\nodata&\nodata\\
NGC\,6921      & 0.028 &\nodata& 0.027 & 0.035 &\nodata&\nodata\\
NGC\,6946      & \itshape 0.167 &\nodata& 0.219--0.531 & 0.030 &\nodata&\nodata\\
IC\,5179       & PM &\nodata& 0.079 &\nodata&\nodata&\nodata\\
NGC\,7331      & 0.095 & 0.040 & 0.070--0.096 & 0.007 &\nodata&\nodata\\
NGC\,7479      & 0.038 &\nodata& 0.040 &\nodata&\nodata&\nodata\\
IR\,23365+3604 & \itshape 0.015 &\nodata& 0.010 & 0.001 &\nodata&\nodata\\
Mrk\,331       & 0.028 & 0.025 & 0.028 &\nodata&\nodata&\nodata\vspace{2pt}
\enddata
\tablenotetext{a}{GBT measurements shown in \textit{italics} differ
  from the measurement range defined by \citet{Araya2004},
  \citet{Baan1990}, and NED measurements by more than 25\%.}
\tablenotetext{b}{PM $\equiv$ Poor measurement due to insufficient total
  power stability.}
\tablenotetext{c}{The Arecibo 4.8~GHz continuum levels were obtained 
from spectral baselines by \citet{Araya2004}, their Table~\ref{tab:galaxies}.}
\tablenotetext{d}{The NRAO 140$^\prime$ 4.8 and 14.5~GHz continuum
  levels from \citet{Baan1990}, their Table~\ref{tab:galaxies}.}
\end{deluxetable*}

\section{Analysis}
\label{Analysis}

\subsection{Comparison to Previous Measurements}
\label{Comparison}

In the following we list previous measurements of the H$_2$CO emission
in galaxies where we have detected the H$_2$CO $1_{10}-1_{11}$ or
$2_{11}-2_{12}$ transitions.  We also present a re-evaluation of our
reported detection 
of H$_2$CO $1_{10}-1_{11}$ emission in UGC\,05101 \citep{Mangum2008}
which we now believe is a non-detection.  Where available, we also
list measurements of high spatial density molecular tracers in these
galaxies.

\noindent
{\bf NGC\,253:}  As noted in \cite{Mangum2008}, the 4.8 and 14.5~GHz
H$_2$CO transitions have previously been detected in absorption
\citep{Gardner1974,Baan1990}.  Our GBT detection of the
$1_{10}-1_{11}$ transition
(Figure~\ref{fig:NGC253NGC660Maffei2IC342FormSpec}) is 
significantly different in shape from the line profile reported by
\citet{Gardner1974}.  \citet{Baan1997} have used the Very Large Array
to image the $1_{10}-1_{11}$ transition, measuring an absorption
structure with FWHM size $39\arcsec\times12\arcsec$.  The peak
integrated flux measured by \citet{Baan1997} of $-1.72$ Jy beam$^{-1}$
km~s$^{-1}$ is about 60\% of the integrated flux we measure,
suggesting that some extended structure detected in our GBT
observations is missing in these interferometric measurements.
Previous CS, H$_2$CO, NH$_3$ and high-J CO measurements
\citep{Baan1990,Baan1997,Huettemeister1997,Ott2005,Guesten2006,Martin2006,Bayet2009}
estimate spatial densities in the range $10^{4-5}$\,cm$^{-3}$, which 
compares favorably to the $10^{5.03\pm0.03}$\,cm$^{-3}$ we measure
(Table~\ref{tab:lvg}).

\noindent
{\bf NGC\,520:}  \citet{Araya2004} first detected the H$_2$CO
$1_{10}-1_{11}$ transition in absorption, in good agreement with the
GBT transition properties (Figure~\ref{fig:ExgalFormSingles-OnePage}).  
Our non-detection of H111$\alpha$ (RMS $\sim 0.25$\,mJy in
$\Delta$v $\sim 40$\,km\,s$^{-1}$) is consistent with the $\sim
0.3$\,mJy detection of H110$\alpha$ reported by \citet{Araya2004}. 
Interferometric measurements of the CO J=$1-0$ emission from NGC\,520
by \citet{Sanders1988} and \citet{Yun2001} reveal kpc-scale structures
over a region $\sim 7--12^{\prime\prime}$ in extent.

\noindent
{\bf NGC\,660:}  Our GBT detection of the H$_2$CO $1_{10}-1_{11}$
transition in 
absorption (Figure~\ref{fig:NGC253NGC660Maffei2IC342FormSpec}) is similar in
both intensity and velocity extent as that 
reported by \citet{Baan1986}, while our detection of the
H$_2$CO $2_{11}-2_{12}$ transition is the first reported.
\citet{Aalto1995} measured the 
$^{12}$CO J=$1-0$, $2-1$; $^{13}$CO J=$1-0$, $2-1$; C$^{18}$O 
J=$1-0$; and HCN J=$1-0$ emission, while \citet{Israel2009} measured
the $^{12}$CO J=$1-0$, $2-1$, $3-2$, $4-3$ and $^{13}$CO
J=$1-0$, $2-1$, and $3-2$ emission from this galaxy.  From CO
J=$1-0$ \citep{Aalto1995} and $3-2$ \citep{Israel2009} images these
studies derived a source diameter of $\lesssim 15^{\prime\prime}$.

\noindent
{\bf NGC\,891:}  Our $1_{10}-1_{11}$ absorption measurement was
reported in \citet{Mangum2008}
(Figure~\ref{fig:ExgalFormSingles-OnePage}).
\citet{GarciaBurillo1995} measured the CO J=$2-1$ and $1-0$ emission
toward this edge-on galaxy.  The emission is extended along the
galactic disk to spatial 
scales $\gtrsim 120^{\prime\prime}\times20^{\prime\prime}$, but is mainly
concentrated in a nuclear condensation $\sim
20^{\prime\prime}\times10^{\prime\prime}$ in extent.

\noindent
{\bf Maffei\,2:} Images of the CO J=$1-0$, $2-1$, and CS
J=$2-1$ emission \citep{Kuno2008}, $^{13}$CO and
C$^{18}$O J=$1-0$ and $2-1$ and HCN J=$1-0$ emission
\citep{Meier2008}, and NH$_3$ (1,1) and (2,2) emission
\citep{Lebron2011} all point to a $5^{\prime\prime}\times 30^{\prime\prime}$
two- to four-component structure of the central bar in this nearby
spiral galaxy.  These studies indicate dense gas structures with
spatial densities and kinetic temperatures $\sim 10^4$ cm$^{-3}$ and
T$_K \sim 30-40$ K, respectively.  High-excitation NH$_3$ (6,6)
measurements \citep{Mauersberger2003} have uncovered a
high-temperature (T$_K \sim 130$ K) component in this galaxy.  Our
detections of H$_2$CO $1_{10}-1_{11}$ and $2_{11}-2_{12}$ absorption
(Figure~\ref{fig:NGC253NGC660Maffei2IC342FormSpec}) 
are the first reported measurements of H$_2$CO in this galaxy.  The
K-doublet H$_2$CO spectra are clearly composed of two velocity
components at $-98$ (component ``C1'') and $+25$\,km\,s$^{-1}$
(component ``C2''), consistent with previous studies
\citep{Mauersberger2003}.

\noindent
{\bf NGC\,1144:} \citet{Gao2004a} reported broad yet weak CO and HCN
J=$1-0$ emission (T$_{mb} \sim$ 50 and 2\,mK, respectively) toward
this Seyfert galaxy.  Our detection of broad H$_2$CO $1_{10}-1_{11}$
absorption (Figure~\ref{fig:ExgalFormSingles-OnePage}) 
which decomposes into three velocity components is consistent with the
HCN J=$1-0$ spectra reported by \citet{Gao2004a}.  There do not appear
to be any imaging studies of the dense gas in this galaxy.

\noindent
{\bf NGC\,1365:} \citet{Baan2008} report single-position J=$1-0$ CO, HCN, HNC,
HCO$^+$, N=$1-0$ CN, J=$2-1$ CO, J=$3-2$ CS, and N=$2-1$ CN observations
with resolutions ranging from $13^{\prime\prime}$ to
$57^{\prime\prime}$ from this barred spiral galaxy.  \citet{Perez2007}
report HCN J=$3-2$ 
emission in addition to the J=$1-0$ and J=$2-1$ transitions of CO,
HCN, HNC, and N=$1-0$ and N=$2-1$ transitions of CN.  \citet{Gao2004a}
report relatively bright (160 and 
8\,mK, respectively) and extended CO and HCN J=$1-0$ emission from
this galaxy, while \citet{Sandqvist1999} mapped the CO J=$3-2$
emission and found extended ($\theta_s \simeq 20^{\prime\prime}$ FWHM)
structure dominated by a circumnuclear molecular torus ($\theta_{torus}
\simeq 13^{\prime\prime}$).  Our detection of weak
H$_2$CO $1_{10}-1_{11}$ absorption
(Figure~\ref{fig:ExgalFormSingles-OnePage}) is consistent with these previous
molecular tracer measurements.

\noindent
{\bf IC\,342:}  The absorption measurements in H$_2$CO $1_{10}-1_{11}$
and $2_{11}-2_{12}$ reported in \citet{Mangum2008} were new detections of these
K-doublet transitions
(Figure~\ref{fig:NGC253NGC660Maffei2IC342FormSpec}).  Previous 
millimeter-wavelength H$_2$CO measurements
\citep{Huettemeister1997,Meier2005,Meier2011}
estimate spatial densities in the range $10^{4-6}$ cm$^{-3}$.
High-excitation NH$_3$ (6,6) emission toward this galaxy
\citep{Mauersberger2003} points to the presence of high kinetic
temperatures.  HCN J=$1-0$ images \citep{Downes1992} resolve IC\,342
into five condensations with sizes ranging from $3^{\prime\prime}$
to $6^{\prime\prime}$ over a region $\sim
10^{\prime\prime}\times15^{\prime\prime}$ in extent.

\noindent
{\bf NGC\,2146:} The relatively intense CO and HCN J=$1-0$ emission
(800 and 
30\,mK, respectively) measured by \citet{Gao2004a} is consistent with
our detections of H$_2$CO $1_{10}-1_{11}$ and $2_{11}-2_{12}$
absorption (Figure~\ref{fig:NGC2146M82M82SWNGC3079FormSpec}), the first reported
detection of this molecule in this 
galaxy.  \citet{Greve2006} imaged the $^{12}$CO J=$1-0$ and J=$2-1$ and
$^{13}$CO J=$1-0$ emission from this barred spiral galaxy, measuring a
disk-like structure $\sim 5^{\prime\prime}\times20^{\prime\prime}$ in
extent.

\noindent
{\bf UGC\,05101:}  This was reported as a new detection of the
$1_{10}-1_{11}$ transition in emission in \citet{Mangum2008}.  Additional
measurements, though, have thrown this assignment into question as we
do not see consistent results between our 2006 \citep[reported in
][]{Mangum2008} and our new 2013 measurements.  We now believe that 
neither the $1_{10}-1_{11}$ nor the $2_{11}-2_{12}$ transitions have
been detected in this galaxy.

\noindent
{\bf M\,82:}  The morphology of M\,82's molecular ISM is characterized
by a double-lobed structure
\citep{Mao2000}. \cite{Seaquist2006} imaged the CO J=$6-5$
emission from this galaxy, finding emission extended over a disk-like
structure $40^{\prime\prime}\times15^{\prime\prime}$ in
extent. H$_2$CO was previously detected in absorption in the 
$1_{10}-1_{11}$ line by \citet{Graham1978}, in emission in the 
$2_{11}-2_{12}$ and $3_{03}-2_{02}$ (218~GHz) lines by
\citet{Baan1990}, and in emission in the $2_{02}-1_{01}$ (146~GHz),
$3_{03}-2_{02}$ and $3_{22}-2_{21}$ / $3_{21}-2_{20}$ (218~GHz) lines
by \citet{Muehle2007}.  
We detect both the 4.8 and 14.5 GHz lines in absorption with high
confidence (Figure \ref{fig:NGC2146M82M82SWNGC3079FormSpec}).  Investigation of our LVG
model predictions of the relative intensities of the 4.8 and 14.5~GHz
emission (see \S\ref{LVG}) cannot reproduce the 14.5~GHz emission
observed by \citet{Baan1990}. The H110$\alpha$ and H111$\alpha$
lines, as well as other higher frequency hydrogen radio recombination
lines, have been detected in M\,82
\citep{Seaquist1981,Bell1984,RodriguezRico2004}.  We re-detect the
H111$\alpha$ line and confirm the two components identified by
\citet{Bell1984}.  Previous CS and H$_2$CO measurements
\citep{Baan1990,Huettemeister1997} estimate spatial densities in the
range $10^{4-5}$ cm$^{-3}$.

Recent \textit{Herschel}-SPIRE imaging spectroscopic measurements of the
molecular emission toward M\,82 \citep{Kamenetzky2012} have
characterized the physical conditions in this galaxy using
measurements of $^{12}$CO and $^{13}$CO emission over a wide range of
molecular excitation.  By using a LVG radiative transfer model with
Bayesian likelihood analysis of the J=$4-3$ through J=$13-12$
$^{12}$CO and $^{13}$CO emission toward M\,82 \cite{Kamenetzky2012}
confirmed the existence of two temperature components
\citep{Panuzzo2010}: A cool component with median T$_{K}$ = 35\,K
(1-$\sigma$ range 12--385\,K) and n(H$_2$) = $10^{3.44}$\,cm$^{-3}$
(1-$\sigma$ range $10^{2.48}$--$10^{5.15}$\,cm$^{-3}$) and a warm
component with median T$_{K}$ = 436\,K (1-$\sigma$ range 344--548\,K)
and n(H$_2$) = $10^{3.58}$\,cm$^{-3}$ (1-$\sigma$ range
$10^{3.17}$--$10^{3.96}$\,cm$^{-3}$).  Our H$_2$CO measurements
clearly correspond to the cool component measured by
\cite{Kamenetzky2012}, with our derived spatial density of
n(H$_2$) = $10^{5.05\pm0.08}$\,cm$^{-3}$ overlapping at
the high-density limit of their derived 1-$\sigma$ range.

We also measure a position offset of ($-12$,$-4$) arcsec from the
nominal M\,82 position in the H$_2$CO $2_{11}-2_{12}$ transition only,
which we refer to as M\,82SW (Figure~\ref{fig:NGC2146M82M82SWNGC3079FormSpec}).  This
is the ``P3'' HCN J=$1-0$ peak 
position noted by \cite{Brouillet1993}, and corresponds to within
$2^{\prime\prime}$ of the ``southwestern molecular lobe'' where NH$_3$
emission was detected in this galaxy 
\citep{Weiss2001}.  M\,82SW appears to be $\sim50\%$ stronger in
H$_2$CO emission than the nominal galaxy center position
(Table~\ref{tab:h2comeasurements}), but note that the spatial
resolution of our H$_2$CO measurements ($\theta_B = 153^{\prime\prime}$
and $51^{\prime\prime}$) allows for some sampling of both components in
all of our M\,82 measurements.  In the NH$_3$ measurements of M\,82
\citep{Mangum2013} only the M\,82SW position possesses detectable
NH$_3$ emission, and is the source of the kinetic temperature assumed
in our LVG analysis (\S\ref{LVG}).

\noindent
{\bf NGC\,3079:}  Reported as a nondetection in \citet{Mangum2008}, additional
integration resulted in a detection of the $1_{10}-1_{11}$ line
(Figure~\ref{fig:NGC2146M82M82SWNGC3079FormSpec}), consistent with the absorption
line reported by \citet[][$-2.1$~mJy]{Baan1986}.  \citet{Perez2007}
detect CO J=$1-0$ and J=$2-1$, CN N=$1-0$ and $2-1$, and HCN and HNC
J=$1-0$ and J=$3-2$ emission from 
the nucleus of this galaxy.  From the HCN J=$1-0$ image presented by
\citet{Kohno2000} \citet{Perez2007} derive a source size of
$5^{\prime\prime}\times 5^{\prime\prime}$. 450\,$\mu$m, 850\,$\mu$m
\citep{Stevens2000}, and 1200\,$\mu$m continuum imaging reveals an
unresolved core $\sim20\arcsec$ in extent.

\noindent{NGC\,3079} is one of two galaxies in our sample which shows
complex absorption and emission structure in its H$_2$CO
$1_{10}-1_{11}$ spectrum.  The $1_{10}-1_{11}$ and $2_{11}-2_{12}$
transitions are dominated by velocity components at $\sim 1010$ and
$\sim 1115$~km~s$^{-1}$ (which is the systemic velocity), with the
$1_{10}-1_{11}$ transition 
exhibiting a tentative third broad component at $\sim
1500$~km~s$^{-1}$.  The two velocity components which are apparent in
both H$_2$CO transitions 
are detected in CH$_3$OH, OH, and HI \citep{Impellizzeri2008}.  The
excitation characteristics of these three velocity components differ
dramatically:
\begin{itemize}
\item \textit{1010\,km\,s$^{-1}$ Component (C1):} $1_{10}-1_{11}$
  emission with $2_{11}-2_{12}$ absorption suggests high spatial
  density.  In subsequent discussion we refer to this as component 1
  (C1). 
\item \textit{1115\,km\,s$^{-1}$ Component (C2):} Complex
  $1_{10}-1_{11}$ emission 
  and absorption spectral structure suggesting that this
  component has a spatial density and kinetic temperature which lies
  near the point where the $1_{10}-1_{11}$ transition goes from
  absorption to emission.  This $1_{10}-1_{11}$ spectral structure
  with $2_{11}-2_{12}$ absorption suggests lower spatial density than
  the 1010~km~s$^{-1}$ component.  In subsequent discussion we refer to this as
  component 2 (C2). 
\item \textit{1500\,km\,s$^{-1}$ Component (C3):} Only detected in
  broad $1_{10}-1_{11}$ absorption at 4.5$\sigma$ confidence in peak
  intensity.  In subsequent discussion we refer 
  to this as component 3 (C3).  We are not aware of corresponding
  spectral line detections of the 1500\,km\,s$^{-1}$ component in
  other high density molecular tracers (though the CO emission toward
  NGC\,3079 includes this velocity; Figure~\ref{fig:NGC2146M82M82SWNGC3079FormSpec}).
\end{itemize}

\noindent
{\bf NGC\,3628:} The $1_{10}-1_{11}$ transition was detected in absorption by 
\citet{Baan1986,Baan1990} and confirmed by \citet{Araya2004}, while
the $2_{11}-2_{12}$ transition was not detected by \citet{Baan1990}.
The GBT $1_{10}-1_{11}$ line properties show good agreement with those
measured by both \citet[][ S = $-2.5$~mJy]{Baan1986} and \citet[][ S =
  $-2.97\pm0.21$~mJy]{Araya2004} to within the calibration
uncertainties (Figure~\ref{fig:NGC3628IC860M83IR15107FormSpec}).  \citet{Israel2009}
measured the $^{12}$CO J=$1-0$, J=$2-1$, J=$3-2$, J=$4-3$ and $^{13}$CO
J=$1-0$, J=$2-1$, and J=$3-2$ emission from this galaxy.  From their
CO J=$3-2$ map one notes that the source size is $<15^{\prime\prime}$.
450\,$\mu$m and 850\,$\mu$m continuum imaging \citep{Stevens2005}
expose a $\sim30^{\prime\prime}$ core.

\noindent
{\bf IC\,860:}  The $1_{10}-1_{11}$ emission line was detected by 
\citet{Baan1993} and \citet{Araya2004}, and the GBT detection shows
good agreement in line properties (Figure~\ref{fig:NGC3628IC860M83IR15107FormSpec}).
Our discovery of H$_2$CO $2_{11}-2_{12}$ absorption is a new result.
$15^{\prime\prime}$ resolution CO J=$2-1$ and $1-0$ emission
measurements \citet{Yao2003} provide an estimate of the maximum source
size.

\noindent
{\bf NGC\,5194 (M51):} The ``Whirlpool Galaxy'' is a well-studied Sbc
galaxy with a 
weak AGN.  The inner 2.$^\prime$5 of this galaxy have been imaged in a
variety of molecular tracers, including CO, $^{13}$CO, and C$^{18}$O
J=$1-0$ (\citet{Kohno1996}; \citet{Aalto1999}), CO J=$2-1$
\citep{Schinnerer2010}, and HCN and HCO$^+$ J=$1-0$
(\citet{Kohno1996}; \citet{Schinnerer2010}).  These dense gas studies
have revealed structure associated with the nearly face-on spiral arms
down to spatial scales $\sim120$ pc.  Our H$_2$CO $1_{10}-1_{11}$
absorption measurement (Figure~\ref{fig:ExgalFormSingles-OnePage}) is
the first reported detection of H$_2$CO in this galaxy.  

\noindent
{\bf M\,83:}  Both H$_2$CO transitions are detected in absorption,
with the 14.5~GHz transition showing a significantly larger optical 
depth than the 4.8~GHz transition (Figure~\ref{fig:NGC3628IC860M83IR15107FormSpec}).
CO J=$2-1$ and $3-2$ emission imaging of this barred starburst galaxy
\citep{Sakamoto2004} indicates a
$45^{\prime\prime}\times15^{\prime\prime}$ structure.

\noindent
{\bf Mrk\,273:} Imaged in CO J=$3-2$ by \citet{Wilson2008}.  Based on
these \citet{Wilson2008} measurements \citet{Iono2009} derive a
deconvolved CO J=$3-2$ source size of
$0.4^{\prime\prime}\times0.3^{\prime\prime}$.  \citet{GraciaCarpio2008}
measured moderately intense HCN and HCO$^+$ J=$1-0$ and $3-2$, along with CO
J=$1-0$ emission from this ULIRG.  Furthermore, \textit{Herschel} has
detected OH emission due to outflows from this galaxy
\citep{Fischer2010}.  Our H$_2$CO $1_{10}-1_{11}$ absorption
measurement is the first reported measurement of H$_2$CO in this
galaxy (Figure~\ref{fig:ExgalFormSingles-OnePage}).

\noindent
{\bf IR\,15107+0724:}  Our $1_{10}-1_{11}$ measurement is consistent
with that reported by \citet{Baan1993} and \citet{Araya2004}.
We also detect the $2_{11}-2_{12}$ line in absorption
(Figure~\ref{fig:NGC3628IC860M83IR15107FormSpec}). \citet{Planesas1991} imaged the CO
J=$1-0$ emission from this galaxy, deriving a compact nuclear source
with size $\sim3^{\prime\prime}$. 

\noindent
{\bf Arp\,220/IC\,4553:} The H$_2$CO $1_{10}-1_{11}$ transition was
detected in emission by \citet{Baan1986}, but the $2_{11}-2_{12}$ line
was not subsequently detected \citep{Baan1990}.  We find good
agreement between our GBT measurement (Figure~\ref{fig:Arp220NGC6946FormSpec})
and the total integrated intensity of the $1_{10}-1_{11}$
sub-arcsecond resolution interferometric measurements reported by
\citet{Baan1995}.  Our 4.8 and 14.5~GHz GBT spectra are consistent
with those reported by \citet{Araya2004}.  Note 
also that previous studies \citep{Baan1986} have suggested that
H$_2$CO $1_{10}-1_{11}$ emission is produced by maser amplification in
this galaxy.  In \citet{Mangum2008} we argued that, to the contrary, the
$1_{10}-1_{11}$ emission in Arp\,220 is simply the signature of a dense
gas component in this ULIRG.

\noindent{\citet{Mangum2008}} summarized the previous molecular
spectral line measurements of Arp\,220.  Of particular note is the
dichotomy between the relative intensities of various dense gas
tracers from the two nuclei which comprise this merging system.  As
noted by \citet{Greve2009}, this dichotomy reflects differing spatial
densities in these two components: the western component possessing
lower spatial densities, the eastern component possessing higher 
spatial densities.  Our H$_2$CO measurements partially reflect this
trend.  Both H$_2$CO transitions can be fit with velocity components
at $\sim 5330$ and $\sim 5460$~km~s$^{-1}$, likely corresponding to
the western and eastern velocity components noted previously.  We also
find an indication of a third component at $\sim 5600$~km~s$^{-1}$.
We are not aware of corresponding spectral line detections of this
$\sim 5600$~km~s$^{-1}$ velocity component in other molecular species.

\noindent{Recent} \textit{Herschel}-SPIRE imaging spectroscopic
measurements of the 
molecular emission toward Arp\,220 \citep{Rangwala2011} have
characterized the physical conditions in this galaxy using
measurements of $^{12}$CO and HCN emission over a wide range of
molecular excitation.  By using a radiative transfer model with
Bayesian likelihood analysis of the J=$4-3$ through J=$13-12$
$^{12}$CO emission toward Arp\,220 \cite{Rangwala2011} derived two
temperature components: A cool component with median T$_{K}$ = 50\,K
(1-$\sigma$ range 34--67\,K) and n(H$_2$) = $10^{2.8}$\,cm$^{-3}$
(1-$\sigma$ range $10^{2.6}$--$10^{3.2}$\,cm$^{-3}$) and a warm
component with median T$_{K}$ = 1343\,K (1-$\sigma$ range
1247--1624\,K) and n(H$_2$) = $10^{3.2}$\,cm$^{-3}$ (1-$\sigma$ range
$10^{3.0}$--$10^{3.2}$\,cm$^{-3}$).  A similar LVG radiative transfer
and Bayesian likelihood analysis of the HCN J=$12-11$ through J=$17-16$
emission yielded a single-temperature fit with T$_K \gtrsim 320$\,K
and n(H$_2$) $\gtrsim 10^{6.3}$\,cm$^{-3}$.  Our H$_2$CO-derived
spatial density (n(H$_2$) = $10^{4.09\pm0.09}$\,cm$^{-3}$) and
NH$_3$-derived kinetic temperature (T$_k$ = $234\pm52$\,K) correspond most 
closely to the spatial density and kinetic temperature derived from the
\cite{Rangwala2011} HCN measurements, though are slightly lower in
both quantities (see Table~\ref{tab:lvg}).  The cool and warm
components derived from the \cite{Rangwala2011} CO analysis do not
correspond to any physical components in our H$_2$CO or NH$_3$
measurements, suggesting that the contribution to the CO emitting gas
in Arp\,220 from high spatial density components is small.

\noindent
{\bf NGC\,6240:}  Our GBT detection of the $1_{10}-1_{11}$ transition
in absorption (Figure~\ref{fig:ExgalFormSingles-OnePage}; 5.2$\sigma$
in a single smoothed 20 km~s$^{-1}$ channel) is not consistent with
the \textit{emission} reported by \citet{Baan1993}.  As noted by
\citet{Mangum2008}, our $1_{10}-1_{11}$ spectrum peaks near 7300
km~s$^{-1}$, slightly blueshifted relative to the systemic velocity
of 7359 km~s$^{-1}$ \citep{Greve2009}.  CO J=$1-0$
images \citep{Bryant1999} indicate a molecular gas source size of
$10^{\prime\prime}$.  NGC\,6240 has also been imaged in CO J=$3-2$ and 
HCO$^+$ J=$4-3$ by \citet{Wilson2008}.  Based on the
\citet{Wilson2008} measurements \citet{Iono2009} derive a deconvolved
CO J=$3-2$ source size of
$0.9^{\prime\prime}\times0.6^{\prime\prime}$.  The HCO$^+$ J=$4-3$
image presented by \citet{Wilson2008} suggests a source size
$<0.5^{\prime\prime}$.

\noindent
{\bf NGC\,6921:} \citet{Gao2004a} measure very weak CO and HCN J=$1-0$
($\sim10$ and 1\,mK, respectively) emission toward this spiral galaxy.
We are not aware of any spectral line imaging measurements of this
galaxy at radio wavelengths.  Somewhat surprisingly, given the low
intensity of the HCN J=$1-0$ measurement, we detect significant
H$_2$CO $1_{10}-1_{11}$ absorption toward this galaxy
(Figure~\ref{fig:ExgalFormSingles-OnePage}).  As the 4.8 GHz 
continuum flux is measured to be only 28 mJy toward this galaxy,
H$_2$CO absorption of a strong background continuum source does not
provide a mechanism for producing significant H$_2$CO absorption.  The
H$_2$CO $1_{10}-1_{11}$ absorption we measure is clearly absorbing the
cosmic microwave background.

\noindent
{\bf NGC\,6946:}  In \citet{Mangum2008} we reported the discovery of
the H$_2$CO $1_{10}-1_{11}$ transition in absorption.  We add to this
the H$_2$CO $2_{11}-2_{12}$ transition, also in absorption
(Figure~\ref{fig:Arp220NGC6946FormSpec}).  \citet{Schinnerer2006} 
and \citet{Schinnerer2007} imaged the CO J=$1-0$, J=$2-1$, and HCN
J=$1-0$ emission from this galaxy, detecting a compact nuclear source
size of $\sim2^{\prime\prime}$ and a ``nuclear spiral'' structure
$5^{\prime\prime}\times10^{\prime\prime}$ in size.

\subsection{H$_2$CO Apparent Optical Depth Calculations}
\label{OpticalDepth}

The apparent peak optical depths \citep[$\tau$, see Equation~2 of
][]{Mangum2008} for our H$_2$CO absorption measurements
(Table~\ref{tab:h2comeasurements}) have been calculated using the GBT
continuum emission intensities listed in Table~\ref{tab:contmeas} and
assuming T$_{ex} \ll$ T$_c$.  With the exception of NGC\,6921, which has
an optical depth of $\sim0.06$, all of the other galaxies detected in
H$_2$CO in our sample have optical depths $\leq 0.007$, indicating
optically-thin H$_2$CO emission.

\subsection{Spatial Density and Column Density Derivation Using LVG Models}
\label{LVG}

As was done in \citet{Mangum2008}, to derive the H$_2$ spatial density
(number density) and H$_2$CO column density of the dense gas in our
galaxy sample, we use a Large Velocity Gradient (LVG) model
\citep{Sobolev1960} to calculate the radiative transfer properties of
the H$_2$CO transitions.  The detailed properties of our
implementation of the LVG approximation are described in
\citet{Mangum1993}.  One 
important point regarding this implementation of the LVG model is our
scaling of the calculated ortho-H$_2$CO/He excitation rates to those
appropriate for collisions with H$_2$.  Following the recommendation
of \cite{Green1991}, we scale the calculated He rates by a factor of
2.2 to account for (1) the reduced collision velocity of He relative
to H$_2$, which scales as the inverse-square-root of the masses
of He and H$_2$ and (2) the larger cross section of the H$_2$ molecule
($\sim 1.6$; \cite{Nerf1975}) relative to He.

Radiative transfer models of the molecular emission in astrophysical
environments are dependent upon collisional excitation rates for the
molecule(s) under study.  As noted by \citet{Mangum1993} and
\citet{Mangum2008}, the uncertainty associated with the collisional
excitation rates must be considered in any analysis of the physical
conditions derived from radiative transfer modelling.
\citet{Green1991} suggests that the total collisional excitation rate
for a given H$_2$CO transition is accurate to $\sim 20\%$.  This
implies that the physical conditions derived from our LVG modelling
are limited to an accuracy of no better than 20\%.

For the 13 galaxies and one galaxy offset position (M\,82SW) where
both the H$_2$CO $1_{10}-1_{11}$ and $2_{11}-2_{12}$ transitions were
detected we can derive a unique solution to the (ensemble average)
spatial density and H$_2$CO column density \textit{for an assumed gas
  kinetic temperature}.  This unique solution to the physical
conditions is derived by fitting to the intercept between the H$_2$CO
$1_{10}-1_{11}$ and $2_{11}-2_{12}$ transition ratio and the H$_2$CO
$1_{10}-1_{11}$ transition intensity at the assumed kinetic
temperature.  Absorption line measurements afford the possibility of
using the measured apparent optical depths in the LVG model fitting
procedure.  For our
galaxy sample, where with the exception of M\,82 $T_c \lesssim
T_{cmb}$, this method has the disadvantage of involving an estimate of
the excitation temperature (T$_{ex}$) in order to derive the apparent
optical depth.
Since T$_{ex}$ is dependent upon the input physical conditions that we
are attempting to derive, we do not believe that fitting to measured
apparent H$_2$CO optical depths is advantageous.  In fact, 
as we showed in \S~5.3.3 of \citet{Mangum2008}, these two approaches to
estimating the excitation of H$_2$CO absorption lines in the presence
of the cosmic background radiation and weak ambient continuum emission
are roughly equivalent for the measurements presented in this paper.
Furthermore, we showed in \citet{Mangum2008} that for M\,82, which
possesses strong continuum emission, ignoring contributions due to
background continuum emission produces a derived spatial density which
is a lower-limit to the true average spatial density in this galaxy.
 
As was done in \citet{Mangum2008}, a model grid of predicted H$_2$CO transition
intensities over a range and step size in spatial density, ortho-H$_2$CO column
density per velocity gradient, and kinetic temperature (n(H$_2$),
N(ortho-H$_2$CO)/$\Delta v$, T$_K$) = ($10^{4.0}-10^{7.0}$\,cm$^{-3}$,
$10^{10.0}-10^{14.0}$\,cm$^{-2}$/km\,s$^{-1}$, 20--300\,K) and
($\Delta\log$(n(H$_2$)), $\Delta\log$(N(ortho-H$_2$CO)/$\Delta v$),
$\Delta$T$_K$) = (0.03\,(cm$^{-3}$), 0.04\,(cm$^{-2}$/km\,s$^{-1}$), 5\,K) was
calculated. Note that the model grid used in the present calculations is
significantly larger in all three modeled parameters than that used in
\citet{Mangum2008}.  The predicted transition intensities were then compared to
our measured H$_2$CO $1_{10}-1_{11}$ and $2_{11}-2_{12}$ transition
intensities (Table~\ref{tab:h2comeasurements}).  

Table~\ref{tab:lvg} lists the derived LVG model best-fit physical
conditions for all of the H$_2$CO-detected galaxies in our sample.
There are three different sets of assumptions used to derive these
physical conditions dependent upon the information available:
\begin{itemize}
\item \textit{Both Transitions Detected:} For the 13 galaxies and one
  galaxy offset position where we
  have detected both the H$_2$CO $1_{10}-1_{11}$ and $2_{11}-2_{12}$
  transitions, we calculate explicit values for the spatial density
  and ortho-H$_2$CO column density assuming T$_{K}$ = T$_{dust}$, with
  T$_{dust}$ from Table~\ref{tab:galaxies}.  For the 12 galaxies where
  we also have NH$_3$-based kinetic temperature measurements
  \citep{Mangum2013}, we also derive spatial density and column
  density assuming these gas-derived kinetic temperatures.  As NGC\,253
  and IC\,342 possess two NH$_3$-derived temperature components LVG
  model fits assuming both of these kinetic temperatures are listed.
  For the 
  LVG model fits with NH$_3$-derived kinetic temperatures, kinetic
  temperature uncertainties \citep{Mangum2013} are included.  By
  using NH$_3$-derived kinetic temperatures to constrain our
  H$_2$CO-derived spatial densities we are assuming cospatiality for
  these two dense molecular gas tracers.  As the spectral line
  profiles from NH$_3$ and H$_2$CO are similar in all but one galaxy
  (NGC\,660; see \S\ref{NGC660}), the assumption of cospatiality seems
  plausible.
\item \textit{One Transition Detected and Limit to Other Transition:}
  For the 8 galaxies and one galaxy velocity component where we have
  detected only one of the two K-doublet 
  transitions, but also have a limit to the nondetected transition, we
  make two estimates of the physical conditions:
  \begin{itemize}
  \item We derive limits to both the spatial density and ortho-H$_2$CO
    column density assuming a kinetic temperature of 40~K, a $3\sigma$
    intensity limit, and an equivalent line width for the nondetected
    transition.  Furthermore, since our LVG modeling results imply
    that the $1_{10}-1_{11}$ transition goes into emission before the
    $2_{11}-2_{12}$ transition as the spatial density is increased
    (for a fixed kinetic temperature), a detection of absorption in
    the $1_{10}-1_{11}$ transition means that the (undetected)
    $2_{11}-2_{12}$ transition \textit{must} also be in absorption.
    On the other hand, a detection of the $1_{10}-1_{11}$ transition
    in emission implies that the $2_{11}-2_{12}$ transition can either
    be in emission or absorption.  These results are listed as limits
    in columns 3 and 4 of Table~\ref{tab:lvg}.
  \item As was done in \citet{Mangum2008}, we calculate the LVG model
    derived column density assuming n(H$_2$) = $10^5$ cm$^{-3}$ (for
    absorption sources) or $10^{5.65}$ cm$^{-3}$ (for emission
    sources).  These assumed values for the spatial density are
    reasonable averages from our H$_2$CO galaxy sample where spatial
    density was derived.  We also
    assume\footnote{We assumed T$_K$ = 40\,K in \citet{Mangum2008}.
      Note, though, that T$_{dust} \simeq$ 40\,K for most galaxies in
      our sample.} T$_K$ = T$_{dust}$.  This line of
    reasoning allows us to derive an actual value rather than a limit
    for N(ortho-H$_2$CO)/$\Delta$v that we can use in a calculation of
    the dense gas mass.  The column densities derived using these
    assumptions are listed in column 3 of Table~\ref{tab:lvg}.
  \end{itemize}
\item \textit{Neither Transition Detected:} For the 35 galaxies where
  we derive only limits to the $1_{10}-1_{11}$ and/or the
  $2_{11}-2_{12}$ transition intensities we assume a $3\sigma$ limit
  to the measured yet undetected intensities.  This allows us to
  calculate a limit to the ortho-H$_2$CO column density assuming
  n(H$_2$) = $10^5$ cm$^{-3}$ and T$_K$ = T$_{dust}$.  Note that we
  conservatively do not assume a line width in these limits, which
  would decrease the intensity limit by the square-root of the number
  of channels over which the line is integrated.  
\end{itemize}
For all of our LVG calculations we assume T$_{cmb}$ = 2.73 K and negligible
contribution due to any background continuum emission (T$_c$ = 0).  In
\S~5.3.3 of \citet{Mangum2008} we showed that our results represent a
lower-limit to the true spatial density when contributions due to
strong background continuum sources (S $>$ 1\,Jy) are neglected.  

%
%

%
\begin{deluxetable*}{lccc}
\tablewidth{325pt}
\tablecolumns{4}
\tablecaption{Derived Physical Conditions\label{tab:lvg}}
\tablehead{
\colhead{Galaxy} & 
\colhead{T$_K$\tablenotemark{a}} &
\colhead{log(N(ortho-H$_2$CO)/$\Delta$v)\tablenotemark{b,c}} & 
\colhead{log(n(H$_2$))\tablenotemark{b}} \\
& \colhead{(K)} &
\colhead{(cm$^{-2}$/km s$^{-1}$)} & 
\colhead{(cm$^{-3}$)}
}
\startdata
NGC\,55        & T$_{dust}$ = 40\tablenotemark{d} & $<11.31$ & \nodata \\
NGC\,253       & T$_{dust}$ = 34 & $12.14\pm0.04$ & $5.11\pm0.02$ \\
              & T$_K$(NH$_3$) = $78\pm22$ & $12.24\pm0.06$ & $5.03\pm0.06$ \\
              & T$_K$(NH$_3$) $>150$ & $>12.54$ & $<4.82$ \\
IC\,1623       & T$_{dust}$ = 39 & $<10.71$ & \nodata \\
NGC\,520       & T$_{dust}$ = 38 & $<10.84, 11.13\pm0.08$ & $<4.58$ \\
NGC\,598       & T$_{dust}$ = 40\tablenotemark{d} & $<10.87$ & \nodata \\
NGC\,660       & T$_{dust}$ = 37 & $11.53\pm0.16$ & $5.34\pm0.04$ \\
              & T$_K$(NH$_3$) = $160\pm96$\tablenotemark{e} & $11.94\pm0.48$ & $4.88\pm0.48$ \\
IR\,01418+1651 & T$_{dust}$ = 40\tablenotemark{d} & $<10.72$ & \nodata \\
NGC\,695       & T$_{dust}$ = 34 & $<10.88$ & \nodata \\
Mrk\,1027      & T$_{dust}$ = 37 & $<11.07$ & \nodata \\
NGC\,891       & T$_{dust}$ = 28 & $<10.66, 10.95\pm0.10$ & $<4.50$ \\
NGC\,925       & T$_{dust}$ = 40\tablenotemark{d} & $<11.12$ & \nodata \\
NGC\,1022      & T$_{dust}$ = 39 & $<11.05$ & \nodata \\
NGC\,1055      & T$_{dust}$ = 29 & $<10.88$ & \nodata \\
Maffei\,2      & T$_{dust}$ = 40\tablenotemark{d} & $11.17\pm0.12$ & $4.93\pm0.09$ \\
Maffei\,2C1    & T$_K$(NH$_3$) = $62\pm25$ & $11.27\pm0.87$ & $5.00\pm0.50$ \\
Maffei\,2C2    & T$_K$(NH$_3$) = $64\pm24$ & $11.00\pm0.50$ & $4.91\pm0.27$ \\
NGC\,1068      & T$_{dust}$ = 40 & $<11.00$ & \nodata \\
UGC\,02369     & T$_{dust}$ = 40\tablenotemark{d} & $<11.00$ & \nodata \\
NGC\,1144      & T$_{dust}$ = 40\tablenotemark{d} & $<10.91, 11.31\pm0.02$ & $<4.16$ \\
NGC\,1365      & T$_{dust}$ = 32 & $<10.71, 10.83\pm0.06$ & $<4.85$ \\
IR\,03359$+$1523 & T$_{dust}$ = 40\tablenotemark{d} & $<11.08$ & \nodata \\
IC\,342        & T$_{dust}$ = 30 & $11.27\pm0.22$ & $5.05\pm0.11$ \\
              & T$_K$(NH$_3$) = $24\pm7$ & $11.33\pm0.28$ & $5.02\pm0.14$ \\
              & T$_K$(NH$_3$) $>150$ & $>11.37$ & $<4.88$ \\
NGC\,1614      & T$_{dust}$ = 46 & $<11.16$ & \nodata \\
VIIZw31       & T$_{dust}$ = 34 & $<11.06$ & \nodata \\
NGC\,2146      & T$_{dust}$ = 38 & $11.25\pm0.12$ & $5.33\pm0.03$ \\
NGC\,2623      & T$_{dust}$ = 40\tablenotemark{d} & $<10.97$ & \nodata \\
Arp\,55        & T$_{dust}$ = 36 & $<12.42$ & \nodata \\
NGC\,2903      & T$_{dust}$ = 29 & $<10.59$ & \nodata \\
UGC\,5101      & T$_{dust}$ = 36 & $<10.89$ & \nodata \\
M\,82          & T$_{dust}$ = 45 & $11.80\pm0.02$ & $4.95\pm0.02$ \\
              & T$_K$(NH$_3$) = $58\pm19$ & $11.84\pm0.06$ & $4.92\pm0.05$ \\
M\,82SW        & T$_{dust}$ = 45 & $11.86\pm0.08$ & $5.09\pm0.03$ \\
              & T$_K$(NH$_3$) = $58\pm19$ & $11.86\pm0.08$ & $5.05\pm0.08$ \\
NGC\,3079C1    & T$_{dust}$ = 32 & $12.34\pm0.28$ & $5.56\pm0.02$ \\
              & T$_K$(NH$_3$) $>100$ & $>12.46$ & $<5.39$ \\
NGC\,3079C2    & T$_{dust}$ = 32 & $11.83\pm0.10$ & $5.46\pm0.02$ \\
              & T$_K$(NH$_3$) $>150$ & $>12.22$ & $<5.09$ \\
NGC\,3079C3    & T$_{dust}$ = 32 & $<11.07, 10.48\pm0.08$ & $<5.33$ \\
IR\,10173+0828 & T$_{dust}$ = 40\tablenotemark{d} & $<12.36$ & \nodata \\
NGC\,3227      & T$_{dust}$ = 40\tablenotemark{d} & $<12.33$ & \nodata \\
NGC\,3627      & T$_{dust}$ = 30 & $<10.90$ & \nodata \\
NGC\,3628      & T$_{dust}$ = 30 & $11.19\pm0.10$ & $4.65\pm0.14$ \\
NGC\,3690      & T$_{dust}$ = 40\tablenotemark{d} & $<10.86$ & \nodata \\
NGC\,4631      & T$_{dust}$ = 30 & $<10.92$ & \nodata \\
NGC\,4736      & T$_{dust}$ = 40\tablenotemark{d} & $<11.33$ & \nodata \\
Mrk\,231       & T$_{dust}$ = 47 & $<11.02$ & \nodata \vspace{2pt}
\enddata
\end{deluxetable*}
 
\addtocounter{table}{-1}
\begin{deluxetable*}{lccc}[H]
\tablewidth{325pt}
\tablecolumns{4}
\tablecaption{\vspace{-8.3pt} \hspace{80pt} --- {\it Continued}}
\tablehead{
\colhead{Galaxy} & 
\colhead{T$_K$\tablenotemark{a}} &
\colhead{log(N(ortho-H$_2$CO)/$\Delta$v)\tablenotemark{b,c}} & 
\colhead{log(n(H$_2$))\tablenotemark{b}} \\
& \colhead{(K)} &
\colhead{(cm$^{-2}$/km s$^{-1}$)} & 
\colhead{(cm$^{-3}$)}
}
\startdata
NGC\,5005      & T$_{dust}$ = 28 & $<11.25$ & \nodata \\
IC\,860        & T$_{dust}$ = 40\tablenotemark{d} & $12.32\pm0.06$ &
$5.70\pm0.02$ \\
              & T$_K$(NH$_3$) = $206\pm79$ & $12.44\pm0.18$ & $4.79\pm0.54$ \\
NGC\,5194      & T$_{dust}$ = 40\tablenotemark{d} & $<10.67, 10.70\pm0.06$ & $<4.89$ \\
M\,83          & T$_{dust}$ = 31 & $11.45\pm0.32$ & $5.29\pm0.10$ \\
              & T$_K$(NH$_3$) = $56\pm15$ & $11.49\pm0.40$ & $5.27\pm0.15$ \\
Mrk\,273       & T$_{dust}$ = 48 & $<11.09, 11.13\pm0.12$ & $<4.86$ \\
NGC\,5457      & T$_{dust}$ = 40\tablenotemark{d} & $<10.94$ & \nodata \\
IR\,15107+0724 & T$_{dust}$ = 40\tablenotemark{d} & $12.46\pm0.12$ &
$5.65\pm0.02$ \\
               & T$_K$(NH$_3$) = $189\pm57$ & $12.71\pm0.12$ & $4.92\pm0.35$ \\
Arp\,220       & T$_{dust}$ = 44 & $12.63\pm0.02$ & $5.64\pm0.02$ \\
              & T$_K$(NH$_3$) = $234\pm52$ & $12.79\pm0.08$ & $4.09\pm0.09$ \\
NGC\,6240      & T$_{dust}$ = 40\tablenotemark{d} & $<10.61, 10.75\pm0.10$ & $<4.80$ \\
IR\,17208-0014 & T$_{dust}$ = 46 & $<11.05$ & \nodata \\
IR\,17468+1320 & T$_{dust}$ = 40\tablenotemark{d} & $<11.11$ & \nodata \\
NGC\,6701      & T$_{dust}$ = 32 & $<10.63$ & \nodata \\
NGC\,6921      & T$_{dust}$ = 34 & $<10.65, 11.03\pm0.06$ & $<4.27$ \\
NGC\,6946      & T$_{dust}$ = 30 & $11.01\pm0.36$ & $5.05\pm0.20$ \\
              & T$_K$(NH$_3$) = $47\pm8$ & $10.99\pm0.34$ & $5.06\pm0.21$ \\
 IC\,5179       & T$_{dust}$ = 33 & $<12.28$ & \nodata \\
NGC\,7331      & T$_{dust}$ = 28 & $<10.81$ & \nodata \\
NGC\,7479      & T$_{dust}$ = 36 & $<11.03$ & \nodata \\
IR\,23365+3604 & T$_{dust}$ = 45 & $<10.89$ & \nodata \\
Mrk\,331       & T$_{dust}$ = 41 & $<11.05$ & \nodata \vspace{2pt}
\enddata
\tablenotetext{a}{~T$_{dust}$ from Table~\ref{tab:galaxies};
  T$_K$(NH$_3$) from \citet{Mangum2013} and \citet{Ao2011} for NGC\,1068.} 
\tablenotetext{b}{~See \S\ref{LVG} for assumptions used in calculating
  these quantities.}
\tablenotetext{c}{~LVG-derived column densities listed as limits for
  unconstrained density fits, followed by column density derived
  assuming a fixed density.  See \S\ref{LVG} for details.}
\tablenotetext{d}{~Assumed value.}
\tablenotetext{e}{~Kinetic temperature for the polar ring component
  (see \S\ref{NGC660}).}
\end{deluxetable*}

\subsubsection{The Multiple Dense Gas Components of NGC\,660}
\label{NGC660}

Comparison of the NH$_3$ and H$_2$CO K-doublet spectra toward NGC\,660
suggests that the bulk of the emission from these two molecules
originates from different spatial components.  The NH$_3$ (J,K) =
(1,1) through (7,7) (for J=K) and (2,1) rotation-inversion transitions
have been detected toward NGC\,660 with multiple absorption components
comprising a total FWZI $\sim 50-100$\,km\,s$^{-1}$ centered near
V$_{hel} \sim 840$\,km\,s$^{-1}$ \citep{Mangum2013}.  The V$_{hel}$
and FWZI values determined from the NH$_3$ spectra correspond to the
narrow absorption peak embedded 
within the H$_2$CO $2_{11}-2_{12}$ absorption component
(Figure~\ref{fig:NGC253NGC660Maffei2IC342FormSpec}).  Analysis of the (narrow) NH$_3$
emission within NGC\,660 \citep{Mangum2013} strongly suggests that it
contains four velocity components, two of which originate in the disk
of the galaxy.  The bulk of the 
H$_2$CO absorption traces the nuclear region of this galaxy.  In spite
of the partial non-cospatiality of the NH$_3$ and H$_2$CO absorption
in this galaxy, we present physical condition calculations assuming
both the dust- and NH$_3$-derived kinetic temperature for H$_2$CO
components associated with the disk component.

\subsubsection{LVG Model Dependence on Kinetic Temperature}
\label{LVGTk}

It is difficult to determine kinetic temperatures in extragalactic
molecular clouds. The most common kinetic temperature probe is
interstellar dust, which is available for a large number of
galaxies.  However, dust and gas kinetic temperatures, even
though being about 10 K in local dark clouds, are not always the same.
Good coupling is normally only achieved at densities in excess of
$10^5$ cm$^{-3}$. In the case of substantial gas heating through
cosmic rays \citep{Papadopoulos2010}, even high densities
cannot prevent a significant discrepancy between dust and gas
temperatures. 

Ammonia (NH$_3$) is a more direct probe of the gas.  Its temperature
sensitive inversion transitions trace molecular gas in star forming as
well as in quiescent regions
\citep[\eg,][]{Benson1983,Mauersberger2003}. As is the case with many 
molecules, NH$_3$ shows some chemical peculiarities in star formation
regions in our own Galaxy, which have to be kept in mind.  NH$_3$ is
easily destroyed in photon dominated regions and shows particularly
high abundances in hot cores, where temperatures of 100 K or more lead
to dust grain mantle evaporation
\citep[\eg,][]{Mauersberger1987}. Differences in abundances in
individual Galactic regions 
can amount to two orders of magnitude. This is also the case when
comparing NH$_3$ abundances between M\,82 and other detected
galaxies. Unlike dust, ammonia has been measured in only a limited
number of galaxies. 

Formaldehyde is another direct probe of kinetic temperature
\citep{Mangum1993}.  The first application of the technique described by
\citet{Mangum1993}, using the H$_2$CO $3_{03}-2_{02}$,
$3_{22}-2_{21}$, and $3_{21}-2_{20}$ transitions, has been applied to
the kinetic temperature measurement of M\,82 by \citet{Muehle2007}.

\begin{figure}
\resizebox{\hsize}{!}{
\includegraphics[scale=0.80,trim=30pt 110pt 20pt 150pt,clip=true]{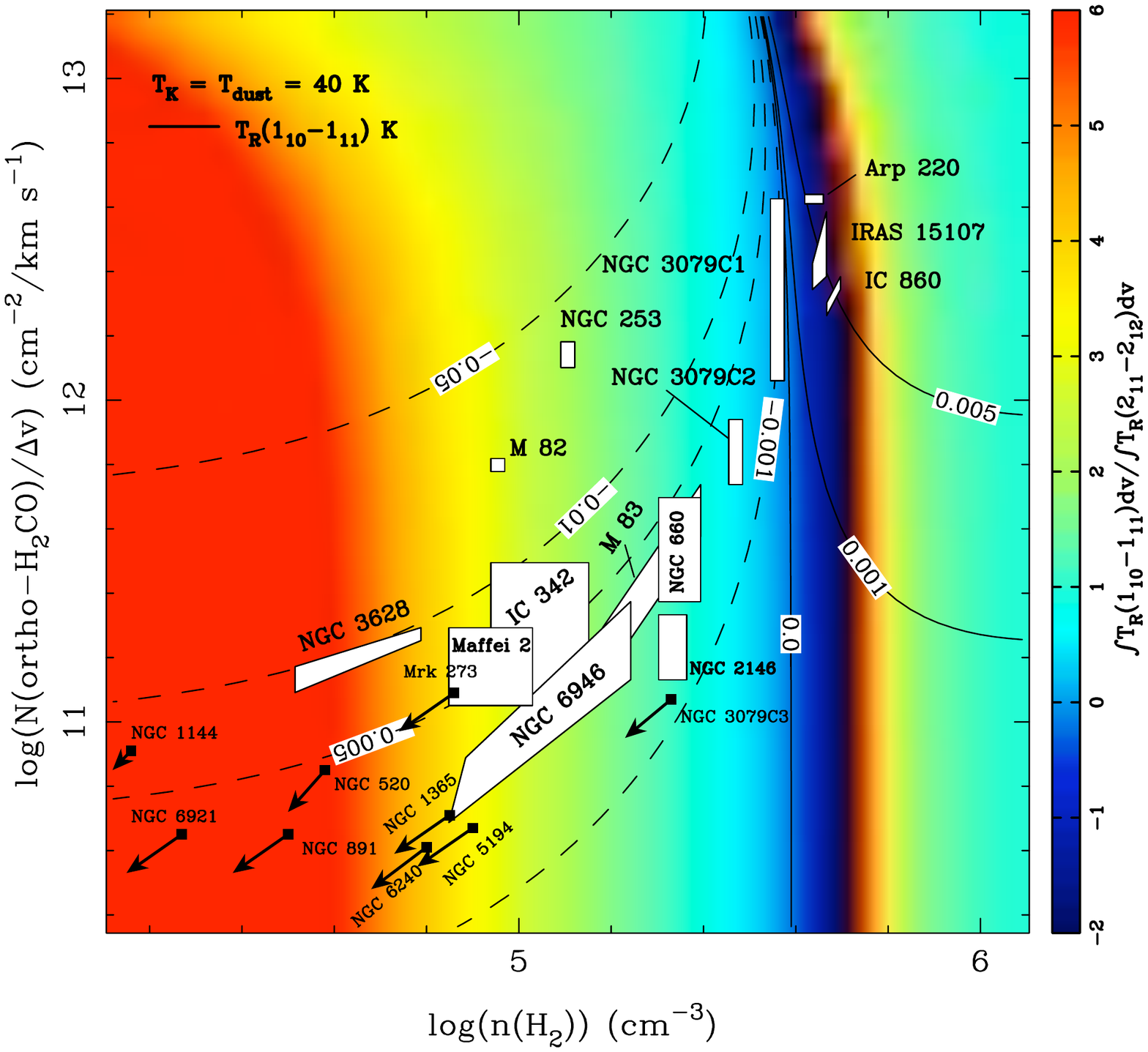}}  
\caption{LVG model predictions for the spatial density (n(H$_2$)) and
  ortho-H$_2$CO column density per velocity gradient
  (N(ortho-H$_2$CO)/$\Delta$v) for the 11 galaxies with a total of 14
  velocity components in our sample with at least one detection of either the
  H$_2$CO $1_{10}-1_{11}$ or $2_{11}-2_{12}$ transition where we have
  assumed that T$_K$ = T$_{dust}$.  Arrows indicate limits for
  galaxies with only one detected transition.  Fit results are
  overlain on model line ratios (color) and intensities (solid
  (positive) and dashed (negative) contours, in K) for an assumed
  kinetic temperature of 40 K and no background continuum emission
  (T$_c$ = 0).}
\label{fig:ExgalH2COModPlotTd}
\end{figure}

As noted in \citet{Mangum2008}, our LVG model-based derivations of the spatial
density and ortho-H$_2$CO column density are dependent upon the
assumed kinetic temperature used in these models.  As we did not have
a complete set of gas kinetic temperature measurements for the sample
of starburst galaxies presented in \citet{Mangum2008}, we assumed T$_K$ =
T$_{dust}$, which were generally in the range 30--40\,K.  An update of
Figure~10 from \citet{Mangum2008}, showing a graphical representation of our LVG
model derivations of the spatial density and ortho-H$_2$CO column
density assuming T$_K$ = T$_{dust}$ is shown in
Figure~\ref{fig:ExgalH2COModPlotTd}.  These results of \citet{Mangum2008}
naturally lead to the conclusion that spatial density is an important
factor in the development and evolution of starbursts in galaxies.
The higher mean spatial densities of $\gtrsim 10^{5.6}$~cm$^{-3}$
found in the LIRG and ULIRG (Arp\,220, IR\,15107+0724)
measurements of \citet{Mangum2008} contrast with the lower mean 
spatial densities of $10^4-10^{5.6}$~cm$^{-3}$ found in the normal
starburst galaxies of \citet{Mangum2008}.

\begin{figure*}
\centering
\includegraphics[scale=0.75,trim=30pt 110pt 80pt 165pt,clip=true]{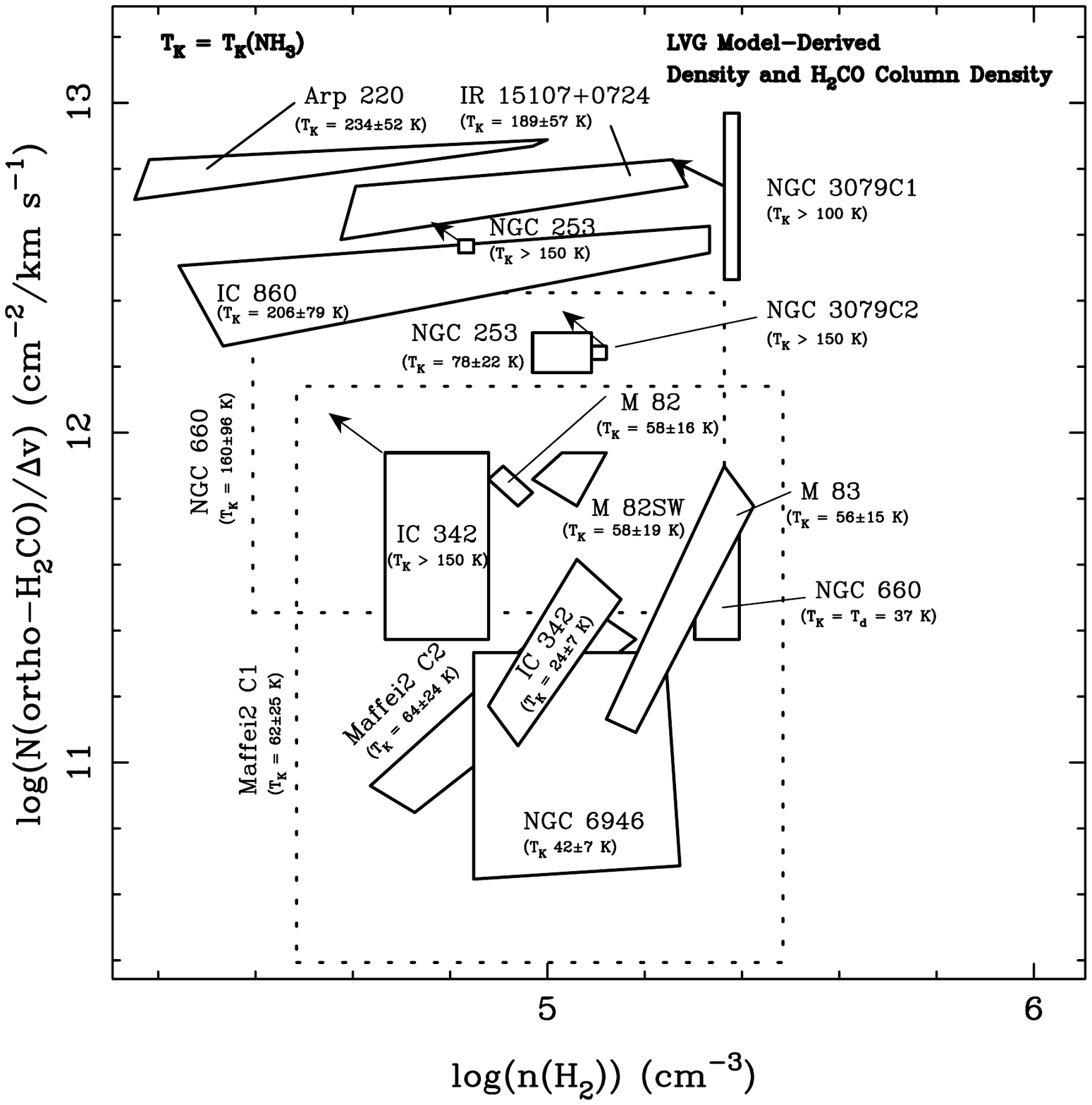}  
\caption{LVG model predictions for the spatial density (n(H$_2$)) and
  ortho-H$_2$CO column density per velocity gradient
  (N(ortho-H$_2$CO)/$\Delta$v) for the 12 galaxies and galaxy velocity
  components in our sample with measured H$_2$CO $1_{10}-1_{11}$
  and $2_{11}-2_{12}$ and NH$_3$-derived T$_K$ \citep{Mangum2013}.
  For each galaxy or galaxy component the NH$_3$-measured T$_K$ used
  in the LVG model fit is listed.  Galaxies with a lower limit to the
  kinetic temperature are displayed as a range in (n,N) at the
  lower-limit kinetic temperature, with an arrow indicating the trend
  in (n,N) as the kinetic temperature is raised.  As the fit range is
  so large for Maffei\,2C1 (due to low signal-to-noise) and the
  high-temperature fit to NGC\,660, the fit ranges
  for these galaxy components are shown dotted.  The need for the two
  fit ranges shown for NGC\,660 is described in \S\ref{NGC660}.}
\label{fig:ExgalH2COModFits}
\end{figure*}

\begin{figure}
\resizebox{\hsize}{!}{
\includegraphics[scale=0.50,trim=30pt 110pt 80pt 165pt,clip=true]{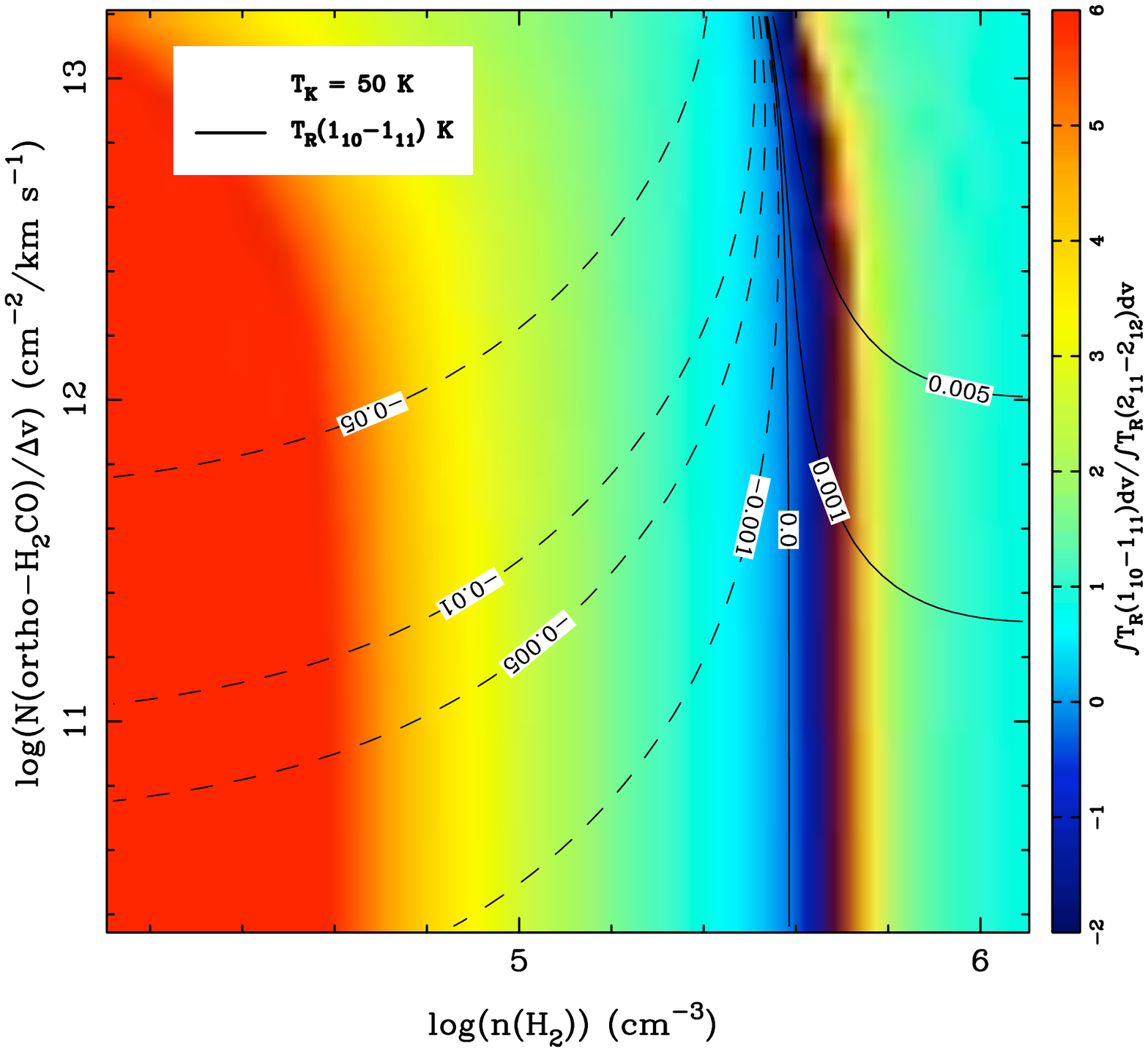}}\\ 
\resizebox{\hsize}{!}{
\includegraphics[scale=0.50,trim=30pt 110pt 80pt 165pt,clip=true]{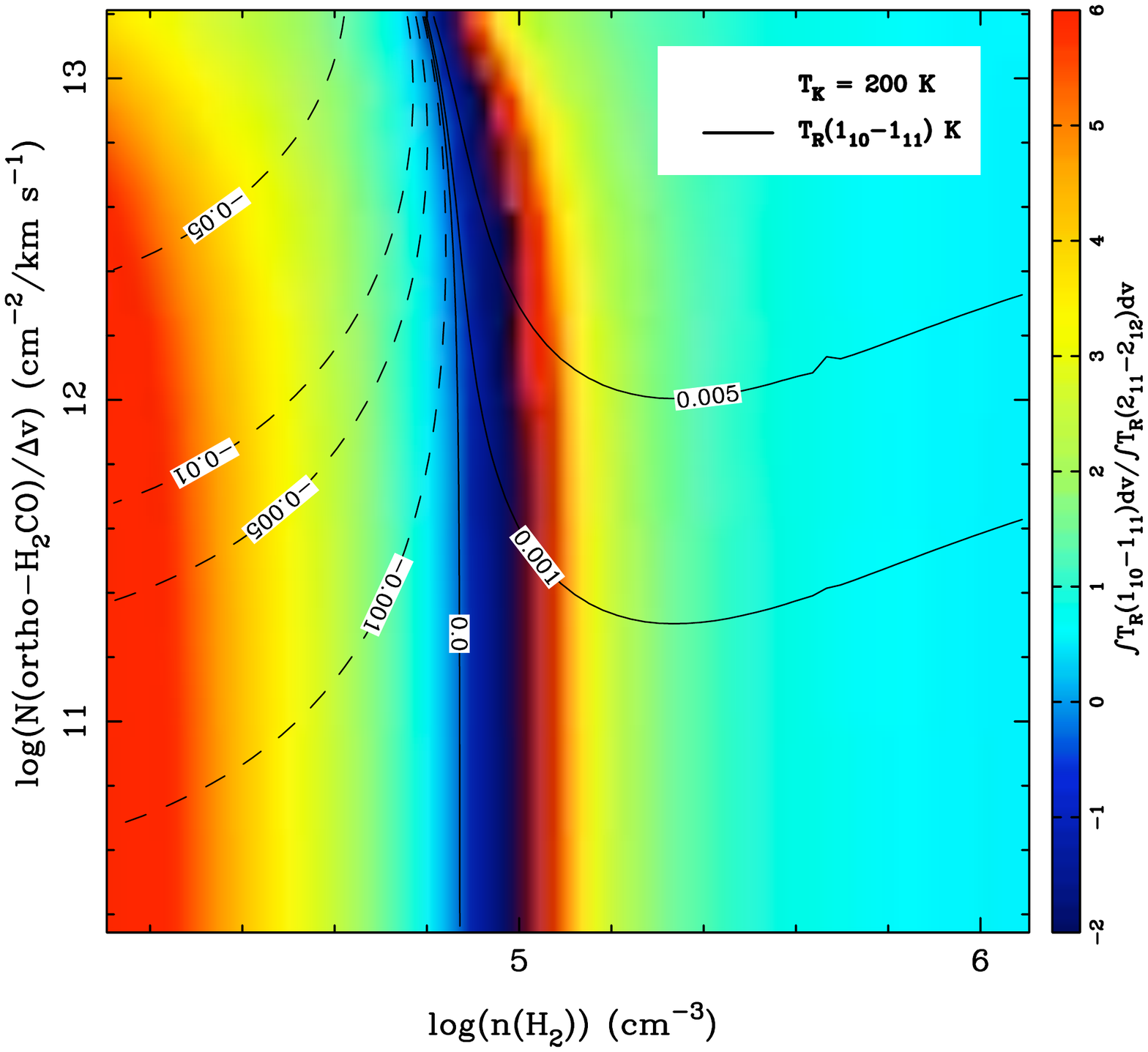}}  
\caption{LVG model H$_2$CO $1_{10}-1_{11}$ and $2_{11}-2_{12}$ line
  ratio (color) and H$_2$CO $1_{10}-1_{11}$ intensity (solid
  (positive) and dashed (negative) contours, in K) predictions as functions
  of the spatial density (n(H$_2$)) and ortho-H$_2$CO column density
  per velocity gradient (N(ortho-H$_2$CO)/$\Delta$v).  LVG model
  predictions at kinetic temperatures of 50~K (top) and 200~K
  (bottom) are shown (T$_c$ = 0).  Note how the predicted spatial
  density for a given H$_2$CO line ratio decreases (increases) as the
  kinetic temperature increases (decreases).}
\label{fig:ExgalH2COModPlotTk}
\end{figure}

 As we now have NH$_3$ measurements of the dense gas kinetic
temperature in 11 galaxies with a total of 14 velocity components
\citep{Mangum2013}, sampled over similar though slightly smaller
($\theta_B \simeq 30^{\prime\prime}$) spatial scales to our H$_2$CO
measurements, we have used these values 
to constrain our LVG models and derive revised 
spatial densities and ortho-H$_2$CO column densities.
Figure~\ref{fig:ExgalH2COModFits} shows the LVG model-predicted spatial
densities and ortho-H$_2$CO column densities for these 11 galaxies (and
their individual velocity components, as appropriate).  In all 11
galaxies with a total of 14 velocity components the measured kinetic
temperature is significantly higher than 
the previously-assumed T$_{dust}$, thus driving the derived spatial
densities significantly lower.  The correlation between higher kinetic
temperature and lower spatial density is shown in Figure
\ref{fig:ExgalH2COModPlotTk}.  The range of best-fit values for
spatial density and column density for each galaxy listed in
Table~\ref{tab:lvg} and shown in Figure~\ref{fig:ExgalH2COModFits} are
driven by signal-to-noise limitations of our H$_2$CO measurements and
the uncertainties in our derived kinetic temperature measurements.

\citet{Mangum2008} note, based on LVG model-derived spatial densities
which assumed T$_K$ = T$_{dust}$, that there appeared to be a
correlation between infrared luminosity and spatial density.  This
was purported to be another representation of the $L_{IR}-M_{dense}$
correlation \citep{Gao2004b}.  The revised spatial density
measurements, which include appropriate dense gas kinetic temperature
measurements, now point to a relatively narrow range in spatial
density of $10^{4.5}-10^{5.5}$~cm$^{-3}$ in our starburst galaxy
sample. We should stress that while the assumed kinetic temperature
influences the spatial density that we derive from our H$_2$CO
measurements, the density-kinetic temperature anti-correlation is
dramatically smaller in parameter space than most other molecular
tracers; only 1.2 dex in spatial density
(Figure~\ref{fig:ExgalH2COModFits}).  In \S\ref{DensityIndependent} we
analyze the implications of the narrow range in spatial density
derived for our galaxy sample.

\subsubsection{H$_2$CO Excitation Pumping by Continuum Emission}
\label{Maser}

As was noted in \citet{Mangum2008}, the H$_2$CO $1_{10}-1_{11}$
transition can be driven into a maser excitation state by bright and
compact nonthermal continuum emission at frequencies near 4.8\,GHz
\citep{Baan1986}, producing anomalously bright H$_2$CO emission.
Analysis of the potential for maser amplification
of the H$_2$CO $1_{10}-1_{11}$ transition in Arp\,220, in light of the
fact that the $2_{11}-2_{12}$ transition is observed in absorption,
suggest either that the $1_{10}-1_{11}$ transition is not masing or
that there are differences in the small-scale structure traced by the
two K-doublet H$_2$CO transitions.  Using the simplest possible
interpretation, we have so far relied on beam averaged
quantities.  While such an approach is reasonable for first order
estimates, it ignores the potential impact of masers, distorting the
picture outlined above. With our LVG models, all data could be
successfully simulated without having to adopt population
inversion. However, low-spatial resolution data alone do not provide a
tool to discriminate safely between quasi-thermal and maser
emission. Higher spatial resolution (\ie\ using the VLA)
measurements will make it possible to clarify the influence of maser
emission on the few extragalactic sources of H$_2$CO K-doublet
emission.

\subsection{H$_2$CO Luminosity and Dense Gas Mass}
\label{H2COLuminosityandMass}

The correlation between the infrared continuum and molecular spectral
line luminosity can be used to characterize the infrared emission power
source in these objects.  HCN \citep{Gao2004a,Gao2004b} and HCO$^+$
\citep{GraciaCarpio2006} extragalactic emission surveys have shown
this correlation to be very good, suggesting that star formation is
the main power source for the large infrared luminosities observed in
these galaxies.  Following the formalism for calculating molecular
spectral line luminosity and dense gas mass described in
\citet{Mangum2008}, we have derived L$_{IR}$ \citep[][
  Equation~12]{Mangum2008} from existing infrared measurements and
M$_{dense}$(H$_2$CO) \citep[][ Equation~11]{Mangum2008} from our
H$_2$CO measurements (Table~\ref{tab:luminosity}):

\begin{align}
& M_{dense} = N_{mol} \Omega_s D^2_A \frac{\mu m_{H_2}}{X_{mol}} \nonumber \\ 
         &= 1.125\times10^5~\pi\theta^2_s(arcsec)\times \nonumber \\
         & \qquad\qquad\qquad \frac{D^2_L(Mpc) N_{mol}(cm^{-2}) \mu
    m_{H_2}(gm)}{4\ln(2) (1+z)^{2} X_{mol}}~M_{\odot} \nonumber \\ 
         &= \frac{5.79~\theta^2_s(arcsec) D^2_L(Mpc)
                 N_{mol}(\times 10^{10} cm^{-2})}{(1+z)^{2}
                 X_{mol}(\times 10^{-9})}~M_{\odot}
\label{eq:mdense}
\end{align}

\noindent{where} $\Omega_s$ is the solid angle of a gaussian source
convolved with the gaussian telescope beam, $D_A$ is the angular size
distance to the galaxy, $m_{H_2}$ is the mass of molecular hydrogen,
$\mu$ accounts for the mass fraction due to He (1.36), $X_{mol}$
is the abundance (relative to H$_2$) of the molecule, and uniform
filling of the molecular volume is assumed.  Where two
calculations of M$_{dense}$(H$_2$CO) are listed the two calculations
of N(ortho-H$_2$CO)/$\Delta$v listed in Table~\ref{tab:lvg} have been
used.

We also list 
M$_{dense}$(HCN) derived from HCN measurements \citep{Gao2004a}.  Note
that our M$_{dense}$(H$_2$CO) calculations assume X(H$_2$CO) =
$10^{-9}$ and include a conservative 50\% uncertainty to our
assumed source sizes and H$_2$CO abundances.  Due to the lack of
H$_2$CO source structure information our H$_2$CO-derived M$_{dense}$
values are only estimates to be compared with other molecule-derived
dense gas masses.

\begin{deluxetable*}{lllll}
\tabletypesize{\scriptsize}
\tablewidth{0pt}
\tablecolumns{5}
\tablecaption{Infrared Luminosities and Dense Gas Masses\label{tab:luminosity}}
\tablehead{
\colhead{Galaxy} &
\colhead{Size\tablenotemark{a}} & 
\colhead{L$_{IR}$\tablenotemark{b}} &
\colhead{M$_{dense}$(HCN)\tablenotemark{c}} &
\colhead{M$_{dense}$(H$_2$CO)\tablenotemark{d,e}} \\
& \colhead{(arcsec)} &
\colhead{($10^{10} L_\odot$)} & 
\colhead{($10^8 M_\odot$)} &
\colhead{($10^8 M_\odot$)}
}
\startdata

NGC\,55         & $\sim15$ & 0.05$\pm$0.01 & \nodata & $<0.03$ \\

NGC\,253        & $39\times12$ & 3.41$\pm$0.24 & 2.7 & 9.68$\pm$6.97 (T$_K$ = 78$\pm$22\,K) \\
                & $39\times12$ & \nodata & \nodata & $<19.31$ (T$_K$ $>150$\,K) \\

IC\,1623        & $\sim2$ & 45.10$\pm$3.20 & 0.85 & $<0.37$ \\

NGC\,520        & $7.1\times2.9$ & 8.22$\pm$0.57 & \nodata & $<0.80; 1.56\pm1.14$ \\

NGC\,598        & $\sim15$ & 0.13$\pm$0.05 & \nodata & $<0.004$ \\

NGC\,660        & $\sim15$ & 3.03$\pm$0.22 & $>2.6$ & 82.04$\pm$124.62
(T$_K$ = 160$\pm$96\,K) \\

                & $\sim15$ & 3.03$\pm$0.22 & $>2.6$ & 31.92$\pm$25.57
(T$_K$ = T$_d$ = 37\,K)  \\

IR\,01418+1651  & $\sim 15$ & 36.33$\pm$2.72 & \nodata & $<38.97$ \\

NGC\,695        & $\sim15$ & 42.01$\pm$3.03 & 43.0 & $<78.58$ \\

Mrk\,1027       & $\sim15$ & 23.66$\pm$1.92 & 18.9 & $<105.23$ \\

NGC\,891        & $20\times10$ & 2.25$\pm$0.16 & 2.5 & $<1.01; 1.98\pm1.47$ \\

NGC\,925        & $\lesssim15$ & 0.26$\pm$0.02 & \nodata & $<0.73$ \\

NGC\,1022       & $<60$ & 2.17$\pm$0.15 & 2.0 & $<42.70$ \\

NGC\,1055       & $<60$ & 1.72$\pm$0.12 & $<3.7$ & $<14.10$ \\

Maffei\,2       & $\sim 15$ & 0.31$\pm$0.05 & \nodata & 0.33$\pm$0.25 \\

Maffei\,2\,C1   & $\sim 15$ & \nodata & \nodata & 0.42$\pm$1.55 \\

Maffei\,2\,C2   & $\sim 15$ & \nodata & \nodata & 0.22$\pm$0.36 \\

NGC\,1068       & $\sim 15$ & 22.72$\pm$1.64 & 36 & $<1.49$ \\

UGC\,02369      & $<60$ & 39.15$\pm$2.85 & \nodata & $<1529.50$ \\

NGC\,1144       & $\sim 15$ & 23.86$\pm$2.16 & 27 & $<807.37; 2028.01\pm1437.06$ \\

NGC\,1365       & $\sim 15$ & 14.31$\pm$1.00 & 31 & $<6.08; 8.01\pm5.78$ \\

IR\,03359+1523  & $<60$ & 29.49$\pm$2.61 & \nodata & $<2360.18$ \\

IC\,342         & $10\times15$ & 1.01$\pm$0.07 & 4.7 & 0.14$\pm$0.12 \\

                & $10\times15$ & \ldots & \ldots & 0.16$\pm$0.16 (T$_K$ = 24$\pm$7\,K) \\

                & $10\times15$ & \ldots & \ldots & $<0.17$ (T$_K >$ 150\,K) \\

NGC\,1614       & $1.0\times0.8$ & 40.36$\pm$2.84 & 12.6 & $<0.17$ \\

VIIZw31        & $\sim15$ & 82.62$\pm$6.08 & 98.0 & $<328.43$ \\

NGC\,2146       & $5\times20$ & 12.56$\pm$1.36 & 8 & 6.19$\pm$4.70 \\

NGC\,2623       & $0.5\times0.5$ & 35.13$\pm$2.49 & \nodata & $<0.04$ \\

Arp\,55         & $0.5\times0.5$ & 46.99$\pm$3.59 & 38 & $<4.76$ \\

NGC\,2903       & $5\times10$ & 1.23$\pm$0.09 & $>0.8$ & $<0.03$ \\

UGC\,5101       & $1.1\times0.8$ & 88.88$\pm$6.32 & 100 & $<0.56$ \\

M\,82           & $40\times15$ & 15.69$\pm$1.11 & 3.0 & 19.10$\pm$13.76 \\	

M\,82SW         & $40\times15$ & 15.69$\pm$1.11 & 3.0 & 18.68$\pm$13.66 \\	

NGC\,3079C1     & $5\times5$ & 6.93$\pm$0.50 & $\sim10$ & $<4.88$ \\

NGC\,3079C2     & $5\times5$ & 6.93$\pm$0.50 & $\sim10$ & $<10.27$ \\

NGC\,3079C3     & $5\times5$ & 6.93$\pm$0.50 & $\sim10$ & $<0.77; 0.20\pm0.15$ \\

IR\,10173+0828  & $<6$ & 61.69$\pm$4.74 & \nodata & $<926.90$ \\

NGC\,3227       & $<60$ & 1.41$\pm$0.10 & \nodata & $<911.22$ \\

NGC\,3627       & $<60$ & 1.00$\pm$0.07 & $>0.8$ & $<3.51$ \\

NGC\,3628       & $<15$ & 1.27$\pm$0.09 & 2.4 & 2.78$\pm$2.07 \\

NGC\,3690       & $<15$ & 76.96$\pm$5.40 & \nodata & $<10.87$ \\

NGC\,4631       & $<15$ & 1.60$\pm$0.11 & $\sim0.9$ & $<0.31$ \\

NGC\,4736       & $<15$ & 0.53$\pm$0.04 & \nodata & $<0.32$ \\

Mrk\,231        & $0.5\times0.3$ & 319.46$\pm$22.45 & \nodata & $<0.14$ \\

NGC\,5005       & $<60$ & 3.23$\pm$0.23 & 6.5 & $<68.60$ \\

IC\,860         & $<15$ & 11.95$\pm$0.93 & \nodata & 1604.96$\pm$1325.28 \\

NGC\,5194       & $\sim 15$ & 2.89$\pm$0.20 & \nodata & $<1.10; 1.18\pm0.85$ \\

M\,83           & $45\times15$ & 1.57$\pm$0.11 & 3.5 & 1.77$\pm$2.25 \\

Mrk\,273        & $0.4\times0.3$ & 138.79$\pm$9.71 & 152 & $<0.23; 0.26\pm0.20$ \\

NGC\,5457       & $<15$ & 1.36$\pm$0.10 & \nodata & $<0.22$ \\

IR\,15107+0724  & $\sim3$ & 19.98$\pm$1.44 & \nodata & 195.60$\pm$148.75 \\

Arp\,220        & $\sim1$ & 167.11$\pm$11.70 & 30--370 & $62.85\pm45.94$ \\

NGC\,6240       & $0.9\times0.6$ & 74.36$\pm$5.22 & 80--260 & $<0.26; 0.35\pm0.26$ \\

IR\,17208-0014  & $<1.0$ & 244.25$\pm$17.23 & 376 & $<1.00$ \\

IR\,17468+1320  & $\sim15$ & 10.08$\pm$1.67 & \nodata & $<44.61$ \\

NGC\,6701       & $<60$ & 11.80$\pm$0.82 & 28 & $<151.26$ \\

NGC\,6921       & $<60$ & 12.42$\pm$0.88 & $\sim28$ & $<1098.72; 2635.65\pm1899.15$ \\

NGC\,6946       & $5\times10$ & 1.51$\pm$0.11 & 4.9 & 0.12$\pm$0.13 \\

IC\,5179        & $\sim15$ & 15.45$\pm$1.08 & 34 & $<288.96$ \\

NGC\,7331       & $\sim15$ & 3.61$\pm$0.25 & $>4$ & $<0.87$ \\

NGC\,7479       & $\sim15$ & 6.60$\pm$0.48 & 11.2 & $<7.80$ \\

IR\,23365+3604  & $\sim15$ & 127.81$\pm$11.06 & 150 & $<307.49$ \\

Mrk\,331        & $\sim15$ & 28.10$\pm$1.97 & 33.5 & $<39.53$ \vspace{2pt}

\enddata
\tablenotetext{a}{~See \S\ref{Comparison} for galaxies detected in
  H$_2$CO.  Other source size references and measurements used are
  NGC\,55, NGC\,1022, NGC\,1055, NGC\,3627, NGC\,5005, NGC\,6701:
  \citet{Gao2004a}, HCN J=$1-0$; IC\,1623: \citet{Imanishi2007}, HCN
  and HCO$^+$ J=$1-0$; NGC\,598: \citet{Rosolowsky2007}, CO J=$1-0$;
  NGC\,925, NGC\,4736: \citet{Leroy2009}, CO J=$2-1$; UGC\,02369: this
  work; IR\,03359+1523: \citet{Sanders1991}, CO J=$1-0$; NGC\,1614,
  NGC\,2623, Arp\,55, IR\,17208-0014: \citet{Iono2009}, CO J=$3-2$ and
  HCO$^+$ J=$4-3$; VIIZw31, NGC\,3690: \citet{GraciaCarpio2008}, HCN J=$1-0$;
  IR\,10173+0828: \citet{Planesas1991}, CO J=$1-0$; NGC\,3227:
  \citet{Lisenfeld2008}, CO J=$1-0$; NGC\,4631: \citet{Israel2009}, CO
  J=$2-1$ and $3-2$; NGC\,5457: \citet{Helfer2003}, CO J=$1-0$.}
\tablenotetext{b}{~Luminosities from \citet{Sanders2003}, derived from
  IRAS fluxes over 8 to 1000\,$\mu$m.}
\tablenotetext{c}{~Except for Arp\,220 and NGC\,6240 \citep{Greve2009},
  from \cite{Gao2004a}.}
\tablenotetext{d}{~Limits assume $3\sigma$ in nondetected H$_2$CO
  transition(s).} 
\tablenotetext{e}{~When two values for M$_{dense}$(H$_2$CO) are listed
  they are derived from the two calculations of
  N(ortho-H$_2$CO)/$\Delta$v described in \S\ref{LVG} for galaxies
  with only one H$_2$CO transition.}
\end{deluxetable*}

Our M$_{dense}$ estimates range from $<4\times10^5 M_\odot$ (NGC\,598)
to $\sim 1\times10^{10} M_\odot$ (NGC\,660, NGC\,1144,
IR\,15107+0724, NGC\,6921).  In general, M$_{dense}$
derived from our H$_2$CO measurements agrees to within an
order-of-magnitude with $M_{dense}$ derived from HCN measurements
(excluding NGC\,1144, which differs by a factor of $\sim 100$).  There is an
apparent trend, though, for M$_{dense}$(H$_2$CO) $<$
M$_{dense}$(HCN).  Note that the uncertainties in many of the
quantities which go into the calculation of M$_{dense}$ limit its
accuracy to no better than an order-of-magnitude.

\citet{Gao2004a} found that the values of $M_{dense}$(HCN) are factors
of 5--200 smaller than $M(H_2)$ derived from CO measurements for their
sample of spiral, LIRG, and ULIRG galaxies, many of which are included
in our sample.  Our $M_{dense}$(H$_2$CO) values are generally smaller
than those derived using HCN.  The progression from high to low masses
derived from dense gas tracers such as CO, HCN, and H$_2$CO reflects
the hierarchical structure of the giant molecular clouds in these
galaxies \citep{Gao2004a,Gao2004b,Greve2009}.  H$_2$CO traces a
denser, more compact, component of the giant molecular clouds in our 
galaxy sample than low-excitation transitions of CO or HCN. This
result is consistent with high-resolution studies of the K-doublet
H$_2$CO emission in our own Galaxy \citep{Mangum1993a}.

\begin{figure}
\resizebox{\hsize}{!}{
\includegraphics[angle=-90,scale=0.63,trim=60pt 30pt 50pt 90pt,clip=true]{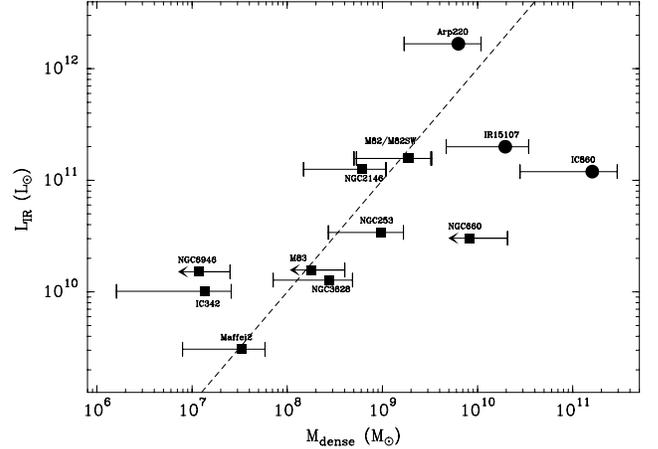}}
\caption{$L_{IR}$ versus $M_{dense}$ for the 11 galaxies and one galaxy
  offset position (M\,82 and M\,82SW overlap) where we have derived
  values for N(ortho-H$_2$CO)/$\Delta$v from our LVG model fits, with
  T$_K$ derived from NH$_3$ measurements \citep{Mangum2013}.
  $M_{dense}$ has been derived using Equation~\ref{eq:mdense} assuming
  $X(H_2CO) = 10^{-9}$ and uncertainties calculated assuming 50\%
  errors in source size and $X(H_2CO)$.  Filled circles and squares
  are used to indicate H$_2$CO-detected galaxies at distances $>$ and
  $\leq$ 50 Mpc, respectively.  Limit arrows indicate lower-limits
  which extend to zero.  The dotted line represents the log-linear
  relation $\log(L_{IR}) = 2 + 1.0 \log(M_{dense})$ and is not a fit
  to these data.}
\label{fig:lirmdense}
\end{figure}

In Figure~\ref{fig:lirmdense} we show $M_{dense}$ versus $L_{IR}$ for
the 11 galaxies and one galaxy offset position where we have derived
values for N(ortho-H$_2$CO)/$\Delta$v from our LVG model fits (rather
than simply limits to the ortho-H$_2$CO column density) with T$_K$
derived from NH$_3$ measurements \citep{Mangum2013}.  We also show 
in Figure~\ref{fig:lirmdense} the log-linear relation $\log(L_{IR}) =
2+1.0\log(M_{dense})$.  Note that given the uncertainties regarding
the size of the dense gas distribution we have not performed a formal
fit of this relation.  This correlation, first noted by
\citet{Solomon1992}, reflects the correlation between the infrared
luminosity and the amount of material to form stars in galaxies.  This
in turn leads directly to the Kennicutt-Schmidt laws, which
relate the star formation rate to the mass of gas available to produce
stars \citep{Schmidt1959,Schmidt1963} in galaxies.  Since $M_{dense}
\propto L_{mol}$, the suggestion of a linear correlation between
$L_{IR}$ and $M_{dense}(H_2CO)$ implies that, similar to HCN, H$_2$CO
traces the dense star-forming gas in starburst galaxies.

As was alluded to in \citet{Mangum2008}\footnote{Equation 14 in
  \citet{Mangum2008} presents a similar relation for M$^\prime_{dense}$ that has
  two errors: a missing factor of $\frac{1}{(1+z)^2}$ and an incorrect
  exponent to D$_L$.  Equation 14 in \citet{Mangum2008} is also not properly
  normalized.}, the dense gas mass can also be 
calculated using the spatial density n(H$_2$) rather than the
molecular column density N$_{mol}$:

\begin{align}
& M^\prime_{dense} = n(H_2) h \Omega_s D^2_A m_{mol} \nonumber \\ 
       &= 3.47\times10^{23}~\pi\theta^2_s(arcsec)\times\nonumber \\
       & \qquad\quad \frac{D^2_L(Mpc)
          n(H_2)(cm^{-3}) h(pc) \mu m_{H_2}(gm)}{4\ln(2)
          (1+z)^{2}}~M_{\odot} \nonumber \\
       &= \frac{1.31~\theta^2_s(arcsec) D^2_L(Mpc)
          n(H_2)(cm^{-3}) h(pc)}{(1+z)^{2}}~M_{\odot}
\label{eq:mdenseprime}
\end{align}

\noindent{where} we have assumed that the volume of emitting gas is a
uniformly-filled face-on disk with height $h$, $\mu$ = 1.36, and
$m_{mol}$ is the mass of an individual molecule.  Comparing the mass
derived from 
Equation~\ref{eq:mdense} (M$_{dense}$) with M$^\prime_{dense}$ allows
for a consistency check of our dense gas mass calculation.  Comparing
the terms in Equations~\ref{eq:mdense} and \ref{eq:mdenseprime}, this
consistency check amounts to a check of our assumptions regarding the
H$_2$CO abundance (X$_{mol}$) and the dense gas emitting volume
thickness $h$.  Assuming $h = 10$~pc (similar in size to the GMC
structures identified in IC\,342 by \cite{Meier2011}) for all galaxies
we find that for 
the ten galaxies which have both N$_{mol}$ and n(H$_2$) measurements
that our calculations of M$^\prime_{dense}$ and M$_{dense}$
(Table~\ref{tab:mdensecompare}) agree to
within a factor of $\sim 2$ for NGC\,253, NGC\,660, M\,82, and
IR\,15107+0724.  For the other seven galaxies M$^\prime_{dense}$ and
M$_{dense}$ agree to within a factor of $\sim 20$, with
M$^\prime_{dense}$ larger than M$_{dense}$ in all but one galaxy
(Arp\,220).  To bring these two 
dense gas mass estimates into agreement we can either lower the dense
gas emitting volume thickness $h$ or lower the H$_2$CO abundance by a
factor of $\sim 20$.  As our estimate for $h = 10$\,pc is consistent
with the GMC-scale structures imaged at high-spatial resolution in
several nearby starburst galaxies (\eg\ IC\,342; \cite{Meier2011},
M\,82; \cite{Carlstrom1991}), the more plausible option appears to be
to lower the H$_2$CO abundance to $\sim 10^{-10}$ (recall that our
assumed X$_{mol}$ for all calculations above is $10^{-9}$).  Note that
models of molecular abundances in starburst galaxies with low to
moderate cosmic ray ionization rates (the local Milky Way value of
$2\times 10^{-17}~s^{-1}$ to $5\times 10^{-16}$\,s$^{-1}$) predict H$_2$CO
abundances in the range $10^{-10} - 10^{-9}$ \citep{Bayet2011} and
$10^{-9} - 10^{-13}$ \citep[][; R.~Meijerink 2011, priv.~comm.]{Meijerink2011}.  In
these models as the cosmic ray ionization rate is increased the
H$_2$CO abundance decreases.  If mechanical heating is included
\citep{Meijerink2011} a more complicated prediction of the H$_2$CO
abundance with increasing cosmic ray ionization rate is predicted
which lies in the range $10^{-8}$--$10^{-13}$.

\begin{deluxetable}{lll}
\tabletypesize{\scriptsize}
\tablewidth{0pt}
\tablecolumns{3}
\tablecaption{Comparison of M$_{dense}$ and M$^\prime_{dense}$\label{tab:mdensecompare}}
\tablehead{
\colhead{Galaxy} &
\colhead{M$_{dense}$(H$_2$CO)\tablenotemark{a}} & 
\colhead{M$^\prime_{dense}$(H$_2$CO)\tablenotemark{b}}\\
& \colhead{($10^8 M_\odot$)} &
\colhead{($10^8 M_\odot$)}
}
\startdata

NGC\,253 (T$_K$ = 78$\pm$22\,K)  & 9.68$\pm$6.97 & 12.68$\pm$6.57 \\

NGC\,660 (T$_K$ = 160$\pm$97\,K) & 82.04$\pm$124.62 & 45.12$\pm$123.47 \\

Maffei\,2       & 0.33$\pm$0.25 & 3.31$\pm$2.18 \\

IC\,342 (T$_K$ = $24\pm7$\,K) & 0.16$\pm$0.16 & 4.05$\pm$3.37 \\

NGC\,2146       & 6.19$\pm$4.70 & 105.22$\pm$55.59 \\

M\,82           & 19.10$\pm$13.76 & 31.21$\pm$17.47 \\	

M\,82SW         & 18.68$\pm$13.66 & 42.10$\pm$26.53 \\

NGC\,3628       & 2.78$\pm$2.07 & 12.90$\pm$10.71 \\

M\,83           & 1.77$\pm$2.25 & 36.39$\pm$31.65 \\

IR\,15107+0724  & 110.00$\pm$83.65 & 49.92$\pm$46.51 \\

Arp\,220        & 62.85$\pm$45.94  & 1.46$\pm$0.96 \\

NGC\,6946       & 0.12$\pm$0.13 & 3.11$\pm$3.50 \vspace{2pt}

\enddata
\tablenotetext{a}{Derived from Equation~\ref{eq:mdense}.}
\tablenotetext{b}{Derived from Equation~\ref{eq:mdenseprime}.}
\end{deluxetable}

\section{Density-Independent Star Formation}
\label{DensityIndependent}

As was noted in \S\ref{LVGTk}, the measured mean spatial density
within the starburst environments of our sample of 13 galaxies
with measured H$_2$CO $1_{10}-1_{11}$ and $2_{11}-2_{12}$ emission or
absorption lies within a rather limited range of
$10^{4.5}-10^{5.5}$~cm$^{-3}$.  These galaxies span a considerable variety in
kinetic temperature (T$_K$ $\simeq 30-300$\,K), infrared luminosity
($L_{IR} \simeq 10^{9.5} - 10^{12.5} L_\odot$), and dense gas mass
($M_{dense} \simeq 10^7 - 10^{11} M_\odot$), representing a wide range
of starburst environments.  The physical size scales over which these
measurements apply range from $<0.3$ to $\sim 20$\,kpc based on our spatial
resolution of 51$^{\prime\prime}$ to 153$^{\prime\prime}$, sample
galaxy distances of 1 to 200\,Mpc (see Table~\ref{tab:galaxies}) and
unknown single dish beam filling factors.  The relatively narrow range
of the mean spatial 
density within our starburst galaxy sample suggests that spatial
density is not a driver of the star formation process in the most
luminous starburst galaxies.  Furthermore, a relatively constant mean
spatial density implies that the Schmidt-Kennicutt relation between
L$_{IR}$ and M$_{dense}$ is a measure of the dense gas mass
\textit{reservoir} available to form stars, and not a reflection of a
higher spatial density in the most luminous and massive starburst
galaxies. 

We should point out that the relatively narrow range of spatial
densities derived in our starburst galaxy sample is not due to
limitations of our densitometry technique.  The H$_2$CO K-doublet
transitions we have used in this analysis absorb the cosmic microwave
background at n($H_2$) $\lesssim 10^{5.5}$ cm$^{-3}$ and appear in
emission at higher spatial densities (see \S\ref{H2coProbe}).  For
typical H$_2$CO abundances the H$_2$CO K-doublet transitions are
excited at n($H_2$) $\gtrsim 10^{4}$ cm$^{-3}$.  As there is no upper
limit to the spatial density sensitivity of these H$_2$CO transitions,
biased sensitivity to spatial density does not affect our spatial
density measurements.  Observational support for this lack of bias to
spatial density exists for both lower \citep{Zeiger2010} and
higher \citep{McCauley2011,Ginsburg2011} spatial densities than those
derived in this study.

\section{Conclusions}
\label{Conclusions}

Measurements of the $1_{10}-1_{11}$ and $2_{11}-2_{12}$
K-doublet transitions of H$_2$CO toward a sample of 56 starburst
galaxies have been used to study the physical conditions within the
active starburst environments of these galaxies.  By applying
detections of at least one of these K-doublet H$_2$CO transitions
toward 21 of these starburst galaxies an estimate to the spatial
density (n(H$_2$)) and dense gas mass within each of these starburst
galaxies has been derived.  For the 13 galaxies where both K-doublet
H$_2$CO transitions have been detected we have derived accurate
\textit{measurements} of the mean spatial density.  By applying an
appropriate measurement of the dense gas kinetic temperature in 11 
of the galaxies in our sample we have improved the spatial density 
measurement procedure presented in \citet{Mangum2008}.  Furthermore,
all of our K-doublet H$_2$CO measurements can be fit to our LVG model
assuming quasithermal exciation.  This fact does not entirely exclude
potential maser emission, as noted in \cite{Mangum2008}.

Using measured kinetic temperatures has narrowed the range of derived
spatial densities presented in \citet{Mangum2008} to 10$^{4.5}$ to 10$^{5.5}$
cm$^{-3}$ in our starburst galaxy sample.  This improvement to the
spatial density measurement in our starburst galaxy sample has removed
the trend between $L_{IR}$ and our derived $n(H_2)$ noted in
\citet{Mangum2008}. 
That trend, which suggested that starburst galaxies with higher
spatial densities also possessed higher IR luminosities, was
indicative of another representation of the $L_{IR}$-$M_{dense}$
correlation.  Instead, our results now imply that the
Schmidt-Kennicutt relation between L$_{IR}$ and M$_{dense}$:
\begin{itemize}
\item Is a measure of the dense gas mass reservoir available to form
  stars, and
\item Is not directly dependent upon a higher average density driving
  the star formation process in the most luminous starburst galaxies.
\end{itemize}
This extension of the characterization of the spatial density in
starburst galaxies presented in \citet{Mangum2008} produces the most
accurate measurements of this important physical quantity in starburst
galaxies made to-date.

As was done in \citet{Mangum2008}, we have also used our H$_2$CO
measurements to derive a measure of the dense gas mass which ranges from
$0.14-110\times10^8 M_\odot$.  By comparing this traditional measure
of the dense gas mass to that derived using our spatial density
measurements we find agreement to 
within a factor of $\sim 20$.  The most extreme differences between
these two methods of calculating the dense gas mass using H$_2$CO can
best be reconciled with a modification of the assumed H$_2$CO
abundance of X(H$_2$CO) = $10^{-9}$ toward several of the galaxies in
our sample.  Recent 
modeling of the molecular abundances in starburst galaxies suggest
that low to moderate cosmic ray ionization rates \citep{Bayet2011} and
mechanical heating \citep{Meijerink2011} can affect the
H$_2$CO abundance in starburst environments.  Furthermore, comparison
of our measures of the dense gas mass using H$_2$CO with those derived
using HCN suggest that the H$_2$CO K-doublet transitions trace a
denser, more compact, component of the giant molecular clouds in our
starburst galaxy sample than low-excitation transitions of CO or HCN.

\acknowledgments

The GBT staff were characteristically helpful
and contributed significantly to the success of our observing
program.  Ben Zeiger provided assistance with a preliminary analysis
of these data.  We also thank our anonymous referee for providing
several very good comments and suggestions which significantly
improved this presentation.  
We acknowledge the support of the NSF through grant AST-0707713.
This research has made use of the NASA/IPAC Extragalactic Database (NED) 
which is operated by the Jet Propulsion Laboratory, California Institute 
of Technology, under contract with the National Aeronautics and Space
Administration.

\textit{Facilities:} \facility{GBT}


\end{document}